\documentclass[12pt, draftclsnofoot, onecolumn]{IEEEtran}

\usepackage{graphicx}
\usepackage{mathrsfs}
\usepackage{stfloats}
\usepackage{dsfont}
\usepackage{setspace}
\usepackage{array}
\usepackage{bbm}
\usepackage{hyperref}
\usepackage{here}
\usepackage{amsmath,amsthm,amsfonts,amssymb}
\usepackage{color}
\usepackage{accents}

\usepackage{epstopdf}
\usepackage[centerlast,small]{caption}
\usepackage{subcaption}
\usepackage{bbm}
\usepackage{booktabs}
\usepackage{siunitx}
\usepackage{multirow}

\newtheorem{definition}{Definition}
\newtheorem{theorem}{Theorem}
\newtheorem{corollary}{Corollary}
\newtheorem{proposition}{Proposition}
\newtheorem{lemma}{Lemma}
\newtheorem{remark}{Remark}

\newcommand\mystepA{\mathrel{\stackrel{\makebox[0pt]{\mbox{$(a)$}}}{=}}}
\newcommand\mystepB{\mathrel{\stackrel{\makebox[0pt]{\mbox{$(b)$}}}{=}}}
\newcommand\mystepC{\mathrel{\stackrel{\makebox[0pt]{\mbox{$(c)$}}}{=}}}

\def\SS{\mathcal{S}}
\def\A{\mathbf{A}}
\def\sign{{\operatorname{sign}}}
\newcommand\myapproxA{\mathrel{\stackrel{\makebox[0pt]{\mbox{$(a)$}}}{\approx}}}
\newcommand\myapproxB{\mathrel{\stackrel{\makebox[0pt]{\mbox{$(b)$}}}{\approx}}}
\newcommand\myapproxC{\mathrel{\stackrel{\makebox[0pt]{\mbox{$(c)$}}}{\approx}}}
\newcommand\myapproxD{\mathrel{\stackrel{\makebox[0pt]{\mbox{$(d)$}}}{\approx}}}
\newcommand{\vect}[1]{\mathbf{#1}}

\DeclareMathOperator{\Hessian}{Hess}
\DeclareMathOperator{\sinc}{sinc}

\begin{document}
	
	\title{\Large On the Path-Loss of Reconfigurable Intelligent Surfaces:\\ An Approach Based on Green's Theorem Applied to Vector Fields \vspace{-0.25cm}}
	
	\author{\normalsize F.~H.~Danufane, \normalsize M.~Di~Renzo,~\IEEEmembership{\normalsize Fellow,~IEEE}, \normalsize J. de Rosny, and S. Tretyakov,~\IEEEmembership{\normalsize Fellow,~IEEE} \vspace{-2cm}
		
		%\thanks{Manuscript received July 13, 2020. F. H. Danufane and M. Di Renzo are with Universit\'e Paris-Saclay, CNRS and CentraleSup\'elec, Laboratoire des Signaux et Syst\`emes, Gif-sur-Yvette, France. (e-mail: marco.direnzo@centralesupelec.fr). J. de Rosny is with Paris Sciences \& Lettres, CNRS, Institut Langevin, France. S. Tretyakov is with Aalto University, Finland.}}
		\thanks{Manuscript received July 26, 2020. F. H. Danufane, M. Di Renzo are with Universit\'e Paris-Saclay \& CNRS. (e-mail: marco.direnzo@centralesupelec.fr). J. de Rosny is with PSL \& CNRS, France. S. Tretyakov is with Aalto University, Finland.}} 
%	\markboth{Transactions on Wireless Communications} {F. H. Danufane et al.: On the Path-Loss of Reconfigurable Intelligent Surfaces: An Approach Based on Green's Theorem Applied to Vector Fields}
	
	\maketitle

\begin{abstract} \vspace{-0.5cm}
In this paper, we introduce a physics-based analytical characterization of the free-space path-loss of a wireless link in the presence of a reconfigurable intelligent surface. The proposed approach is based on the vector generalization of Green's theorem. The obtained path-loss model can be applied to two-dimensional homogenized metasurfaces, which are made of sub-wavelength scattering elements and that operate either in reflection or transmission mode. The path-loss is formulated in terms of a computable integral that depends on the transmission distances, the polarization of the radio waves, the size of the surface, and the desired surface transformation. Closed-form expressions are obtained in two asymptotic regimes that are representative of far-field and near-field deployments. Based on the proposed approach, the impact of several design parameters and operating regimes is unveiled.
\end{abstract}
	
	\vspace{-0.5cm}
	\begin{IEEEkeywords} \vspace{-0.5cm}
	Smart radio environments, reconfigurable intelligent surfaces, path-loss, Green's theorems.
	\end{IEEEkeywords}

	\vspace{-0.5cm}
	\section{Introduction}\label{sec:Introduction} \vspace{-0.25cm}
Reconfigurable intelligent surfaces (RISs) are an emerging transmission technology for application to wireless communications \cite{Marco-JSAC}. Compared with, e.g., phased arrays, multi-antenna transmitters, and relays, RISs require the largest number of scattering elements, but each of them needs to be backed by the fewest and least costly components. Also, no power amplifiers are usually needed. For these reasons, RISs constitute an emerging and promising software-defined architecture that can be realized at reduced cost, size, weight, and power (C-SWaP design). 

Motivated by recent experiments on the realization of unobtrusive transparent glasses that implement anomalous reflections and transmissions \cite{Docomo_Glass}, we aim to characterize the free-space path-loss of a planar metamaterial-based RIS whose scattering elements have sizes and inter-distances much smaller than the wavelength. Under these conditions, the RIS is homogenizable and can be modeled as a continuous surface through appropriate functions, e.g., susceptibilities, impedances. Interested readers are referred to \cite{Marco-JSAC} for further information on homogenized RISs.

Given the importance of modeling the path-loss in wireless networks in order to make appropriate link budget predictions, a few authors have recently conducted research on modeling the path-loss of RIS-aided wireless communications \cite{Tang-RIS}-\cite{Marco-SPAWC}. With the exception of our companion conference paper \cite{Marco-SPAWC} and \cite{Davide}, the available contributions are applicable to RISs made of large arrays of inexpensive antennas that are usually spaced half of the wavelength apart, and, therefore, are not homogenizable. In \cite{Tang-RIS}, the authors perform a measurement campaign in an anechoic chamber and show that the power reflected from an RIS follows a scaling law that depends on many parameters, including the size of the RIS, the mutual distances between the transmitter/receiver and the RIS (i.e., near-field vs. far-field), and whether the RIS is used for beamforming or broadcasting. In \cite{Bucheli-RIS-bridging}, the authors employ antenna theory to compute the electric field in the near-field and far-field of a finite-size RIS, and prove that an RIS is capable of acting as an anomalous mirror in the near-field of the array. The results are obtained numerically and no explicit analytical formulation of the received power as a function of the distance is given. Similar results are obtained in \cite{Ellingson-Path-loss}. In \cite{Khawaja-Coverage}, the power measured from passive reflectors in the millimeter-wave frequency band is compared against ray tracing simulations. By optimizing the area of the surface that is illuminated, it is shown that a finite-size passive reflector can act as an anomalous mirror. The studies in \cite{Emil} and \cite{Schober} rely on the assumption of plane waves and are applicable in the far-field of the RIS. The model proposed in \cite{Davide} is applicable to continuous RISs, and holds in the near-field and far-field of the RIS. However, the author focuses on charactering the available spatial degrees of freedom of two RISs communicating with each other, rather than on RISs that are utilized for reflection or transmission. In \cite{Marco-SPAWC}, we propose a path-loss model that is applicable only to one-dimensional RISs that are deployed in a two-dimensional space. Also, the approach in \cite{Marco-SPAWC} does not account for the vectorial nature of the electromagnetic waves.

Motivated by the need of accurate but tractable path-loss models in order to quantify the performance of RISs in wireless networks, we propose an approach for calculating the free-space path-loss of an RIS-aided transmission link. The proposed path-loss model leverages the vector generalization of Green's theorem \cite{Green_1828}, and it is formulated in terms of a computable integral that depends on the transmission distances, the polarization of the radio waves, the size of the RIS, and the desired surface transformations. Closed-form expressions are obtained in two asymptotic regimes that are representative of far-field and near-field transmission. Based on the proposed model, the impact of several design parameters is unveiled, and the differences and similarities between the far-field and near-field asymptotic regimes are discussed. Numerical results are illustrated and discussed in order to validate the accuracy and applicability of the asymptotic analytical formulations of the path-loss. Our study shows that the path-loss highly depends on the size of the RIS and the transmission distances, especially in the near-field regime.

The rest of this paper is organized as follows. In Section II, the system model and the modeling assumptions are introduced. In Section III, preliminary results and definitions about the asymptotic regimes of interested are given. In Sections IV and V, RISs that are configured to operate as reflecting and transmitting surfaces are analyzed, respectively. In Section VI, numerical results are illustrated to validate the obtained findings. Finally, Section VII concludes this paper.

	\begin{figure}[!t] 
		\centering
		\begin{subfigure}[b]{0.48\columnwidth}
			\includegraphics[width=0.85\linewidth]{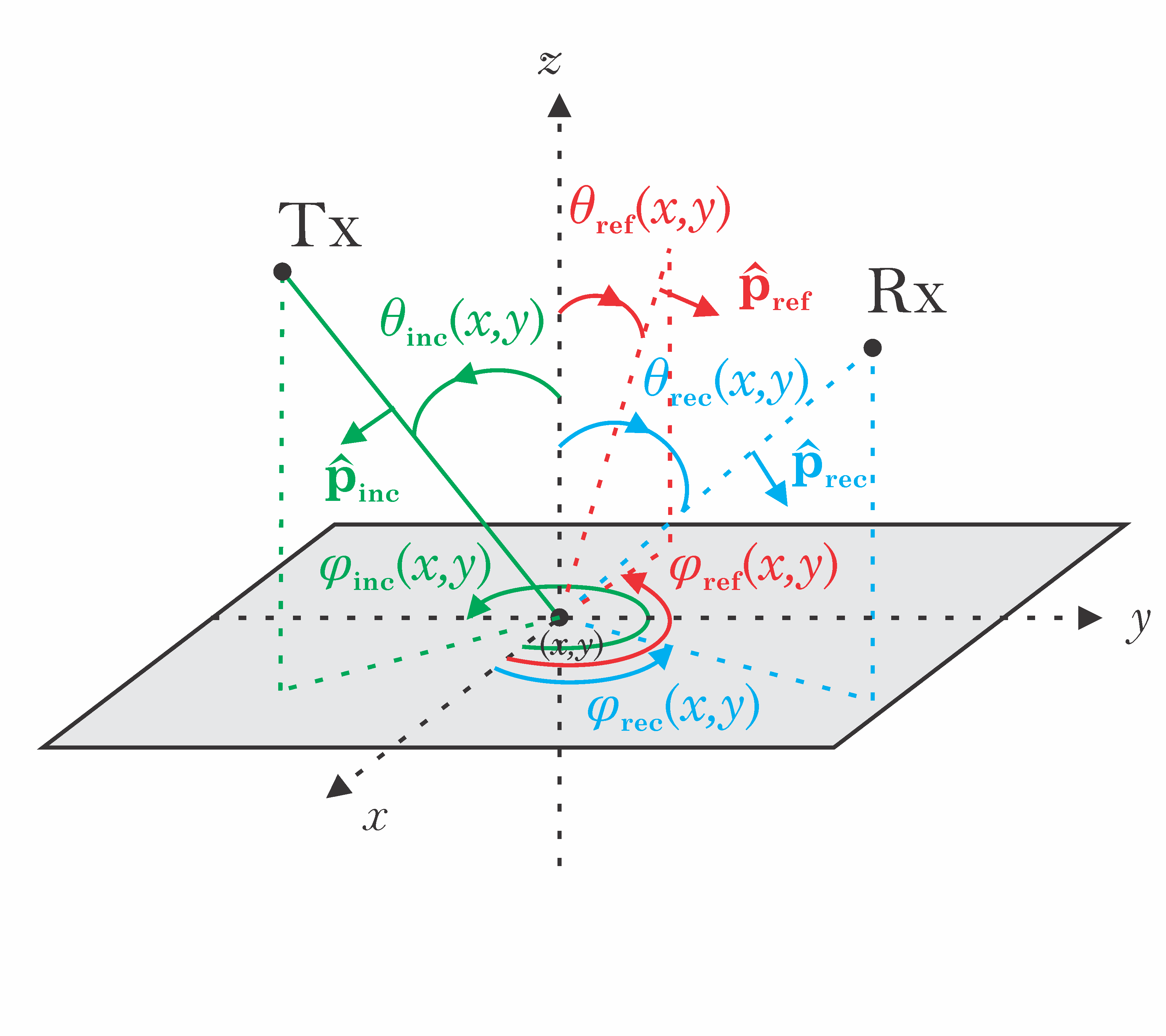}
			\vspace{-0.5cm} \caption{\vspace{-0.25cm} Tx and Rx are on the same side of the surface}
			\label{fig:systemmodelreflection}
		\end{subfigure}
		\begin{subfigure}[b]{0.48\columnwidth}
			\includegraphics[width=0.85\linewidth]{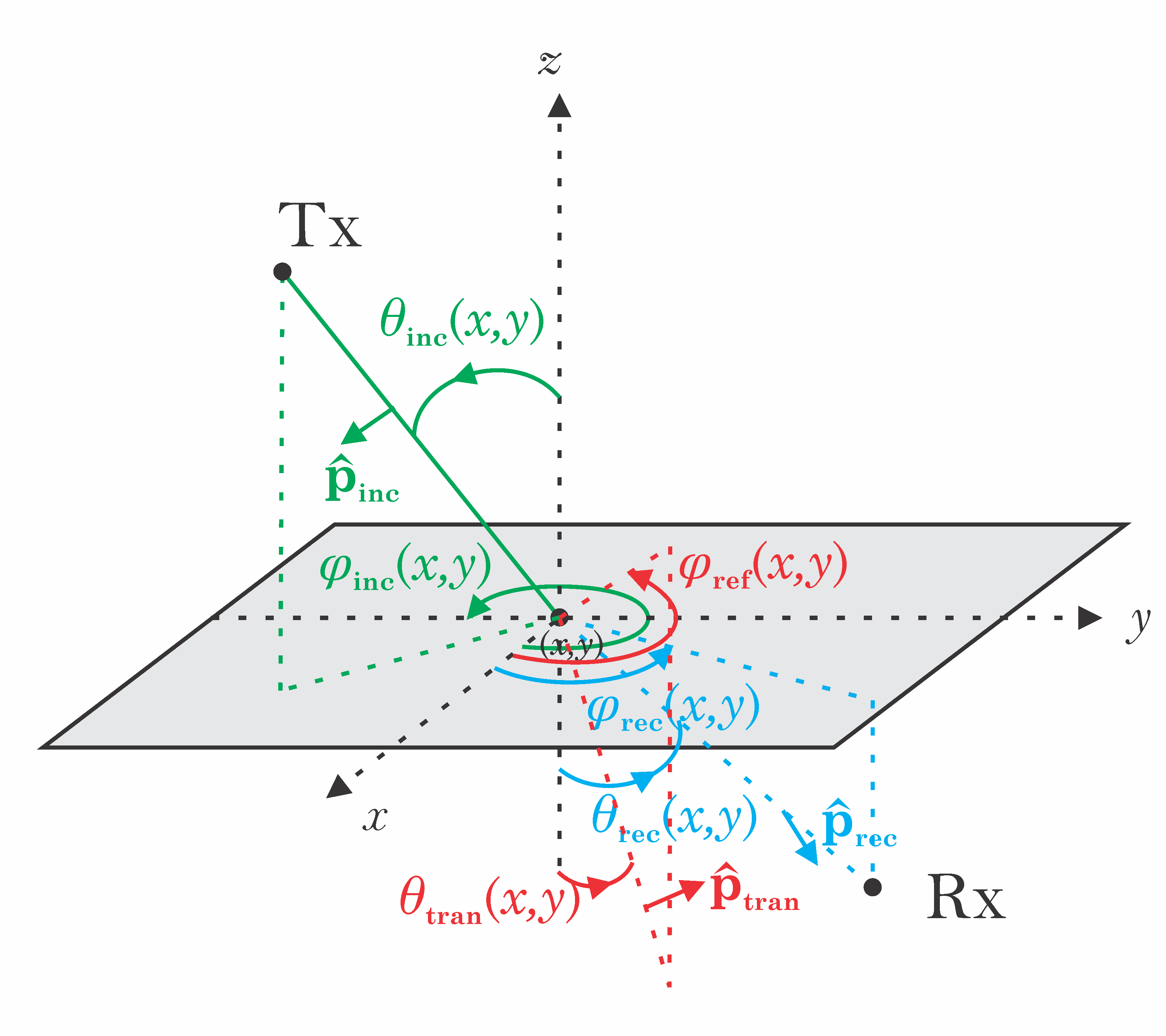}
		\vspace{-0.5cm} 	\caption{\vspace{-0.25cm} Tx and Rx are on opposite sides of the surface}
			\label{fig:systemmodeltransmission}
		\end{subfigure}
		\caption{System model.} \label{fig:systemmodel} \vspace{-0.75cm}
	\end{figure}
	
	\vspace{-0.5cm}
	\section{System Model}\label{section:system-model} \vspace{-0.25cm}
	
	In a three-dimensional (3D) space, we consider a system that
	consists of a transmitter (Tx), a receiver (Rx), and a flat surface
	($\SS$) of zero-thickness. The surface $\SS$ is a rectangle that lies on the $xy$-plane (i.e., $z=0$) whose center is located at the origin. The sides of $\SS$ are parallel to the $x$-axis and $y$-axis and have length $2L_x$ and $2L_y$, respectively. $\SS$ is defined as follows: \vspace{-0.25cm}
	\begin{equation} \label{SurfaceDefinition}
	\SS = \left\{\vect{s} = x\hat{\vect{x}} + y\hat{\vect{y}}:|x|\leq L_x,|y|\leq L_y\right\} \vspace{-0.25cm}
	\end{equation} 
	As shown in Fig. \ref{fig:systemmodel}, Tx and Rx are located at $\vect{r}_{\textup{Tx}} = x_{\textup{Tx}} \hat{\vect{x}} + y_{\textup{Tx}} \hat{\vect{y}} + z_{\textup{Tx}} \hat{\vect{z}}$ and $\vect{r}_{\textup{Rx}} = x_{\textup{Rx}} \hat{\vect{x}} + y_{\textup{Rx}} \hat{\vect{y}} + z_{\textup{Rx}} \hat{\vect{z}}$, respectively. Without loss of generality, we assume $z_{\textup{Tx}} > 0$. As for $z_{\textup{Rx}}$, we consider two cases: (i) $z_{\textup{Rx}}>0$, i.e., Tx and Rx are located on the same side of $\SS$; and (ii) $z_{\textup{Rx}}<0$, i.e., Tx and Rx are located on opposite sides of $\SS$. In the first case, the radio wave scattered by $\SS$ towards Rx is referred to as the \textit{reflected wave}, and, thus, $\SS$ operates as a \textit{reflecting surface}. In the second case, the radio wave scattered by $\SS$ towards Rx is referred to as the \textit{transmitted wave} and, thus, $\SS$ operates as a \textit{transmitting surface}. Tx emits electromagnetic (EM) waves through the vacuum whose permittivity and permeability are $\epsilon_0$ and $\mu_0$, respectively. The EM waves emitted by Tx travel at the speed of light $c = 1/\sqrt{\epsilon_0\mu_0}$. The carrier frequency, the wavelength, and the wavenumber are denoted by $f$, $\lambda=c/f$,  and $k = 2\pi/\lambda$, respectively.
	
	For any point $\vect{s} = x\hat{\vect{x}} + y\hat{\vect{y}} \in \SS$, the Tx-to-$\SS$ and $\SS$-to-Rx distances are denoted by $d_{\textup{Tx}}(x,y) = \sqrt{(x-x_{\textup{Tx}})^2 + (y-y_{\textup{Tx}})^2 + {z_{\textup{Tx}}}^2}$ and 
	$d_{\textup{Rx}}(x,y) = \sqrt{(x_{\textup{Rx}}-x)^2 + (y_{\textup{Rx}}-y)^2 + {z_{\textup{Rx}}}^2}$, respectively. More precisely, $d_{\textup{Tx}}(x,y)$ is the radius of the wavefront of the EM wave that is emitted by Tx and intersects $\SS$ at $\vect{s}$, and $d_{\textup{Rx}}(x,y)$ is the
	radius of the wavefront of the EM wave that originates from $\SS$ at $\vect{s}$ and is observed at Rx. We define $d_{\textup{Tx}0} = d_{\textup{Tx}}(0,0)$ and $d_{\textup{Rx}0} = d_{\textup{Rx}}(0,0)$, i.e., $d_{\textup{Tx}0}$ and $d_{\textup{Rx}0}$ are the distances of Tx and Rx with respect to the center of $\SS$, respectively.
	The polar angle of the incident wave at $\vect{s}$ is denoted by 	$\theta_{\textup{inc}}(x,y) = \cos^{-1}\left(z_\textup{Tx}/d_\textup{Tx}(x,y)\right)$. It represents the smallest angle formed by the $z$-axis and the wavefront of the EM wave that originates from Tx and intersects $\SS$ at $\vect{s}$. The polar angle of the received wave
	at $\vect{r}_{\textup{Rx}}$ is denoted by $\theta_{\textup{rec}}(x,y) =  \cos^{-1}\left(|z_\textup{Rx}|/d_\textup{Rx}(x,y)\right)$. It represents the smallest angle formed by the $z$-axis and the wavefront of the EM wave that is emitted by $\SS$ at $\vect{s}$ and is observed at Rx.	The azimuth angle of incidence and reflection at $\vect{s}$ are denoted by $\varphi_{\textup{inc}}(x,y)$ and $\varphi_\textup{rec}(x,y)$, respectively. In particular, $\varphi_{\textup{inc}}(x,y)$ represents the angle formed by the $x$-axis and the projection of the EM wavefront emitted from Tx towards $\SS$ onto the $xy$-plane, and $\varphi_{\textup{rec}}(x,y)$ represents the angle formed by the $x$-axis and the projection of the EM wavefront emitted from $\SS$ towards Rx onto the $xy$-plane:
	\vspace{-0.25cm}
	\begin{align}
	\sin\varphi_{\textup{inc}}(x,y) = \frac{y_{\textup{Tx}}-y}{\sqrt{(x_{\textup{Tx}}-x)^2+(y_{\textup{Tx}}-y)^2}}, \quad
	\cos\varphi_{\textup{inc}}(x,y) = \frac{x_{\textup{Tx}}-x}{\sqrt{(x_{\textup{Tx}}-x)^2+(y_{\textup{Tx}}-y)^2}}\\
	\sin\varphi_{\textup{rec}}(x,y) = \frac{y_{\textup{Rx}}-y}{\sqrt{(x_{\textup{Rx}}-x)^2+(y_{\textup{Rx}}-y)^2}}, \quad
	\cos\varphi_{\textup{rec}}(x,y) = \frac{x_{\textup{Rx}}-x}{\sqrt{(x_{\textup{Rx}}-x)^2+(y_{\textup{Rx}}-y)^2}}
	\vspace{-0.25cm}
	\end{align}
The polar and azimuth angles of the incident and received
	waves with respect to the center of $\SS$ are denoted by $\theta_{Q0} = \theta_Q(0,0)$ and $\varphi_{Q0} = \varphi_Q(0,0)$, where $Q = \textup{\textup{inc}}$ for the incident wave and $Q = \textup{\textup{rec}}$ for the reflected or transmitted wave, respectively. Further notation is given in Table \ref{Table_Notation}.

	\begin{table*}[!t] \footnotesize
		\centering
		\caption{Main operators ($G(x,y,z)$ is a scalar function, $\vect{F} = F_x \hat{\vect{x}} + F_y \hat{\vect{y}} + F_z \hat{\vect{z}}$ is a vector field with $\vect{F}= \vect{F}(x,y,z)$ and ${F_a}= {F_a}(x,y,z)$ for $a = x, y, z$). Symbols in bold denote vectors. Unit-norm vectors are denoted by $\hat{(\cdot)}$. \vspace{-0.15cm}}
		\newcommand{\tabincell}[2]{\begin{tabular}{@{}#1@{}}#2\end{tabular}}
		\begin{tabular}{l|l} \hline
			\hspace{3.5cm} Operator & \hspace{3cm} Definition \\ \hline
			$\delta(\cdot,\cdot)$, $\Hessian (\cdot)$, $\mod (\cdot)$ & Dirac delta function, Hessian matrix, modulo operator\\
			%\hline
			$|C|$, $\angle C$ & Modulus and argument of complex number $C$ \\
			%\hline
			$\cdot$, $\times$ & Scalar product and vector product \\
			%\hline
			$\nabla^2 G(x,y,z) =\left(\frac{\partial^2 }{\partial x^2} + \frac{\partial^2 }{\partial y^2} + \frac{\partial^2 }{\partial z^2}\right) G(x,y,z) $ &{Laplacian of $G(x,y,z)$} \\
			%\hline
			$\nabla {G(x,y,z)} = \frac{\partial G(x,y,z)}{\partial x} \hat{\vect{x}}+ \frac{\partial G(x,y,z)}{\partial y} \hat{\vect{y}}+\frac{\partial G(x,y,z)}{\partial z}\hat{\vect{z}}$ & {Gradient of $G(x,y,z)$} \\
			%\hline
			$\nabla \times \vect{F}= \left(\frac{\partial F_z}{\partial y} - \frac{\partial F_y}{\partial z}\right) \hat{\vect{x}}+ \left(\frac{\partial F_x}{\partial z} - \frac{\partial F_z}{\partial x}\right) \hat{\vect{y}} + \left(\frac{\partial F_y}{\partial x} - \frac{\partial F_x}{\partial y}\right) \hat{\vect{z}} $ 
			& {Curl of $\vect{F}$} \\
			%\hline
			$\nabla \cdot \vect{F} = \frac{\partial F_x}{\partial x} + \frac{\partial F_y}{\partial y} + \frac{\partial F_z}{\partial z}$ & Divergence of $\vect{F}$\\
			%\hline
			$\vec{\nabla}^2\vect{F} = \nabla^2 F_x \hat{\vect{x}} + \nabla^2 F_y \hat{\vect{y}} + \nabla^2 F_z \hat{\vect{z}}$ & Vector Laplacian of $\vect{F}$\\
			%\hline
			$\nabla^2_{\vect{r}}G(x,y,z)$, $\nabla_{\vect{r}}G(x,y,z)$ & Laplacian and gradient of $G(x,y,z)$ evaluated at $\vect{r}$\\
			%\hline
			$\vec{\nabla}^2_{\vect{r}}\vect{F}$, $\nabla_{\vect{r}}\cdot\vect{F}$ & Vector Laplacian and divergence of $\vect{F}$ evaluated at $\vect{r}$\\
			%\hline
			$G(\vect{r}_1,\vect{r}_2) = \frac{\exp\left(-jk\left|\vect{r}_1-\vect{r}_2\right|\right)}{4\pi\left|\vect{r}_1-\vect{r}_2\right|}$ 	 & Green's function solution of \eqref{eq:Greens-function-Helmholtz} \\
			%\hline
			$f(x)|_{x=x_1}^{x=x_2} = f(x_2) - f(x_1)$ 	 & Shorthand notation\\
			%\hline
			$g(x,y)|_{x=x_1}^{x=x_2}|_{y=y_1}^{y=y_2} = g(x_2,y_2) - g(x_2,y_1) - g(x_1,y_2) + g(x_1,y_1)$ 	 & Shorthand notation \\
			\hline
		\end{tabular}
		\label{Table_Notation} \vspace{-0.75cm}
	\end{table*}

	\vspace{-0.5cm}
	\subsection{Source Modeling} \vspace{-0.25cm}
	\label{subsec:source-modeling}
	
	Tx is characterized by the charge density $\rho(\vect{r},\vect{r}_{\textup{Tx}})$ and the current density $\vect{J}(\vect{r},\vect{r}_{\textup{Tx}})$, where $\vect{r}_{\textup{Tx}}$ is the center location of Tx and $\vect{r}$ is a generic location in the 3D space. We assume that $\rho(\vect{r},\vect{r}_{\textup{Tx}})$ and $\vect{J}(\vect{r},\vect{r}_{\textup{Tx}})$ are non-zero within a volume $V_{\textup{Tx}}$ that contains $\vect{r}_{\textup{Tx}}$ and are zero elsewhere. In particular, $\rho(\vect{r},\vect{r}_{\textup{Tx}})$ and $\vect{J}(\vect{r},\vect{r}_{\textup{Tx}})$ are not independent and fulfill the charge density continuity equation \cite[Sec. IV]{stratton1939diffraction}, i.e., $\nabla_{\vect{r}} \cdot \vect{J}(\vect{r},\vect{r}_{\textup{Tx}}) + j\omega \rho(\vect{r},\vect{r}_{\textup{Tx}}) = 0$, where $\omega = 2 \pi f$. Our proposed analytical framework can be applied to general EM sources, but, to obtain concrete results, we model Tx as a dipole antenna. In this case,  $\rho(\vect{r},\vect{r}_{\textup{Tx}})$ and $\vect{J}(\vect{r},\vect{r}_{\textup{Tx}})$ are \cite[Eq. (15.5.1)]{orfanidis2016electromagnetic}:
	\vspace{-0.25cm}
	\begin{equation}\label{eq:sources}
	\rho(\vect{r},\vect{r}_{\textup{Tx}}) = - \vect{p} \cdot \nabla_{\vect{r}}  \delta (\vect{r},\vect{r}_{\textup{Tx}}), \quad
	\vect{J}(\vect{r},\vect{r}_{\textup{Tx}}) = j \omega \vect{p} \delta (\vect{r},\vect{r}_{\textup{Tx}}) \vspace{-0.25cm}
	\end{equation} 
	where $\vect{p} = p_{\textup{dm}} \hat{\vect{p}}_{\textup{inc}}$ is the electric dipole moment, 
	$p_{\textup{dm}} = |\vect{p}|$ is the modulus of the dipole moment, 
	and $\hat{\vect{p}}_{\textup{inc}} = \tilde{\vect{p}}_{\textup{inc}}e^{j\phi_{\textup{inc}}}$ is the (complex) transmit polarization vector with $\tilde{\vect{p}}_{\textup{inc}}$ being a real unit-norm vector and  $\phi_{\textup{inc}} \in [0,2\pi)$ being the phase of each component of $\hat{\vect{p}}_{\textup{inc}}$. Similar results can be obtained for other source models, e.g., small linear wire antennas \cite[Sec. 15.4]{orfanidis2016electromagnetic}. 
	
	\vspace{-0.5cm}
	\subsection{Metasurface Modeling} \vspace{-0.25cm}
We assume that the surface $\SS$ is a metamaterial-based RIS, which is electrically-large and is made of sub-wavelength reconfigurable scattering elements whose inter-distances are much smaller than the wavelength. As detailed in \cite[Sec. III-E]{Marco-JSAC}, therefore, $\SS$  is homogenizable, i.e., it can be modeled through appropriate continuous surface-averaged functions (e.g., susceptibilities), even though the RIS is made of discrete elements. More specifically, the RIS is regarded as an EM discontinuity, i.e., the total tangential components of the EM fields at the two sides ($z=0^+$ and $z=0^-$) of $\SS$ are discontinuous, and their difference is dictated by constituent equations that are referred to as generalized sheet transition conditions \cite[Fig. 17]{Marco-JSAC}. For a homogenizable metamaterial-based RIS, the relation between the reflected (transmitted) tangential components of the EM fields can be formulated in terms of inhomogeneous functions as stated in \cite[Eq. (50)]{Marco-JSAC}. Each Cartesian component of the reflected (transmitted) EM field may be formulated as a weighted linear combination of all the Cartesian components of the incident EM field. By virtue of linearity, we consider, without loss of generality, one term of the linear combination, whose corresponding inhomogeneous function is referred to as (field) local reflection or transmission coefficient if $\SS$ operates as a reflecting surface or as a transmitting surface, respectively.

In particular, the reflection (transmission) coefficient is denoted by $\widetilde \Gamma_{\textup{ref}}(\vect{s})$ ($\widetilde \Gamma_{\textup{tran}}(\vect{s})$), which is a complex function that is appropriately engineered (through the design of surface-averaged susceptibilities) in order to apply specified transformations to the impinging EM waves. Specific examples are provided in further text. As elaborated in \cite[Fig. 14]{Marco-JSAC} and detailed in further text, the surface equivalent theorem dictates that the EM field scattered by $\SS$ at any point in a 3D space can be formulated in terms of only the incident fields, $\widetilde \Gamma_{\textup{ref}}(\vect{s})$, and $\widetilde \Gamma_{\textup{tran}}(\vect{s})$ at $\vect{s} \in \SS$. 

For generality, the RIS is assumed to be capable of modifying the polarization of the impinging radio waves. More precisely, given an incident signal with polarization $\hat{\vect{p}}_{\textup{inc}}$, the polarization of the reflected and transmitted signals are denoted by $\hat{\vect{p}}_{\textup{ref}} = \tilde{\vect{p}}_{\textup{ref}}e^{j\phi_{\textup{ref}}}$ and $\hat{\vect{p}}_{\textup{tran}} = \tilde{\vect{p}}_{\textup{tran}}e^{j\phi_{\textup{tran}}}$, respectively. Similar to the definition of $\hat{\vect{p}}_{\textup{inc}}$, $\tilde{\vect{p}}_{\textup{ref}}$ and $\tilde{\vect{p}}_{\textup{tran}}$ are real unit-norm vectors and $\phi_{\textup{ref}} \in [0,2\pi)$ and $\phi_{\textup{tran}} \in [0,2\pi)$ are the phases of each component of $\hat{\vect{p}}_{\textup{ref}}$ and $\hat{\vect{p}}_{\textup{tran}}$, respectively. 
	
Based on these modeling assumptions, the electric field at any point $\vect{s} \in \SS$ on the reflection side of the RIS (i.e., $z=0^+$) can be formulated as follows:
\vspace{-0.25cm}	
	\begin{equation}\label{eq:E-surface-reflection}
	\vect{E}_{\SS}(\vect{s}) = \vect{E}_{\SS}(\vect{s}, z=0^+) =
	\vect{E}_{\textup{inc}}(\vect{s};\hat{\vect{p}}_{\textup{inc}}) + \widetilde \Gamma_{\textup{ref}}(\vect{s}) \vect{E}_{\textup{inc}}(\vect{s};\hat{\vect{p}}_{\textup{ref}}) \vspace{-0.25cm}
	\end{equation}	
where $\vect{E}_{\textup{inc}}(\vect{s};\hat{\vect{p}}_{\textup{inc}})$ is the incident field at $\vect{s}$ with polarization $\hat{\vect{p}}_{\textup{inc}}$ and	$\widetilde \Gamma_{\textup{ref}}(\vect{s}) = \Gamma_{\textup{ref}}(\vect{s}) \mathcal{E}_{\textup{ref}}(\hat{\vect{p}}_{\textup{inc}}, \hat{\vect{p}}_{\textup{ref}})$ is the reflection coefficient. To make explicit the impact of the change of polarization introduced by $\SS$, $\widetilde \Gamma_{\textup{ref}}(\vect{s})$ is formulated as the product of two terms: (i) $\Gamma_{\textup{ref}}(\vect{s})$ that is polarization-independent; and (ii) $\mathcal{E}_{\textup{ref}}(\hat{\vect{p}}_{\textup{inc}}, \hat{\vect{p}}_{\textup{ref}})$ that denotes the efficiency of the change of polarization from $\hat{\vect{p}}_{\textup{inc}}$ to $\hat{\vect{p}}_{\textup{ref}}$. In addition, $\vect{E}_{\textup{inc}}(\vect{s};\hat{\vect{p}}_{\textup{ref}})$ denotes the reflected electric field whose polarization is $\hat{\vect{p}}_{\textup{ref}}$, which is formally the same as the incident electric field except for the change of polarization.

Along the same lines and with a similar meaning of the symbols, the electric field at any point $\vect{s} \in \SS$ on the transmission side of the RIS (i.e., $z=0^-$) can be formulated as follows:
\vspace{-0.25cm}	
	\begin{equation}\label{eq:E-surface-transmission}
	\vect{E}_{\SS}(\vect{s}) = \vect{E}_{\SS}(\vect{s}, z=0^-)
	= \widetilde \Gamma_{\textup{tran}}(\vect{s}) \vect{E}_{\textup{inc}}(\vect{s};\hat{\vect{p}}_{\textup{tran}})=	
	\Gamma_{\textup{tran}}(\vect{s}) \mathcal{E}_{\textup{tran}}(\hat{\vect{p}}_{\textup{inc}}, \hat{\vect{p}}_{\textup{tran}}) \vect{E}_{\textup{inc}}(\vect{s};\hat{\vect{p}}_{\textup{tran}}) \vspace{-0.25cm}	
	\end{equation}
where we have taken into account that at $z=0^-$ there is no incident field \cite[Eq. (6)]{Marco-JSAC}. 
	
We emphasize, as detailed in \cite[Fig. 29]{Marco-JSAC}, that \eqref{eq:E-surface-reflection} and \eqref{eq:E-surface-transmission} are applicable in the far-field of the RIS microstructure, i.e., at distances from $\SS$ at which the presence of possible evanescent fields that are excited to realize RISs with high reflection and transmission efficiency can be safely ignored. In the next sections, we assume $k \gg 1/d_{\textup{Tx}}(x,y)$ and $k \gg 1/d_{\textup{Rx}}(x,y)$ that are typically fulfilled for wireless applications and allow us to ignore the presence of possible evanescent fields. The far-field of the RIS microstructure encompasses the near-field and the far-field of the RIS. These two regimes are analyzed, in detail, in further text.

\vspace{-0.5cm}	
\section{Preliminaries}\label{section:preliminary} \vspace{-0.25cm}	
In this section, we introduce a general formulation of the received EM field at Rx in the presence of $\SS$. The proposed approach adheres to the principles of physical optics and overcomes the limitations of geometric optics \cite[Sec. 8.2.1]{osipov2017modern}. Also, we introduce methods for computing recurrent integrals that characterize the EM filed scattered by reflecting and transmitting surfaces.

\vspace{-0.5cm}	
	\subsection{Received Field at Rx} \vspace{-0.25cm}	
Assuming the universal time-dependency $e^{j\omega t}$, the electric field, $\vect{E}(\vect{r})$, and magnetic field, $\vect{H}(\vect{r})$, at any location $\vect{r} \in \mathbb{R}^3$ in vacuum satisfy the differential equations \cite[Eqs. (6), (7)]{stratton1939diffraction}: 
\vspace{-0.25cm}	
	\begin{align}
	\nabla_{\vect{r}} \times \left( \nabla_{\vect{r}} \times \vect{E}(\vect{r}) \right) &= k^2 \vect{E}(\vect{r}) - j\omega\mu_0 \vect{J}(\vect{r},\vect{r}_{\textup{Tx}}) \label{eq:Helmholtz-equation-E}\\
	\nabla_{\vect{r}} \times \left( \nabla_{\vect{r}} \times \vect{H}(\vect{r}) \right) &= k^2 \vect{H}(\vect{r}) + \nabla_{\vect{r}} \times \vect{J}(\vect{r},\vect{r}_{\textup{Tx}}) \label{eq:Helmholtz-equation-H} 
	\end{align} \vspace{-1.00cm}	
	
The solutions of \eqref{eq:Helmholtz-equation-E} and \eqref{eq:Helmholtz-equation-H} are related through the relation $\vect{H}(\vect{r}) = -\nabla_{\vect{r}} \times \vect{E}(\vect{r})/(j\omega\mu_0)$. Therefore, the complete characterization of the EM field can be given only through $\vect{E}(\vect{r})$. 

In the absence of the RIS, the solution of \eqref{eq:Helmholtz-equation-E}, i.e., $\vect{E}(\vect{r})$, observed at $\vect{r}_{\textup{Rx}}$ boils down, by definition, to the incident electric field with polarization $\hat{\vect{p}}_{\textup{inc}}$. This latter electric field is denoted by $	\vect{E}_{\textup{inc}}(\vect{r}_{\textup{Rx}};\hat{\vect{p}}_{\textup{inc}})$. Using the notation in Section \ref{subsec:source-modeling}, it can be formulated as \cite[Eq. (15.3.10)]{orfanidis2016electromagnetic}:
\vspace{-0.50cm}	
	\begin{equation}\label{eq:incident-E}
	\vect{E}_{\textup{inc}}(\vect{r}_{\textup{Rx}};\hat{\vect{p}}_{\textup{inc}}) = \int_{V_{\textup{Tx}}} 	\left(- j\omega\mu_0 \vect{J}(\vect{r},\vect{r}_{\textup{Tx}})G\left(\vect{r}_{\textup{Rx}}, \vect{r}\right) + \frac{\rho(\vect{r},\vect{r}_{\textup{Tx}})}{\epsilon_0}\nabla_{\vect{r}} G\left(\vect{r}_{\textup{Rx}}, \vect{r}\right) \right)d\vect{r} \vspace{-0.25cm}	
	\end{equation}
where	$G\left(\vect{r}_{\textup{Rx}}, \vect{r}\right)$ is the Green function defined as follows \cite[Eq. (18.10.2)]{orfanidis2016electromagnetic}:
\vspace{-0.35cm}	
	\begin{equation}\label{eq:Greens-function-Helmholtz}
	\nabla_{\vect{r}_{\textup{Rx}}}^2 G(\vect{r}_{\textup{Rx}},\vect{r}) + k^2 G(\vect{r}_{\textup{Rx}},\vect{r})  = -\delta(\vect{r}_{\textup{Rx}},\vect{r}) \vspace{-0.35cm}	
	\end{equation}	

In the presence of $\SS$, $\vect{E}(\vect{r}_{\textup{Rx}})$, at any point $\vect{r}_{\textup{Rx}}$ in a volume $V \subseteq \mathbb{R}^3$, does not have a simple formulation as in \eqref{eq:incident-E}. Under the assumptions of physical optics \cite[Sec. 8.2.1]{osipov2017modern}, the field $\vect{E}(\vect{r}_{\textup{Rx}})$ solution of \eqref{eq:Helmholtz-equation-E} in the presence of $\SS$ can be characterized by using the Stratton-Chu formula \cite{stratton1939diffraction}.

	\begin{figure}[!t]
		\centering
		\begin{subfigure}[b]{0.48\columnwidth}
			\includegraphics[width=0.85\linewidth]{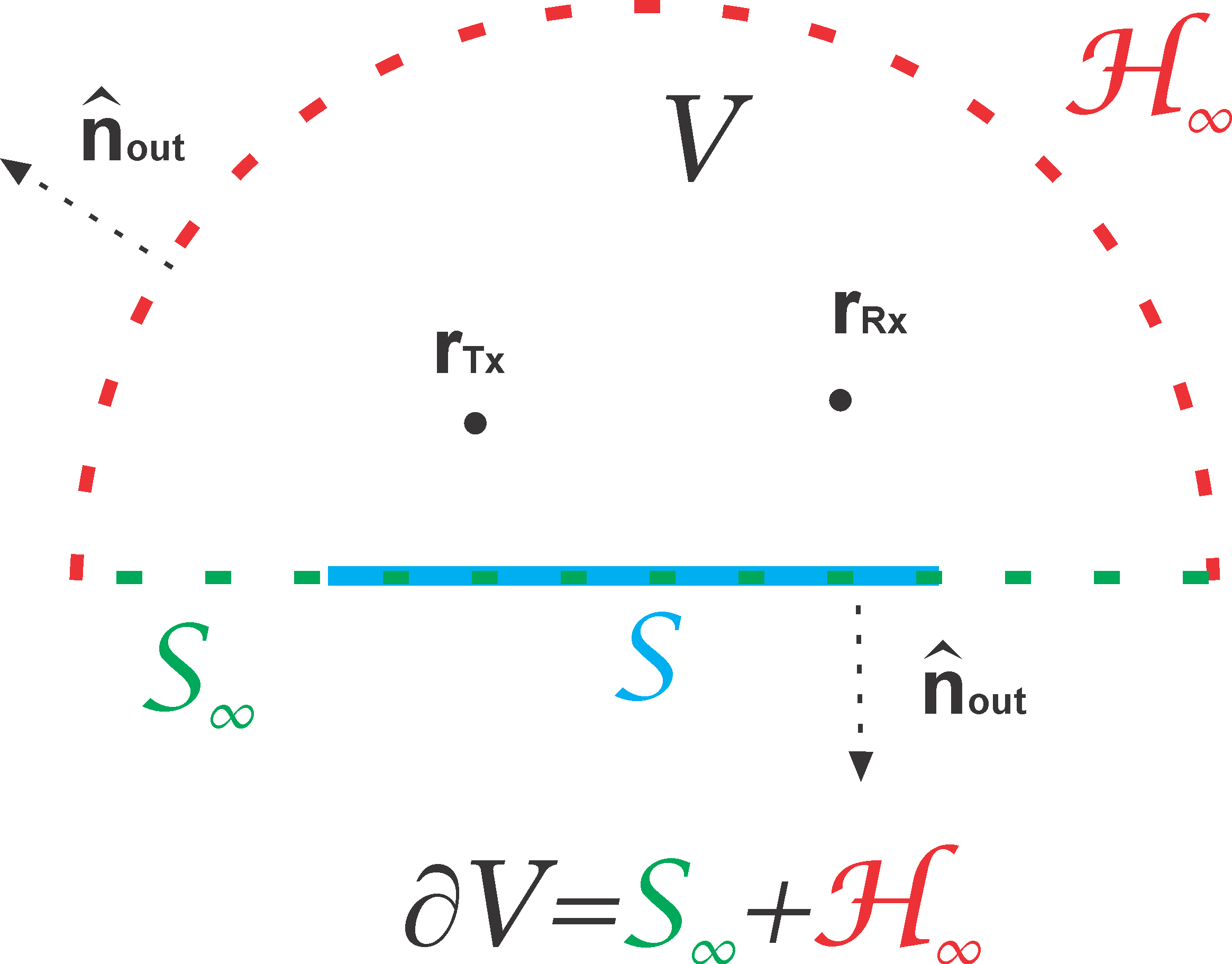}
			\caption{}
			\label{fig:kirchhoffreflection} \vspace{-0.25cm}
		\end{subfigure}
		\begin{subfigure}[b]{0.48\columnwidth}
			\centering
			\includegraphics[width=0.85\linewidth]{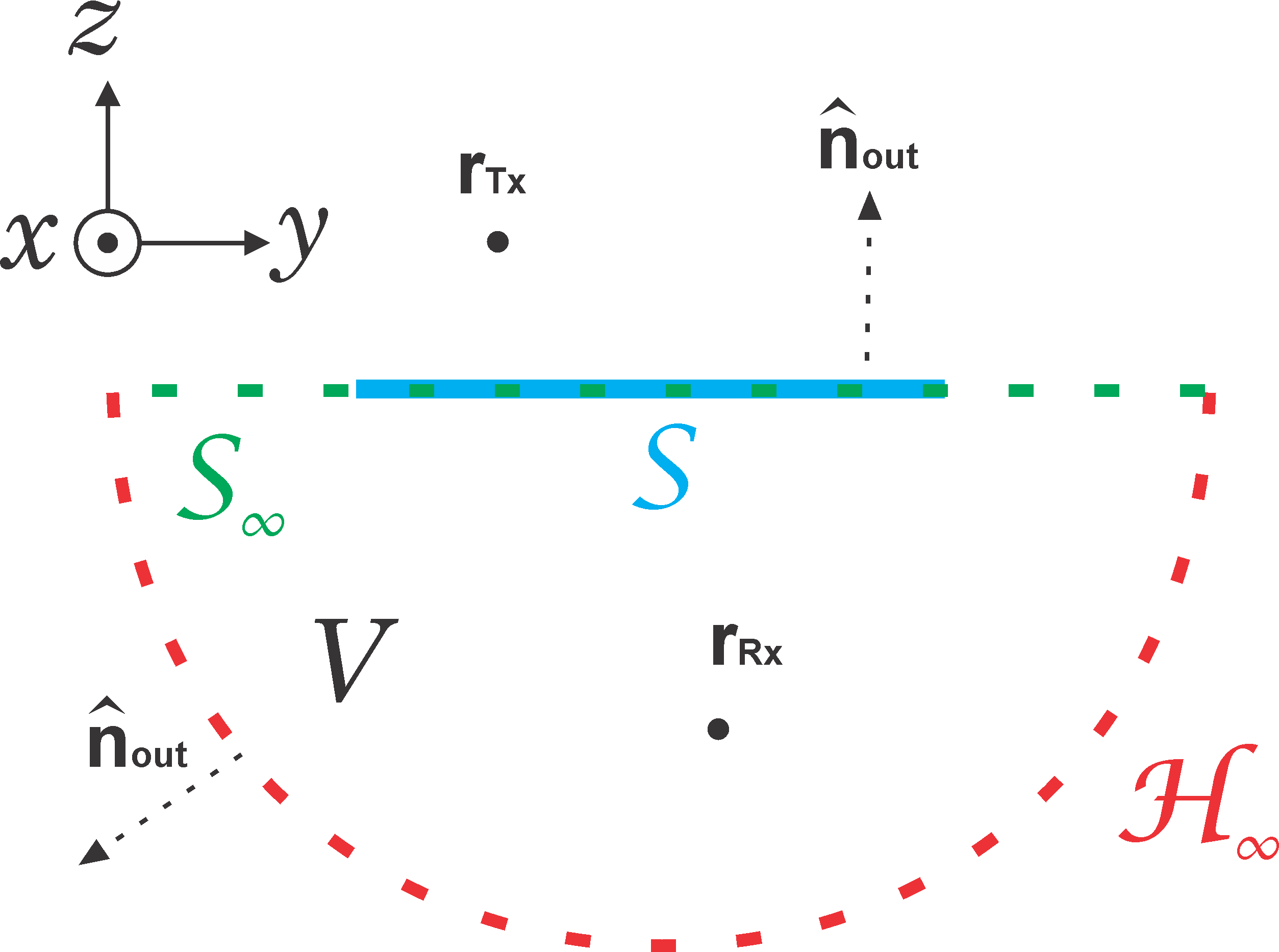}
			\caption{}
			\label{fig:kirchhofftransmission} \vspace{-0.25cm}
		\end{subfigure}
		\caption{Volume $V$ and closed boundary $\partial V$ for a reflecting (a) and transmitting (b) surface. 
		}
		\label{fig:kirchhoffboundary} \vspace{-0.75cm}
	\end{figure}
	
 \vspace{-0.35cm}
	\begin{lemma}
Let $\vect{r}_{\textup{Rx}}$ be the observation point of interest in a generic volume $V \subseteq \mathbb{R}^3$. Let $\partial V$ be a generic closed boundary of $V$ such that (see Fig. \ref{fig:kirchhoffboundary}): (i) Rx is always located inside the volume, i.e., $\vect{r}_{\textup{Rx}} \in V$; (ii) $\SS$ is part of the boundary, i.e., $\SS \in \partial V$; and (iii) Tx is located inside the volume, i.e., $\vect{r}_{\textup{Tx}} \in V$, in the reflection case and outside the volume, i.e.,  $\vect{r}_{\textup{Tx}} \notin V$ in the transmission case, respectively. Let $\vect{r'}$ be a generic point on the closed boundary $\partial V$, i.e., $\vect{r'} \in \partial V$, and let $\vect{E}_{\partial V}(\vect{r'})$ and $\vect{H}_{\partial V}(\vect{r'})$ denote the total electric and magnetic fields at $\vect{r'}$, respectively. Then, $\vect{E}(\vect{r}_{\textup{Rx}})$ solution of \eqref{eq:Helmholtz-equation-E} in the presence of $\SS$ can be formulated as follows: \vspace{-0.25cm}
		\begin{align}
		\vect{E}(\vect{r}_{\textup{Rx}}) 
		&= \mathbbm{1}_{(\vect{r}_{\textup{Tx}} \in V)}\vect{E}_{\textup{inc}}(\vect{r}_{\textup{Rx}};\hat{\vect{p}}_{\textup{inc}})
		- 
		\int_{\partial V} \left[- j\omega\mu_0\left(\hat{\vect{n}}_{\textup{out}}\times\vect{H}_{\partial V}(\vect{r'})\right)G(\vect{r}_{\textup{Rx}},\vect{r'}) \right.
		\label{eq:Stratton-Chu-electric} \nonumber\\
		& \quad
		\left.
		+ \left(\hat{\vect{n}}_{\textup{out}}\cdot\vect{E}_{\partial V}(\vect{r'})\right)\nabla_{\vect{r'}} G(\vect{r}_{\textup{Rx}},\vect{r'}) 
		+ (\hat{\vect{n}}_{\textup{out}}\times \vect{E}_{\partial V}(\vect{r'}))\times \nabla_{\vect{r'}} G(\vect{r}_{\textup{Rx}},\vect{r'}) 
		\right]d\vect{r'} \vspace{-0.5cm}
		\end{align}   
where $\hat{\vect{n}}_{\textup{out}}$ is the normal vector pointing outwards the volume and $\mathbbm{1}_{(\cdot)}$ is the indicator function.
	\end{lemma}
	\vspace{-0.6cm}
	\begin{proof}
		See \cite[Eq. (14)]{stratton1939diffraction}.
	\end{proof}
\vspace{-0.6cm}
\begin{remark}
The choice of $V$ and $\partial V$ is not unique. For convenience, Fig. \ref{fig:kirchhoffboundary} shows an example in which $V$ is the upper or lower half-plane of the 3D space and $\partial V = \mathcal{H}_{\infty} + \SS_{\infty}$, where $\mathcal{H}_{\infty}$ is the hemisphere for $z>0$ or $z<0$ with an infinite radius for a reflecting or transmitting surface, respectively, and $\SS_{\infty}$ is the entire $xy$-plane (including $\SS$). 
\end{remark}
\vspace{-0.6cm}
\begin{remark}
There exist alternative integral expressions for the solution of \eqref{eq:Helmholtz-equation-E} in the presence of $\SS$, e.g., Franz's formula \cite[Eq. (3)]{tai1972kirchhoff}. We choose \eqref{eq:Stratton-Chu-electric} as the basis of our analysis for two reasons: (i) the incident field $\vect{E}_{\textup{inc}}(\vect{r}_{\textup{Rx}};\hat{\vect{p}}_{\textup{inc}})$ explicitly appears in \eqref{eq:incident-E}, which leads to simple interpretations as elaborated in further text; and (ii) the two terms $\hat{\vect{n}}_{\textup{out}}\times \vect{E}_{\partial V}(\vect{r'})$ and $\hat{\vect{n}}_{\textup{out}}\times\vect{H}_{\partial V}(\vect{r'})$ are directly related to the magnetic and electric currents, respectively, that are induced by the incident signal in the scattering elements (i.e., the inclusions) of the metasurface \cite[Eq. (1)]{Marco-JSAC}, which provides us with explicit evidence of the physics-based phenomena that govern the operation of RISs. In particular, \eqref{eq:Stratton-Chu-electric} can be viewed as an instance of the surface equivalent theorem \cite[Fig. 14]{Marco-JSAC}.
\end{remark}

\vspace{-0.25cm}
Even though \eqref{eq:Stratton-Chu-electric} provides us with a computable integral for $\vect{E}(\vect{r}_{\textup{Rx}})$, it does not offer an explicit analytical expression that depends on $\SS$ and that yields insights on the impact of important design parameters. In the sequel, we analyze $\vect{E}(\vect{r}_{\textup{Rx}})$ in detail and compute several equivalent explicit expressions for \eqref{eq:Stratton-Chu-electric} that are useful in wireless applications and that unveil important scaling laws. To this end, we assume, without loss of generality, that Rx is equipped with an antenna whose polarization is $\hat{\vect{p}}_{\textup{rec}} = \tilde{\vect{p}}_{\textup{rec}}e^{j\phi_{\textup{rec}}}$ \cite{treuhaft2011formulating}, where $\tilde{\vect{p}}_{\textup{rec}}$ is a real unit-norm vector and $\phi_{\textup{rec}} \in [0,2\pi)$ is the common phase of the three components of $\hat{\vect{p}}_{\textup{rec}}$. In general, $\vect{E}(\vect{r}_{\textup{Rx}})$ depends on $\hat{\vect{p}}_{\textup{inc}}$, $\hat{\vect{p}}_{\textup{ref}}$ or $\hat{\vect{p}}_{\textup{tran}}$, and $\hat{\vect{p}}_{\textup{rec}}$. To explicitly highlight the impact of $\hat{\vect{p}}_{\textup{rec}}$, we reformulate \eqref{eq:Stratton-Chu-electric} as follows.
\vspace{-0.35cm}	
	\begin{theorem}\label{theorem:Stratton-Chu-equivalence}
The projection of $\vect{E}(\vect{r}_{\textup{Rx}})$ in \eqref{eq:Stratton-Chu-electric} onto the receive polarization vector $\hat{\vect{p}}_{\textup{rec}}$ is:
\vspace{-0.5cm}
	\begin{align}\label{eq:Stratton-Chu-equivalence}
	\vect{E}(\vect{r}_{\textup{Rx}}) \cdot \hat{\vect{p}}_{\textup{rec}} 
	&= \mathbbm{1}_{(\vect{r}_{\textup{Tx}} \in V)}\vect{E}_{\textup{inc}}(\vect{r}_{\textup{Rx}};\hat{\vect{p}}_{\textup{inc}}) \cdot \hat{\vect{p}}_{\textup{rec}}  \\
	&-  \int_{\partial V} \left[(\vect{E}_{\partial V}(\vect{r'}) \cdot \hat{\vect{p}}_{\text{\textup{rec}}}) \nabla_{\vect{r'}} G(\vect{r}_{\textup{Rx}},\vect{r'}) - G(\vect{r}_{\textup{Rx}},\vect{r'}) \nabla_{\vect{r'}} \left(\vect{E}_{\partial V}(\vect{r'})\cdot \hat{\vect{p}}_{\textup{rec}}\right)\right]\cdot \hat{\vect{n}}_{\textup{out}} d\vect{r'} \nonumber \vspace{-0.25cm}	
	\end{align}
	\end{theorem}
\vspace{-0.35cm}
	\begin{proof}
		See Appendix A. % \eqref{appendix:proof-of-Stratton-Chu-equivalence}.
	\end{proof}
\vspace{-0.25cm}
By appropriately choosing $\hat{\vect{p}}_{\textup{rec}}$, $\vect{E}(\vect{r}_{\textup{Rx}})$ along any directions can be retrieved, e.g., $\hat{\vect{p}}_{\textup{rec}} = \hat{\vect{x}}$, $\hat{\vect{p}}_{\textup{rec}} = \hat{\vect{y}}$, and $\hat{\vect{p}}_{\textup{rec}} = \hat{\vect{z}}$. The Stratton-Chu formula in \eqref{eq:Stratton-Chu-equivalence}, however, does not explicitly reveal the impact of $\SS$. Thus, we reformulate \eqref{eq:Stratton-Chu-equivalence} such that $\SS$, instead of $V$ and $\partial V$, appears explicitly.
\vspace{-0.35cm}
\begin{theorem}\label{theorem:Stratton-Chu-equivalence-reduced}
Let $\vect{E}_{\SS}(\vect{s})$ be the surface electric field at point $\vect{s} \in \SS$ in \eqref{eq:E-surface-reflection} and \eqref{eq:E-surface-transmission} for a reflecting and transmitting surface, respectively. Then, \eqref{eq:Stratton-Chu-equivalence} can be equivalently reformulated as follows:
\vspace{-0.25cm}
	\begin{align}\label{eq:Stratton-Chu-equivalence-reduced}
	\vect{E}(\vect{r}_{\textup{Rx}}) \cdot \hat{\vect{p}}_{\textup{rec}}
	=   \vect{E}_{\textup{inc}}(\vect{r}_{\textup{Rx}};\hat{\vect{p}}_{\textup{inc}}) \cdot \hat{\vect{p}}_{\textup{rec}} 
	&-  \int_{\SS} \left[((\vect{E}_{\SS}(\vect{s})-\vect{E}_{\textup{inc}}(\vect{s};\hat{\vect{p}}_{\textup{inc}})) \cdot \hat{\vect{p}}_{\text{\textup{rec}}}) \nabla_{\vect{s}} G(\vect{r}_{\textup{Rx}},\vect{s}) \right.\\
	&\left. 
	- G(\vect{r}_{\textup{Rx}},\vect{s}) \nabla_{\vect{s}} ((\vect{E}_{\SS}(\vect{s})-\vect{E}_{\textup{inc}}(\vect{s};\hat{\vect{p}}_{\textup{inc}})) \cdot \hat{\vect{p}}_{\text{\textup{rec}}})\right]\cdot \hat{\vect{n}}_{\textup{out}} d\vect{s} \nonumber \vspace{-0.25cm}
	\end{align}
	\end{theorem}	
\vspace{-0.5cm}
	\begin{proof}
	See Appendix B. %\eqref{appendix:proof-of-Stratton-Chu-equivalence-reduced}
	\end{proof}
\vspace{-0.25cm}

The reformulation in \eqref{eq:Stratton-Chu-equivalence-reduced} can be applied to any physical source at Tx, i.e., $\rho(\vect{r},\vect{r}_{\textup{Tx}})$ and $\vect{J}(\vect{r},\vect{r}_{\textup{Tx}})$, which determine the incident fields $\vect{E}_{\textup{inc}}(\vect{s};\hat{\vect{p}}_{\textup{inc}})$ and $\vect{E}_{\textup{inc}}(\vect{r}_{\textup{Rx}};\hat{\vect{p}}_{\textup{inc}})$, and to any field transformations applied by the RIS, i.e., $\vect{E}_{\SS}(\vect{s})$. In the following, as a concrete example, we focus our attention on a physical source that corresponds to a dipole antenna \cite[Sec. (15.5)]{orfanidis2016electromagnetic}.
\vspace{-0.35cm}
	\begin{lemma}\label{lemma:incident-E-dipole}
		Let $\hat{\vect{r}}_{\textup{Rx-Tx}}$
		be the unit-norm propagation vector from $\vect{r}_{\textup{Tx}}$ to $\vect{r}_{\textup{Rx}}$. 
		The incident electric field at $\vect{r}_{\textup{Rx}}$ generated by a dipole antenna is $\vect{E}_{\textup{inc}}(\vect{r}_{\textup{Rx}};\hat{\vect{p}}_{\textup{inc}})
		\approx
		\vect{E}_{0,\textup{inc}}\left(\vect{r}_{\textup{Rx}};\hat{\vect{p}}_{\textup{inc}}\right)
		G\left(\vect{r}_{\textup{Rx}},\vect{r}_{\textup{Tx}}\right)$
		where $\vect{E}_{0,\textup{inc}}\left(\vect{r}_{\textup{Rx}};\hat{\vect{p}}_{\textup{inc}}\right) = \frac{k^2p_{\textup{dm}}}{\epsilon_0} \left(\hat{\vect{p}}_{\textup{inc}} - (\hat{\vect{r}}_{\textup{Rx-Tx}}\cdot\hat{\vect{p}}_{\textup{inc}})\hat{\vect{r}}_{\textup{Rx-Tx}}\right) = \frac{k^2p_{\textup{dm}}}{\epsilon_0} \left(\tilde{\vect{p}}_{\textup{inc}} - (\hat{\vect{r}}_{\textup{Rx-Tx}}\cdot\tilde{\vect{p}}_{\textup{inc}})\hat{\vect{r}}_{\textup{Rx-Tx}}\right) e^{j\phi_\textup{inc}}$.
	\end{lemma}
\vspace{-0.6cm}
	\begin{proof}
	The electric field radiated by a dipole antenna is \cite[Eq. (15.5.5)]{orfanidis2016electromagnetic}. The approximation follows from $k \gg 1/{|\vect{r}_{\textup{Rx}} - \vect{r}_{\textup{Tx}}|}$. The proof follows by simplifying the triple vector product.
	\end{proof}
\vspace{-0.5cm}
\begin{remark}
The first addend in \eqref{eq:Stratton-Chu-equivalence-reduced}, i.e., the incident field at $\vect{r}_{\textup{Rx}}$, and the integral that yields the contribution from the RIS sum up, in general, incoherently and, thus, interfere with each other. The phase terms $\angle\Gamma_{\textup{ref}}(x,y)$ and $\angle\Gamma_{\textup{tran}}(x,y)$ of $\SS$ can, however, be optimized in order to make sure that both contributions (incident field and scattered field) add up coherently at $\vect{r}_{\textup{Rx}}$.
\end{remark}
\vspace{-0.25cm}

\vspace{-0.5cm}
	\subsection{Approximations and Asymptotic Regimes}\label{section:approximations} \vspace{-0.25cm}
In Sections \ref{section:reflection} and \ref{section:transmission}, we capitalize on \eqref{eq:Stratton-Chu-equivalence-reduced} in order to derive explicit expressions for the electric field reflected and transmitted by an RIS, and to unveil scaling laws as a function of relevant design parameters. To this end, some recurrent integrals need to be computed and some asymptotic approximations are exploited. In this section, we introduce methods to compute these integrals and we formally introduce the asymptotic operating regimes of interest.

	\subsubsection{Type-1 Integral} 
	Consider the following type of integral: \vspace{-0.25cm}
	\begin{equation}\label{eq:integral-type-1}
	I_1 = \int_{-L_y}^{L_y} \int_{-L_x}^{L_x}  \mathcal{A}_1(d_{\textup{Tx}}(x,y),d_{\textup{Rx}}(x,y))\mathcal{B}_1(x,y)e^{-jk(d_{\textup{Tx}}(x,y)+d_{\textup{Rx}}(x,y)-\mathcal{C}(x,y))}dxdy \vspace{-0.25cm}
	\end{equation}
	where $\mathcal{A}_1(d_{\textup{Tx}}(x,y),d_{\textup{Rx}}(x,y))$, $\mathcal{B}_1(x,y)$, and $\mathcal{C}(x,y)$ are real-valued functions. In particular, we assume the following: (i) $\mathcal{A}_1(d_{\textup{Tx}}(x,y),d_{\textup{Rx}}(x,y))$ depends on $d_{\textup{Tx}}(x,y)$ and $d_{\textup{Rx}}(x,y)$; (ii) $\mathcal{B}_1(x,y)$ is independent of $d_{\textup{Tx}}(x,y)$ and $d_{\textup{Rx}}(x,y)$; and (iii) $\mathcal{C}(x,y)$ is a linear function in $x$ and $y$. 
\vspace{-0.25cm}
	\begin{definition}\label{definition:long-distance-3D}
Define $r_{ES} = 8 (L_x^2 + L_y^2)/\lambda$. Assume that \eqref{eq:Stratton-Chu-equivalence-reduced} is formulated in terms of type-1 integrals. An RIS is said to operate in the electrically-small regime if $d_{\textup{Tx}0} > r_{ES}$ and $d_{\textup{Rx}0} > r_{ES}$.
	\end{definition}
\vspace{-0.35cm}
The electrically-small regime in Definition \ref{definition:long-distance-3D} is analogous to the Fraunhofer far-field \cite[Sec. 4.4.1]{balanis2016antenna}. This can be shown from the Taylor expansion of, e.g., $d_{\textup{Tx}}(x,y)$ around the origin:
\vspace{-0.35cm}
	\begin{equation}\label{eq:Taylor-Tx}
	d_{\textup{Tx}}(x,y) 
	= d_{\textup{Tx}0} -  x\sin\theta_{{\textup{inc}}0}\cos\varphi_{{\textup{inc}}0} - y\sin\theta_{{\textup{inc}}0}\sin{\varphi_{{\textup{inc}}0}} + R_2(x,y) \vspace{-0.35cm}
	\end{equation} 
where $R_2(x,y)$ collects the terms of higher order than the first degree. In general, the Fraunhofer distance is calculated for linear structures, e.g., by assuming $L_y \ll L_x$, and by then replacing the length of the structure ($L=L_x$) with the largest dimension of $\SS$ \cite[Eq. (4.41)]{balanis2016antenna}. Based on \eqref{SurfaceDefinition}, the largest dimension of $\SS$ is its diagonal $D = 2\sqrt{L_x^2 + L_y^2}$. For linear structures, \eqref{eq:Taylor-Tx} reduces to $d_{\textup{Tx}}(x) 
	= d_{\textup{Tx}0} -  x\sin\theta_{{\textup{inc}}0} + R_2(x)$. By definition, the Fraunhofer far-field is the distance $r_{F}$ at which the identity $\max\{R_2(x)\} = \pi/8$ holds true, which gives $r_{F} = 2D^2/\lambda$. Thus, we obtain $r_{F} = r_{ES}$. Notably, $r_{ES}$ can be formulated in terms of the ratio between the surface area and the wavelength, i.e., $r_{ES} = 2\frac{A_{\SS}}{\lambda}{\frac{a_x^2+a_y^2}{a_xa_y}}$, where $A_{\SS}$ is the area of $\SS$ and $L_x = a_xL$, $L_y = a_yL$.
	\vspace{-0.25cm}
	\begin{lemma}\label{lemma:electrically-small-approximation}
		In the electrically-small regime, the integral in \eqref{eq:integral-type-1} can be approximated as: \vspace{-0.25cm}
		\begin{equation}\label{eq:long-distance-approximation-3D}
		I_1
		\approx 
		\mathcal{A}_1(d_{\textup{Tx}0},d_{\textup{Rx}0})e^{-jk(d_{\textup{Tx}0}+d_{\textup{Rx}0})} \int_{-L_y}^{L_y} \int_{-L_x}^{L_x} \mathcal{B}_1(x,y) e^{-jk\left(\mathcal{D}_xx + \mathcal{D}_yy - \mathcal{C}(x,y)\right)}dxdy \vspace{-0.25cm}
		\end{equation}
		where $\mathcal{D}_{x} 
		= \sin\theta_{{\textup{inc}}0}\cos\varphi_{{\textup{inc}}0} + \sin\theta_{{\textup{rec}}0}\cos\varphi_{{\textup{rec}}0}$ and $\mathcal{D}_{y} 
		= \sin\theta_{{\textup{inc}}0}\sin\varphi_{{\textup{inc}}0} + \sin\theta_{{\textup{rec}}0}\sin\varphi_{{\textup{rec}}0}$.
	\end{lemma}
	\vspace{-0.5cm}
	\begin{proof}
		It follows from \eqref{eq:Taylor-Tx} by ignoring $R_2(x,y)$ and noting that $\mathcal{A}_1(\cdot,\cdot) \to \mathcal{A}_1(d_{\textup{Tx}0},d_{\textup{Rx}0})$.
	\end{proof} 
\vspace{-0.5cm}	
	\begin{definition}\label{definition:stationarypoints}
Define $\mathcal{P}(x,y) =d_{\textup{Tx}}(x,y)+d_{\textup{Rx}}(x,y) - \mathcal{C}(x,y)$. The stationary points of $\mathcal{P}(x,y) $ are the points $(x_s,y_s)$ such that $\frac{\partial}{\partial x}\mathcal{P}(x,y)\Huge|_{(x,y)=(x_s,y_s)} = \frac{\partial}{\partial y}\mathcal{P}(x,y)\Huge|_{(x,y)=(x_s,y_s)} = 0$.
	\end{definition}
\vspace{-0.65cm}	
	\begin{definition}\label{definition:short-distance-3D}
Define $\mathcal{P}(x,y) =d_{\textup{Tx}}(x,y)+d_{\textup{Rx}}(x,y) - \mathcal{C}(x,y)$ and let $\Psi$ be the set of its stationary points. Let $D = 2\sqrt{L_x^2 + L_y^2}$ be the diagonal of $\SS$. Assume that \eqref{eq:Stratton-Chu-equivalence-reduced} is formulated in terms of type-1 integrals. An RIS is said to operate in the electrically-large regime if $(2D^2/{\lambda}){({z_{\textup{Tx}}}/{[d_{\textup{Tx}}(x_s,y_s)]^2} + {z_{\textup{Rx}}}/{[d_{\textup{Rx}}(x_s,y_s)]^2})} \gg 1$ for all stationary points $(x_s,y_s) \in \Psi$.
	\end{definition}
\vspace{-0.25cm}	

Similar to the Fraunhofer far-field \cite[Sec. 4.4.1]{balanis2016antenna}, Definition \ref{definition:short-distance-3D} can be justified by starting from a line surface, e.g., by assuming $L_y \ll L_x$ and by then replacing the length of the line ($L=L_x$) with the diagonal $D$ of $\SS$. Consider the line integral $I_\ell = \int_{-L}^{L}  \mathcal{M}(\ell)e^{-jk\mathcal{P}(\ell)}d\ell$ corresponding to \eqref{eq:integral-type-1}, where $\mathcal{M}(\ell)$ is a slowly-varying function in $[-L,L]$, 	$\mathcal{P}(\ell) = d_{\textup{Tx}}(\ell)+d_{\textup{Rx}}(\ell)- \mathcal{C}(\ell)$, $d_{\textup{Tx}}(\ell) = \sqrt{\ell^2 + z_{\textup{Tx}}^2}$, $d_{\textup{Rx}}(\ell) = \sqrt{\ell^2 + z_{\textup{Rx}}^2}$, and $\mathcal{C}(\ell)$ is a linear function in $\ell$. Let $\ell_s \in [-L,L]$ be a stationary point (assuming that it exists) of $\mathcal{P}(\ell)$, i.e., $\frac{\partial}{\partial x}\mathcal{P}(\ell)\Huge|_{\ell=\ell_s} = 0$. Definition \ref{definition:short-distance-3D} can be justified by invoking the stationary phase method to compute $I_\ell$ \cite{Borovikov}. In particular, the integrand of $I_\ell$ oscillates very quickly outside a small region centered at $\ell_s$, and, thus, the contributions outside the small region around $\ell_s$ cancel out when computing the integral. Under these conditions, $I_\ell$ can be well approximated by replacing $\mathcal{P}(\ell)$ with its Taylor approximation at $\ell_s \in [-L,L]$, i.e., $\mathcal{P}(\ell)	\approx \mathcal{P}(\ell_s) + \frac{1}{2}\left( {z_{\textup{Tx}}}/{[d_{\textup{Tx}}(\ell_s)]^2 }	+ {z_{\textup{Rx}}}/{[d_{\textup{Rx}}(\ell_s)]^2 }\right) (\ell-\ell_s)^2$, and by letting the extremes of integration go to infinity, provided that the region around $\ell_s$ that dominates $I_\ell$ is well contained in $[-L,L]$. This is usually true when the minimum of the integrand of $I_\ell$ falls within $[-L,L]$, which occurs if the condition in Definition \ref{definition:short-distance-3D} is fulfilled. Notably, the latter condition can be formulated in terms of ratio between the area of the surface and the wavelength, i.e., ${d_s} \ll {r_{EL}}$ with ${r_{EL}} = \frac{{2{D^2}}}{\lambda }\sqrt {\frac{{{z_{{\rm{Tx}}}}}}{{b_{\rm{Tx}}^2}} + \frac{{{z_{{\rm{Rx}}}}}}{{b_{\rm{Rx}}^2}}}  = \frac{{2{A_{{\SS}}}}}{\lambda }\frac{{a_x^2 + a_y^2}}{{{a_x}{a_y}}}\sqrt {\frac{{{z_{{\rm{Tx}}}}}}{{b_{\rm{Tx}}^2}} + \frac{{{z_{{\rm{Rx}}}}}}{{b_{\rm{Rx}}^2}}}$, ${d_{{\rm{Tx}}}}\left( {{x_s},{y_s}} \right) = {b_{\rm{Tx}}}{d_s}$, ${d_{{\rm{Rx}}}}\left( {{x_s},{y_s}} \right) = {b_{\rm{Rx}}}{d_s}$.
\vspace{-0.9cm}
\begin{remark}
Based on Definition \ref{definition:long-distance-3D} and Definition \ref{definition:short-distance-3D}, the following comments can be made: (i) the terminology electrically-small RIS originates from the conditions $d_{\textup{Tx}0} > r_{ES}$ and $d_{\textup{Rx}0} > r_{ES}$, i.e., the transmission distances (computed with respect to the center of $\SS$) are larger than the electrical size of $\SS$, which is ${A_{\SS}}/{\lambda}$; (ii) the terminology electrically-large RIS originates from the condition ${d_s} \ll {r_{EL}}$, i.e., the transmission distances (computed with respect to the stationary point of the phase term) are smaller than the electrical size, ${A_{\SS}}/{\lambda}$, of $\SS$; and (iii) since, in general, $\sqrt {\frac{{{z_{{\rm{Tx}}}}}}{{b_{\rm{Tx}}^2}} + \frac{{{z_{{\rm{Rx}}}}}}{{b_{\rm{Rx}}^2}}} < 1$, then ${r_{EL}} = {r_{ES}}\sqrt {\frac{{{z_{{\rm{Tx}}}}}}{{b_{\rm{Tx}}^2}} + \frac{{{z_{{\rm{Rx}}}}}}{{b_{\rm{Rx}}^2}}}  < {r_{ES}}$. This implies that the electrically-large regime holds for shorter distances than the radiating near-field regime \cite[Sec. (4.4.2)]{balanis2016antenna}.
\end{remark}
\vspace{-0.15cm}
\vspace{-0.5cm}	
	\begin{lemma}\label{lemma:short-approximation-SPM}
		Define $\A(x_s,y_s) = \Hessian(\mathcal{P}(x,y))|_{(x,y) = (x_s,y_s)}$. 
		Assume that $\Psi$ is not empty and, for $(x_s,y_s)\in\Psi$, $\det(\A(x_s,y_s)) \neq 0$. In the electrically-large regime, \eqref{eq:integral-type-1} can be approximated as: \vspace{-1.10cm}
		\begin{align}\label{eq:SPM}
		I_1
		\approx (2\pi/k)
		\sum\nolimits_{(x_s,y_s) \in\Psi}
		&\mathcal{A}_1(d_{\textup{Tx}}(x_s,y_s),d_{\textup{Rx}}(x_s,y_s))\mathcal{B}_1(x_s,y_s)\left|\det(\A(x_s,y_s))\right|^{-1/2} \nonumber \\
		& \exp\left(-jk\mathcal{P}(x_s,y_s)-{j\pi}\sign\left(\A(x_s,y_s)\right)/4\right) \vspace{-0.25cm}
		\end{align}
where $\sign\left(\A(x_s,y_s)\right) = N^+(\A(x_s,y_s)) - N^-(\A(x_s,y_s))$ is the signature of $\A(x_s,y_s)$, with $N^+(\A(x_s,y_s))$ and $N^-(\A(x_s,y_s))$ the number of positive and negative eigenvalues of $\A(x_s,y_s)$.
	\end{lemma}
\vspace{-0.5cm}
	\begin{proof}
		See Appendix C. %\ref{appendix:proof-of-SPM}.
	\end{proof}
\vspace{-0.5cm}
	\begin{lemma}\label{lemma:short-approximation-nonSPM}
		Let $\Psi$ be empty. In the electrically-large regime, \eqref{eq:integral-type-1} can be approximated as: \vspace{-0.25cm}
		\begin{equation}\label{eq:non-SPM}
		I_1
		\approx \frac{1}{(-jk)^2}  \left[\frac{\mathcal{A}_1(d_{\textup{Tx}}(x,y),d_{\textup{Rx}}(x,y))\mathcal{B}_1(x,y)e^{-jk\mathcal{P}(x,y)}}{\mathcal{P}_x(x,y)\mathcal{P}_y(x,y)}\right]  \Big|^{x=L_x}_{x=-L_x}\Big|^{y=L_y}_{y=-L_y} \vspace{-0.25cm}
		\end{equation} 
		where
		$\mathcal{P}_{x}(x,y) = \frac{\partial}{\partial x}\mathcal{P}(x,y)$ and $\mathcal{P}_{y}(x,y) = \frac{\partial}{\partial y}\mathcal{P}(x,y)$.
	\end{lemma}
	\vspace{-0.6cm}
	\begin{proof}
		See Appendix D. % \ref{appendix:proof-of-short-distance-approximation-nonSPM}
	\end{proof}	
	\vspace{-0.25cm}
	
By comparing Lemmas \ref{lemma:short-approximation-SPM} and \ref{lemma:short-approximation-nonSPM}, we evince that, since $\det(\A(x_s,y_s))$ is independent of $k$, $|I_1|$ is inversely proportional to $k$ if at least one stationary point is contained in $\SS$, and is inversely proportional to $k^2$ if no stationary point lies in $\SS$. For $k \gg 1$, thus, $|I_1|$ is dominated by the contributions from the stationary points. In the rest of this paper, therefore, we focus our attention on the case studies (in the electrically-large regime) in which at least one stationary point exists.
\vspace{-0.65cm}
	\subsubsection{Type-2 Integral}
	Consider the following type of integral: \vspace{-0.25cm}
	\begin{equation}\label{eq:integral-type-2}
	I_2 = \int_{-L_y}^{L_y} \int_{-L_x}^{L_x}  \mathcal{A}_2(d_{\textup{Tx}}(x,y),d_{\textup{Rx}}(x,y))\mathcal{B}_2(x,y)dxdy \vspace{-0.25cm}
	\end{equation}
where $\mathcal{A}_2(d_{\textup{Tx}}(x,y),d_{\textup{Rx}}(x,y))$ is a real-valued function of the distances $d_{\textup{Tx}}(x,y)$ and $d_{\textup{Rx}}(x,y)$, and $\mathcal{B}_2(x,y)$ is a real-valued function that is independent of $d_{\textup{Tx}}(x,y)$ and $d_{\textup{Rx}}(x,y)$.
\vspace{-0.40cm}	
		\begin{lemma}\label{lemma:electrically-small-approximation-type-2}
		Assume $d_{\textup{Tx}0} \gg D$ and $d_{\textup{Rx}0} \gg D$, where $D = 2\sqrt{L_x^2 + L_y^2}$ is the diagonal of $\SS$. The integral in \eqref{eq:integral-type-2} can be approximated as follows:\vspace{-0.25cm}	
		\begin{equation}\label{eq:long-distance-approximation-3D-type-2}
		I_2
		\approx 
		\mathcal{A}_2(d_{\textup{Tx}0},d_{\textup{Rx}0}) \int_{-L_y}^{L_y} \int_{-L_x}^{L_x} \mathcal{B}_2(x,y)dxdy \vspace{-0.25cm}	
		\end{equation}
	\end{lemma}
\vspace{-0.25cm}	
	\begin{proof}
		It follows from \eqref{eq:Taylor-Tx} noting that $d_{\textup{Tx}}(x,y) \approx d_{\textup{Tx}0}$, $d_{\textup{Rx}}(x,y) \approx d_{\textup{Rx}0}$ if $d_{\textup{Tx}0}, d_{\textup{Rx}0} \gg D$.
	\end{proof}
	\vspace{-0.25cm}	

If $d_{\textup{Tx}0} \ll D$ and $d_{\textup{Rx}0} \ll D$, it is not straightforward to compute \eqref{eq:integral-type-2} in general. This case study is analyzed in Sections \ref{sec:focusing-lens-reflection} and \ref{sec:focusing-lens-transmission} for the specific $\mathcal{A}_2(x,y)$ and $\mathcal{B}_2(x,y)$ of interest.

\vspace{-0.25cm}	
\begin{remark}
The asymptotic regime in Lemma \ref{lemma:electrically-small-approximation-type-2} is independent of $\lambda$ and is, in general, different from the asymptotic regime in Definition \ref{definition:long-distance-3D} that depends on $\lambda$. We still refer to it as electrically-small regime, however, since $d_{\textup{Tx}0} \gg D$ and $d_{\textup{Rx}0} \gg D$ implies $D/\lambda \ll d_{\textup{Tx}0}/\lambda$ and $D/\lambda \ll d_{\textup{Rx}0}/\lambda$. Likewise, the regime $d_{\textup{Tx}0} \ll D$ and $d_{\textup{Rx}0} \ll D$ is referred to as electrically-large regime.
	\end{remark}
	\vspace{-0.25cm}

\vspace{-0.5cm}
	\section{Electric Field In the Presence of a Reflecting Surface}\label{section:reflection} \vspace{-0.25cm}
	In this section, we analyze $\vect{E}(\vect{r}_{\textup{Rx}})$ under the assumption that $\SS$ is a reflecting surface according to the definitions and assumptions given in Section \ref{section:system-model} (see Fig. \ref{fig:systemmodelreflection} and Fig. \ref{fig:kirchhoffreflection}).
	\vspace{-0.1cm}
	\begin{proposition}\label{proposition:reflected-field-general}		 \vspace{-0.25cm}
		Let $\hat{\vect{s}}_{(x,y)} = \sin\theta_{\textup{inc}}(x,y)\cos\varphi_{\textup{inc}}(x,y)\hat{\vect{x}} + \sin\theta_{\textup{inc}}(x,y)\sin\varphi_{\textup{inc}}(x,y)\hat{\vect{y}} + \cos\theta_{\textup{inc}}(x,y)\hat{\vect{z}}$, be the  unit-norm propagation vector from $\vect{r}_{\textup{Tx}}$ to $\vect{s} = x \hat{\vect{x}} + y \hat{\vect{y}} \in \SS$. Define
		$\Omega_{\textup{ref}}(x,y;\hat{\vect{p}}_{\textup{ref}},\hat{\vect{p}}_{\textup{rec}}) =
		(k^2/\epsilon_0) p_{\textup{dm}} \left(\tilde{\vect{p}}_{\textup{rec}}\cdot\tilde{\vect{p}}_{\textup{ref}} - \left(\hat{\vect{s}}_{(x,y)}\cdot\tilde{\vect{p}}_{\textup{rec}}\right)\left(\hat{\vect{s}}_{(x,y)}\cdot\tilde{\vect{p}}_{\textup{ref}}\right)\right)\mathcal{E}\left(\hat{\vect{p}}_{\textup{inc}},\hat{\vect{p}}_{\textup{ref}}\right)$.		Under the assumptions stated in Lemma \ref{lemma:incident-E-dipole}, the electric field $\vect{E}(\vect{r}_{\textup{Rx}})$ projected onto $\hat{\vect{p}}_{\textup{rec}}$ can be formulated as follows: 	\vspace{-0.25cm}
		\begin{equation} \label{eq:Ex-surface-contribution-reflection}
		\begin{split}
		&\vect{E}(\vect{r}_{\textup{Rx}})\cdot \hat{\vect{p}}_{\textup{rec}} 
		\approx
		\hat{\vect{p}}_{\textup{rec}} \cdot \vect{E}_{0,\textup{inc}}\left(\vect{r}_{\textup{Rx}};\hat{\vect{p}}_{\textup{inc}}\right)G(\vect{r}_{\textup{Rx}},\vect{r}_{\textup{Tx}})  \\
		& 
		+ jk e^{j(\phi_{\textup{ref}}+\phi_{\textup{rec}})} \int_{\SS}  \Gamma_{\textup{ref}}(\vect{s})\Omega_{\textup{ref}}(x,y;\hat{\vect{p}}_{\textup{ref}},\hat{\vect{p}}_{\textup{rec}})G\left(\vect{s},\vect{r}_{\textup{Tx}}\right)G(\vect{r}_{\textup{Rx}},\vect{s}) \left[\frac{z_{\textup{Rx}}}{|\vect{s} - \vect{r}_{\textup{Rx}}|}  + \frac{z_{\textup{Tx}}}{|\vect{s} - \vect{r}_{\textup{Tx}}|} \right] d\vect{s} \\
		&=
		{\hat{\vect{p}}_{\textup{rec}} \cdot \vect{E}_{0,\textup{inc}}\left(\vect{r}_{\textup{Rx}};\hat{\vect{p}}_{\textup{inc}}\right)G(\vect{r}_{\textup{Rx}},\vect{r}_{\textup{Tx}})+ \mathcal{I}_0 \int_{-L_y}^{L_y}\int_{-L_x}^{L_x} \mathcal{I}_R(x,y)e^{-jk\mathcal{P}_R(x,y)}dxdy }		 \vspace{-0.25cm}
		\end{split}
		\end{equation} 
		where $\mathcal{I}_0 = jk/(16\pi^2)$, and the following shorthand notation is used:\vspace{-0.5cm}
		\begin{align} \label{eq:Ex-surface-contribution-reflection-A}
		\mathcal{P}_R(x,y)
		&= 
		d_\textup{Tx}(x,y) + d_\textup{Rx}(x,y) - (\phi_\textup{rec} + \phi_\textup{ref} + \angle\Gamma_{\textup{ref}}(x,y))/k \\
		\mathcal{I}_R(x,y) &= 
		\frac{\left|\Gamma_{\textup{ref}}(x,y)\right|\Omega_{\textup{ref}}(x,y;\hat{\vect{p}}_{\textup{ref}},\hat{\vect{p}}_{\textup{rec}})}{d_\textup{Tx}(x,y)d_\textup{Rx}(x,y)} \left(\cos\theta_{\textup{inc}}(x,y) + \cos\theta_{\textup{rec}}(x,y)\right) \vspace{-0.25cm}
		\end{align}
	\end{proposition}
	\vspace{-0.25cm}
	\begin{proof}
	See Appendix E. % \ref{appendix:proof-of-reflected-field-general}.
	\end{proof}
	\vspace{-0.5cm}
\begin{remark}
The approximation in \eqref{eq:Ex-surface-contribution-reflection} originates only from the assumptions  $k \gg 1/d_{\textup{Tx}}(x,y)$, $k \gg 1/d_{\textup{Rx}}(x,y)$ (see Section \ref{section:system-model}). This is apparent from the proof in Appendix E. The proof in Appendix E can, however, be readily generalized in order to avoid these assumptions.
\end{remark}
\vspace{-0.25cm}

The electric field in \eqref{eq:Ex-surface-contribution-reflection} is formulated as the sum of the incident electric field in the absence of $\SS$ and the contribution due to the reflection from $\SS$. This latter term is denoted by $F_R(\vect{r}_{\textup{Rx}}) = \mathcal{I}_0 \int_{-L_y}^{L_y}\int_{-L_x}^{L_x} \mathcal{I}_R(x,y)e^{-jk\mathcal{P}_R(x,y)}dxdy$ and is analyzed next to better understand the performance of RISs as a function of important design parameters and configurations for $\SS$, e.g., $\angle\Gamma_{\textup{ref}}(x,y)$. As illustrative examples, we consider case studies that correspond to using phase gradient metasurfaces, which are known to be approximated implementations of perfect anomalous reflectors \cite{Marco-JSAC}. This choice is motivated only for analytical convenience and to shed light on the impact of important design parameters. Proposition \ref{proposition:reflected-field-general} has, in fact, general applicability.

\vspace{-0.5cm}
	\subsection{$\SS$ is Configured for Specular Reflection} \vspace{-0.15cm}
This setup is obtained if $\angle \Gamma_{\textup{ref}}(x,y) = \phi_0$ for $(x,y) \in \SS$, where $\phi_0 \in [0,2\pi)$ is a fixed phase. 
	\vspace{-0.35cm}
	\begin{corollary}\label{corollary:electrically-large-uniform-reflection}
		Let $(x_s,y_s)\in \SS$ be the solution of the following system of equations: \vspace{-0.25cm}
		\begin{equation}
		\frac{(x_s-x_\textup{Tx})}{d_\textup{Tx}(x_s,y_s)} + \frac{(x_s-x_\textup{Rx})}{d_\textup{Rx}(x_s,y_s)} = 0, \quad\quad
		\frac{(y_s-x_\textup{Tx})}{d_\textup{Tx}(x_s,y_s)} + \frac{(y_s-x_\textup{Rx})}{d_\textup{Rx}(x_s,y_s)} = 0 \label{eq:condition-uniform-reflection} \vspace{-0.15cm}
		\end{equation}
		In the electrically-large regime, $F_R(\vect{r}_{\textup{Rx}})$ can be approximated as follows: \vspace{-0.25cm}
		\begin{equation}\label{eq:short-approx-3D-unifrom}
		F_R(\vect{r}_{\textup{Rx}})
		\approx
		\frac{\left|\Gamma_{\textup{ref}}(x_s,y_s)\right|\Omega_{\textup{ref}}(x_s,y_s;\hat{\vect{p}}_{\textup{ref}},\hat{\vect{p}}_{\textup{rec}})}{4\pi(d_\textup{Tx}(x_s,y_s)+d_\textup{Rx}(x_s,y_s))}  e^{-jk\left(d_\textup{Tx}(x_s,y_s)+d_\textup{Rx}(x_s,y_s) - (\phi_0+\phi_{\textup{ref}}+\phi_{\textup{rec}})/k\right)} \vspace{-0.25cm}
		\end{equation}
	\end{corollary}
\vspace{-0.15cm}
	\begin{proof}
		See Appendix F. %\ref{appendix:proof-of-electrically-large-uniform-reflection}
	\end{proof}
\vspace{-0.5cm}
	\begin{remark}
	Assume that Tx and Rx move along directions such that $(x_s,y_s)$, and $\theta_{\textup{inc}}(x_s,y_s)$, $\theta_{\textup{rec}}(x_s,y_s)$, $\varphi_{\textup{inc}}(x_s,y_s)$, $\varphi_{\textup{rec}}(x_s,y_s)$ are kept fixed. 
	From Corollary \ref{corollary:electrically-large-uniform-reflection}, we evince the following.
	\begin{itemize}
		\item Since $\Gamma_{\textup{ref}}(x_s,y_s)$ depends only on $(x_s,y_s)$ and $\Omega_{\textup{ref}}(x_s,y_s;\hat{\vect{p}}_{\textup{ref}},\hat{\vect{p}}_{\textup{rec}})$ depends only on $\tilde{\vect{p}}_{\textup{inc}}$, $\tilde{\vect{p}}_{\textup{rec}}$, $\theta_{\textup{inc}}(x_s,y_s)$, $\theta_{\textup{rec}}(x_s,y_s)$, $\varphi_{\textup{inc}}(x_s,y_s)$, and $\varphi_{\textup{rec}}(x_s,y_s)$, they are both independent	 of the Tx-to-$(x_s,y_s)$ and $(x_s,y_s)$-to-Rx distances.	 In the electrically-large regime, therefore, $\left|{F_R(\vect{r}_{\textup{Rx}})}\right|$ decays as a function of the sum of the Tx-to-$(x_s,y_s)$ and $(x_s,y_s)$-to-Rx distances. 
		\item In the electrically-large regime, $\left|{F_R(\vect{r}_{\textup{Rx}})}\right|$ is independent of the size of $\SS$. This implies that the received power is bounded, even though the size of $\SS$ grows large  (tending to infinity).
		\item The system of equations in \eqref{eq:condition-uniform-reflection} is equivalent to $\varphi_{{\textup{inc}}}(x_s,y_s) = (\varphi_{{\textup{rec}}}(x_s,y_s) + \pi) \mod 2\pi$ and $\theta_{\textup{inc}}(x_s,y_s) = \theta_{\textup{rec}}(x_s,y_s)$. These conditions correspond to the law of reflection.  
		\end{itemize}
	\end{remark}
\vspace{-0.65cm}
	\begin{corollary}\label{corollary:electrically-small-uniform-reflection}
		In the electrically-small regime, $F_R(\vect{r}_{\textup{Rx}})$ can be approximated as follows: \vspace{-0.25cm}
		\begin{align}
		F_R(\vect{r}_{\textup{Rx}})
		&\approx
		\frac{jk\Omega_{\textup{ref}}(0,0;\hat{\vect{p}}_{\textup{ref}},\hat{\vect{p}}_{\textup{rec}})\left(\cos\theta_{{\textup{inc}}0}+\cos\theta_{{\textup{rec}}0}\right)}{16\pi^2\left(d_{\textup{Tx}0}d_{\textup{Rx}0}\right)}e^{-jk\left(d_{\textup{Tx}0}+d_{\textup{Rx}0}-(\phi_0+\phi_{\textup{ref}}+\phi_{\textup{rec}})/k\right)}   \nonumber\\
		&\quad 
		 \int_{-L_y}^{L_y}\int_{-L_x}^{L_x} |\Gamma_{\textup{ref}}(x,y)|e^{jk(\mathcal{D}_xx+\mathcal{D}_yy)}dxdy \vspace{-0.25cm}
		\end{align}
		where $\mathcal{D}_x$ and $\mathcal{D}_y$ are defined in Lemma \ref{lemma:electrically-small-approximation}. Let ${\rm{sinc}}\left( x \right) = \frac{{\sin \left( {\pi x} \right)}}{{\pi x}}$ be the sinc function. If $|\Gamma_{\textup{ref}}(x,y)| = \Gamma_{\textup{ref}} > 0$ for $(x,y) \in \SS$, then $F_R(\vect{r}_{\textup{Rx}})$ can be further simplified as follows: \vspace{-0.15cm}
		\begin{align}\label{eq:electrically-large-uniform-reflection-constant-reflection}
		F_R(\vect{r}_{\textup{Rx}})
		&\approx \frac{jk\Gamma_{\textup{ref}}\Omega_{\textup{ref}}(0,0;\hat{\vect{p}}_{\textup{ref}},\hat{\vect{p}}_{\textup{rec}})L_xL_y\left(\cos\theta_{{\textup{inc}}0}+\cos\theta_{{\textup{rec}}0}\right)}{4\pi^2d_{\textup{Tx}0}d_{\textup{Rx}0}} 
		\sinc\left(kL_x\mathcal{D}_x\right)\sinc\left(kL_y\mathcal{D}_y\right) \nonumber\\
		&\quad 
		e^{-jk\left(d_{\textup{Tx}0}+d_{\textup{Rx}0}-(\phi_0+\phi_{\textup{ref}}+\phi_{\textup{rec}})/k\right)} \vspace{-0.25cm}
		\end{align}
	\end{corollary}
\vspace{-0.5cm}
	\begin{proof}
		If follows directly from \eqref{eq:long-distance-approximation-3D}. 
	\end{proof}
\vspace{-0.5cm}
	\begin{remark}
	Assume that Tx and Rx move along directions such that $\theta_{\textup{inc}0}$, $\theta_{\textup{rec}0}$, $\varphi_{\textup{inc}0}$, $\varphi_{\textup{rec}0}$ are kept fixed. From Corollary \ref{corollary:electrically-small-uniform-reflection}, the following conclusions can be drawn.
		\begin{itemize}
			\item In the electrically-small regime, $\left|{F_R(\vect{r}_{\textup{Rx}})}\right|$ decays as a function of the product of the Tx-to-$(0,0)$ and $(0,0)$-to-Rx distances, where $(0,0)$ is the center of $\SS$.
			\item In the electrically-small regime, $\left|{F_R(\vect{r}_{\textup{Rx}})}\right|$ grows linearly with the area of $\SS$, i.e., $A_{\SS} = 4 L_x L_y$. This does not imply that $\left|{F_R(\vect{r}_{\textup{Rx}})}\right|$ grows unbounded if the size of $\SS$ tends to infinity. If $A_{\SS} \to \infty$, in fact, the RIS does not operate in the electrically-small regime anymore, but in the electrically-large regime. Therefore, the approximation in Corollary \ref{corollary:electrically-small-uniform-reflection} needs to be replaced with the approximation in Corollary \ref{corollary:electrically-large-uniform-reflection}, which does not depend on the size of $\SS$.
			\item In the electrically-small regime, $\left|{F_R(\vect{r}_{\textup{Rx}})}\right|$ attains its maximum for $\mathcal{D}_x = \mathcal{D}_y = 0$. If the angle of incidence $\theta_{\textup{inc}0}$ is fixed, this is fulfilled in correspondence of the angles of observation $\theta_{\textup{inc}0} \hspace{-0.05cm} = \hspace{-0.05cm} \theta_{\textup{rec}0}$ and $\varphi_{\textup{inc}0} \hspace{-0.05cm} = \hspace{-0.05cm} (\varphi_{\textup{rec}0} + \pi) \hspace{-0.05cm} \mod  \hspace{-0.05cm} 2\pi$, which can be interpreted as the law of reflection. Also, the main lobe of $\sinc(kL_x\mathcal{D}_x)$ and $\sinc(kL_y\mathcal{D}_y)$ gets narrower if $L_x$ and $L_y$ increase.
		\end{itemize}
	\end{remark}
\vspace{-0.80cm}
	\subsection{$\SS$ is Configured for Anomalous Reflection} \vspace{-0.15cm}
	This setup is obtained by setting $\angle \Gamma_{\textup{ref}}(x,y) = k(\alpha_R x + \beta_R y) + \phi_0$
	for $(x,y) \in \SS$, where $\alpha_R\in\mathbb{R}$, $\beta_R\in\mathbb{R}$ are design parameters, and $\phi_0 \in [0,2\pi)$ is a fixed phase. As detailed in further text, the direction of anomalous reflection is determined by the specific choice of $\alpha_R$ and $\beta_R$.
	\vspace{-0.35cm}
	\begin{corollary}\label{corollary:electrically-large-anomalous-general-reflection}
	Let $(x_s,y_s)\in \SS$ be the solution of the following system of equations: \vspace{-0.2cm}
	\begin{equation}
	\frac{(x_s-x_\textup{Tx})}{d_\textup{Tx}(x_s,y_s)} + \frac{(x_s-x_\textup{Rx})}{d_\textup{Rx}(x_s,y_s)} = \alpha_R, \quad\quad
	\frac{(y_s-x_\textup{Tx})}{d_\textup{Tx}(x_s,y_s)} + \frac{(y_s-x_\textup{Rx})}{d_\textup{Rx}(x_s,y_s)} = \beta_R \label{eq:condition-anomalous-general-reflection} \vspace{-0.2cm}
	\end{equation}
	Define the shorthand notation $\Theta_{\textup{Q}} = \theta_{\textup{Q}}(x_s,y_s)$ and $\Phi_{\textup{Q}} = \varphi_{\textup{Q}}(x_s,y_s)$ for $\textup{Q} \in \{\textup{inc},\textup{rec}\}$.	 In the electrically-large regime, $F_R(\vect{r}_{\textup{Rx}})$ can be approximated as follows: \vspace{-0.25cm}
	\begin{equation}\label{eq:short-approx-3D-anomalous-general-case}
	F_R(\vect{r}_{\textup{Rx}})
	\approx \frac{\left|\Gamma_{\textup{ref}}(x_s,y_s)\right|\Omega_{\textup{ref}}(x_s,y_s;\hat{\vect{p}}_{\textup{ref}},\hat{\vect{p}}_{\textup{rec}})e^{-jk(d_\textup{Tx}(x_s,y_s)+d_\textup{Rx}(x_s,y_s) - (\alpha_R x_s + \beta_R y_s) -(\phi_0+\phi_{\textup{ref}}+\phi_{\textup{rec}})/k)}}{8\pi{\sqrt{\mathcal{R}_{1} (d_{\textup{Tx}}(x_s,y_s))^2+\mathcal{R}_{2} (d_{\textup{Rx}}(x_s,y_s))^2 + \mathcal{R}_{3} d_{\textup{Tx}}(x_s,y_s)d_{\textup{Rx}}(x_s,y_s)}}} \vspace{-0.20cm}
	\end{equation} 
	where $\mathcal{R}_{1} 
	= \cos^2\Theta_{\textup{rec}}/(\cos\Theta_{{\textup{inc}}} + \cos\Theta_{{\textup{rec}}})^2$, $\mathcal{R}_{2}
	= \cos^2\Theta_{\textup{inc}}/(\cos\Theta_{{\textup{inc}}} + \cos\Theta_{{\textup{rec}}})^2$, and $\mathcal{R}_{3}
	= \left(\cos^2\Theta_{{\textup{inc}}} + \cos^2\Theta_{\textup{rec}} + \sin^2\Theta_{\textup{inc}}\sin^2\Theta_{\textup{rec}}  \sin^2(\Phi_{{\textup{inc}}}-\Phi_{{\textup{rec}}}) \right) /{\left(\cos\Theta_{{\textup{inc}}} + \cos\Theta_{{\textup{rec}}}\right)^2}$.
	\end{corollary}
\vspace{-0.5cm}
	\begin{proof}
		It follows from Lemma \ref{lemma:short-approximation-SPM} along the same lines as the proof of Corollary \ref{corollary:electrically-large-uniform-reflection}. The only difference is that $\mathcal{P}(x,y)= \mathcal{P}_R(x,y)$, $\det(\A(x_s,y_s))$, and $\sign(\A(x_s,y_s))$ depend on $\alpha_R$, $\beta_R$.
	\end{proof}
	\vspace{-0.25cm}	
	The analytical formulation in \eqref{eq:short-approx-3D-anomalous-general-case} does not provide direct design insights. To this end, we introduce an approximation for \eqref{eq:short-approx-3D-anomalous-general-case} in order to unveil scaling laws and performance trends.
	\vspace{-0.25cm}
	\begin{corollary}\label{corollary:electrically-large-anomalous-general-reflection-scaling}
	Consider $\zeta'_1 >0$, $\zeta'_2 > 0$. Define $K_1 
	= (\mathcal{R}_1{\zeta'_1} + \frac{1}{2} \mathcal{R}_3{\zeta'_2})/\sqrt{\mathcal{R}_1{\zeta'_1}^2 + \mathcal{R}_2 {\zeta'_2}^2 + \mathcal{R}_3{\zeta'_1}{\zeta'_2}}$, 
	$K_2 
	= (\mathcal{R}_2{\zeta'_2} + \frac{1}{2} \mathcal{R}_3{\zeta'_1})/\sqrt{\mathcal{R}_1{\zeta'_1}^2 + \mathcal{R}_2 {\zeta'_2}^2 + \mathcal{R}_3{\zeta'_1}{\zeta'_2}} $. Then, 
	\eqref{eq:short-approx-3D-anomalous-general-case} can be approximated as follows: \vspace{-0.75cm}
	\begin{equation}\label{eq:scaling-law-intensity-anomalous-general-reflection}
	F_R(\vect{r}_{\textup{Rx}})
	\approx
	\frac{\left|\Gamma_{\textup{ref}}(x_s,y_s)\right|\Omega_{\textup{ref}}(x_s,y_s;\hat{\vect{p}}_{\textup{ref}},\hat{\vect{p}}_{\textup{rec}})}{8\pi(K_1{d_{\textup{Tx}}(x_s,y_s)} + K_2{d_{\textup{Rx}}(x_s,y_s))}}   e^{-jk(d_\textup{Tx}(x_s,y_s)+d_\textup{Rx}(x_s,y_s) - (\alpha_R x_s + \beta_R y_s) -(\phi_0+\phi_{\textup{ref}}+\phi_{\textup{rec}})/k)} \vspace{-0.25cm}
	\end{equation}
	\end{corollary}
\vspace{-0.5cm}
	\begin{proof}
	For simplicity, let us denote $\zeta_1 = d_{\textup{Tx}}(x_s,y_s)$ and $\zeta_2 = d_{\textup{Rx}}(x_s,y_s)$. Define $f(\zeta_1,\zeta_2) = \sqrt{\mathcal{R}_1\zeta_1^2 + \mathcal{R}_2 \zeta_2^2 + \mathcal{R}_3\zeta_1\zeta_2}$. Consider a generic pair of points $(\zeta'_1,\zeta'_2)$. The function $f(\zeta_1,\zeta_2)$ can be approximated at $(\zeta'_1,\zeta'_2)$ by using the Taylor approximation, which yields $f(\zeta_1,\zeta_2)
	\approx  \sqrt{\mathcal{R}_1{\zeta'_1}^2 + \mathcal{R}_2 {\zeta'_2}^2 + \mathcal{R}_3{\zeta'_1}{\zeta'_2}}+ (\mathcal{R}_{1,3}({\zeta_1}-{\zeta'_1}) + \mathcal{R}_{2,3}({\zeta_2}-{\zeta'_2}))/{\sqrt{\mathcal{R}_1{\zeta'_1}^2 + \mathcal{R}_2 {\zeta'_2}^2 + \mathcal{R}_3{\zeta'_1}{\zeta'_2}}}$,
	where $\mathcal{R}_{1,3} = (\mathcal{R}_1{\zeta'_1} + \mathcal{R}_3{\zeta'_2}/2)$ and $\mathcal{R}_{2,3} = (\mathcal{R}_2{\zeta'_2} + \mathcal{R}_3{\zeta'_1}/2)$. The proof follows with the aid of algebraic steps. The parameters $K_1$ and $K_2$ are independent of the pair $(\zeta'_1,\zeta'_2)$ if $\zeta'_1 = \zeta'_2$.
	\end{proof}
\vspace{-0.20cm}
Given $\Theta_{\textup{inc}}$, $\Theta_{\textup{rec}}$, $\Phi_{\textup{inc}}$, and $\Phi_{\textup{rec}}$,  \eqref{eq:short-approx-3D-anomalous-general-case} and \eqref{eq:scaling-law-intensity-anomalous-general-reflection} coincide only if $K_1 = (\cos\Theta_{\textup{rec}})/(\cos\Theta_{{\textup{inc}}} + \cos\Theta_{{\textup{rec}}})$, $K_2 = (\cos\Theta_{\textup{inc}})/(\cos\Theta_{\textup{inc}} + \cos\Theta_{\textup{rec}})$, and
		$2K_1K_2 = [ \cos^2\Theta_{\textup{inc}} + \cos^2\Theta_{\textup{rec}} + \sin^2\Theta_{\textup{inc}}$ $\sin^2 \Theta_{\textup{rec}} \sin^2(\Phi_{\textup{inc}}-\Phi_{\textup{rec}})]/{(\cos\Theta_{\textup{inc}} + \cos\Theta_{\textup{rec}})^2}$ are satisfied simultaneously. This holds true only if $\SS$ is a uniform surface, i.e., $\alpha_R = \beta_R = 0$, which corresponds to specular reflection. As for anomalous reflection, \eqref{eq:scaling-law-intensity-anomalous-general-reflection} is an approximation for \eqref{eq:short-approx-3D-anomalous-general-case} because Taylor's approximation is used. The approximation in \eqref{eq:scaling-law-intensity-anomalous-general-reflection} depends, in general, on $\zeta'_1$ and $\zeta'_2$. A convenient choice for these parameters is $\zeta'_1=\zeta'_2$, since \eqref{eq:scaling-law-intensity-anomalous-general-reflection} is independent of $\zeta'_1$ and $\zeta'_2$ (i.e., $\zeta'_1$ and $\zeta'_2$ cancel out in \eqref{eq:scaling-law-intensity-anomalous-general-reflection}) if 	$\zeta'_1=\zeta'_2$. With the aid of \eqref{eq:scaling-law-intensity-anomalous-general-reflection}, the impact and scaling laws of key parameters can be unveiled.
\vspace{-1.0cm}
			\begin{remark}		
			Assume that Tx and Rx move along directions such that $(x_s,y_s)$, and $\theta_{\textup{inc}}(x_s,y_s)$, $\theta_{\textup{rec}}(x_s,y_s)$, $\varphi_{\textup{inc}}(x_s,y_s)$, $\varphi_{\textup{rec}}(x_s,y_s)$ are kept fixed. From \eqref{eq:scaling-law-intensity-anomalous-general-reflection}, we evince the following.
		\begin{itemize}
			\item In the electrically-large regime, $\left|{F_R(\vect{r}_{\textup{Rx}})}\right|$ decays as a function of the weighted sum of the Tx-to-$(x_s,y_s)$ and $(x_s,y_s)$-to-Rx distances. Also, $\left|{F_R(\vect{r}_{\textup{Rx}})}\right|$  is independent of the size of $\SS$.
			\item From \eqref{eq:condition-anomalous-general-reflection}, we have
			$\sin\theta_{\textup{inc}}(x_s,y_s)\cos\varphi_{\textup{inc}}(x_s,y_s) + \sin\theta_{\textup{rec}}(x_s,y_s)\cos\varphi_{\textup{rec}}(x_s,y_s) = - \alpha_R $ and $
			\sin\theta_{\textup{inc}}(x_s,y_s)\sin\varphi_{\textup{inc}}(x_s,y_s) + \sin\theta_{\textup{rec}}(x_s,y_s)\sin\varphi_{\textup{rec}}(x_s,y_s) = - \beta_R$. This implies that, in general, the polar and azimuthal angles of incidence and reflection in correspondence of the stationary point $(x_s,y_s)$ are different and depend on $\alpha_R$ and $\beta_R$. This corresponds to the generalized law of reflection. By using \eqref{eq:condition-anomalous-general-reflection}, in particular, $\alpha_R$ and $\beta_R$ can be optimized in order to obtain the desired angle of reflection for a given angle of incidence.
			\item If $\alpha_R = \beta_R = 0$, \eqref{eq:short-approx-3D-anomalous-general-case} and \eqref{eq:scaling-law-intensity-anomalous-general-reflection} reduce, as expected, to \eqref{eq:short-approx-3D-unifrom}.
		\end{itemize}
	\end{remark}
\vspace{-0.25cm}	
\vspace{-0.5cm}
	\begin{corollary}\label{corollary:electrically-small-anomalous-general-reflection}
	In the electrically-small regime, $F_R(\vect{r}_{\textup{Rx}})$ can be approximated as follows:\vspace{-0.25cm}	
	\begin{align}\label{eq:electrically-small-anomalous-general-reflection}	
	F_R(\vect{r}_{\textup{Rx}})
	&\approx
	\frac{jk\Omega_{\textup{ref}}(0,0;\hat{\vect{p}}_{\textup{ref}},\hat{\vect{p}}_{\textup{rec}})\left(\cos\theta_{{\textup{inc}}0}+\cos\theta_{{\textup{rec}}0}\right)}{16\pi^2d_{\textup{Tx}0}d_{\textup{Rx}0}}e^{-jk(d_{\textup{Tx}0}+d_{\textup{Rx}0}-(\phi_0+\phi_{\textup{ref}}+\phi_{\textup{rec}})/k)}  \\
	& \quad
	\int_{-L_y}^{L_y}\int_{-L_x}^{L_x} |\Gamma_{\textup{ref}}(x,y)|e^{jk(\mathcal{D}_{\alpha_R}x + \mathcal{D}_{\beta_R}y)}dxdy \nonumber \vspace{-0.25cm}	
	\end{align}
	where the shorthand notation $\mathcal{D}_{\alpha_R} = \alpha_R + \mathcal{D}_x$ and $\mathcal{D}_{\beta_R} = \beta_R + \mathcal{D}_y$ is used. If $|\Gamma_{\textup{ref}}(x,y)| = \Gamma_{\textup{ref}} > 0$ for $(x,y) \in \SS$, then $F_R(\vect{r}_{\textup{Rx}})$ can be further simplified as follows: \vspace{-0.20cm}	
	\begin{align}\label{eq:electrically-small-anomalous-general-reflection-constant-reflection}
	F_R(\vect{r}_{\textup{Rx}})
	&\approx \frac{jk\Gamma_{\textup{ref}}\Omega_{\textup{ref}}(0,0;\hat{\vect{p}}_{\textup{ref}},\hat{\vect{p}}_{\textup{rec}})L_xL_y\left(\cos\theta_{{\textup{inc}}0}+\cos\theta_{{\textup{rec}}0}\right)}{4\pi^2d_{\textup{Tx}0}d_{\textup{Rx}0}} 
	 \nonumber\\
	&\quad  
	\sinc\left(kL_x\mathcal{D}_{\alpha_R}\right) \sinc\left(kL_y\mathcal{D}_{\beta_R}\right) 
	e^{-jk(d_{\textup{Tx}0}+d_{\textup{Rx}0}-(\phi_0+\phi_{\textup{ref}}+\phi_{\textup{rec}})/k)} \vspace{-0.25cm}	
	\end{align}
	\end{corollary}
	\vspace{-0.5cm}	
	\begin{proof}
		It follows by substituting $\mathcal{C}(x,y) = k(\alpha_R x + \beta_R y) + \phi_0$ in \eqref{eq:long-distance-approximation-3D}.
	\end{proof}
	\vspace{-0.5cm}	
	\begin{remark}
		From \eqref{eq:electrically-small-anomalous-general-reflection-constant-reflection}, conclusions similar to Remark 7 can be drawn with one exception. $\left|{F_R(\vect{r}_{\textup{Rx}})}\right|$ in \eqref{eq:electrically-small-anomalous-general-reflection-constant-reflection} attains its maximum in correspondence of angles of incidence and reflection that fulfill the equalities $\alpha_R = -(\sin\theta_{\textup{inc}0}\cos\theta_{\textup{inc}0} + \sin\theta_{\textup{rec}0}\cos\theta_{\textup{rec}0})$ and $\beta_R = -(\sin\theta_{\textup{inc}0}\sin\theta_{\textup{inc}0} + \sin\theta_{\textup{rec}0}\sin\theta_{\textup{rec}0})$. Thus, $\alpha_R$ and $\beta_R$ can be appropriately optimized for maximizing the reflected signal towards a desired direction, given the angle of incidence with respect to the center of $\SS$. 
	\end{remark}

\vspace{-0.8cm}	
	\subsection{$\SS$ is Configured for Focusing} \vspace{-0.25cm}	
	\label{sec:focusing-lens-reflection}
This setup is obtained by setting $\angle \Gamma_{\textup{ref}}(x,y) = k \left(d_{\textup{Tx}}(x,y) + d_{\textup{Rx}}(x,y)\right) + \phi_0$	for $(x,y) \in \SS$, where $\phi_0 \in [0,2\pi)$ is a fixed phase. With this setup, $F_R(\vect{r}_{\textup{Rx}})$ in \eqref{eq:Ex-surface-contribution-reflection} simplifies as follows: \vspace{-0.25cm}	
	\begin{align}\label{eq:Ex-focusing-lens}
	F_R(\vect{r}_{\textup{Rx}})
	&\approx \frac{jk   e^{j(\phi_0+\phi_\textup{rec}+\phi_\textup{ref})}}{16\pi^2}
	\int_{-L_y}^{L_y}\int_{-L_x}^{L_x} \Omega_{\textup{ref}}(x,y;\hat{\vect{p}}_{\textup{ref}},\hat{\vect{p}}_{\textup{rec}}) \nonumber\\
	&\quad   \frac{|\Gamma_{\textup{ref}}(x,y)|\left(\cos\theta_\textup{inc}(x,y) + \cos\theta_\textup{rec}(x,y)\right)}{d_\textup{Tx}(x,y)d_{\textup{Rx}}(x,y)}dxdy \vspace{-0.25cm}	
	\end{align}

In the electrically-large regime, \eqref{eq:Ex-focusing-lens} cannot be, in general, further simplified, since no fast oscillating term is present in the integrand function, and, hence, the stationary phase method cannot be applied. In this case, therefore, we focus our attention on analyzing an upper-bound for $\left|F_R(\vect{r}_{\textup{Rx}})\right|$, in order to unveil the impact of the size of $\SS$ (e.g., if it tends to infinity).
	\vspace{-0.3cm}	
	\begin{corollary}\label{corollary:bound-Lx-Ly}
		Assume $d_{\textup{P}1}(x,y) \leq d_{\textup{P}2}(x,y)$, where $(\textup{P}1,\textup{P}2) = (\textup{Tx},\textup{Rx})$ or $(\textup{P}1,\textup{P}2) = (\textup{Rx},\textup{Tx})$, $|\Gamma_{\textup{ref}}(x,y)| = \Gamma_{\textup{ref}} > 0$ for $(x,y) \in \SS$, and $z_{\textup{P}1}\neq0$. Define $C_{\textup{ref}} = \frac{2k^3p_{\textup{dm}}\Gamma_{\textup{ref}}\mathcal{E}\left(\hat{\vect{p}}_{\textup{inc}},\hat{\vect{p}}_{\textup{ref}}\right)}{16\pi^2\epsilon_0}$. Then: 	\vspace{-0.15cm}	
		\begin{equation}\label{eq:upperbound-dTx}
		\left|F_R(\vect{r}_{\textup{Rx}})\right|
		\leq
		C_{\textup{ref}} \left(1+\frac{z_{\textup{P}2}}{z_{\textup{P}1}}\right)  \tan^{-1}\left[\frac{(x_{\textup{P}1}-x)(y_{\textup{P}1}-y)}{z_{\textup{P}1}\sqrt{(x_{\textup{P}1}-x)^2+(y_{\textup{P}1}-y)^2+z_{\textup{P}1}^2}}\right]\Bigr|^{x=L_x}_{x=-L_x}\Bigr|^{y=L_y}_{y=-L_y} 	\vspace{-0.15cm}	
		\end{equation}
	\end{corollary}
	\vspace{-0.25cm}	
	\begin{proof}
		Consider $\Omega_{\textup{ref}}(x,y;\hat{\vect{p}}_{\textup{ref}},\hat{\vect{p}}_{\textup{rec}})$ in Proposition \ref{proposition:reflected-field-general}, where $\tilde{\vect{p}}_{\textup{rec}}$, $\tilde{\vect{p}}_{\textup{ref}}$, and $\hat{\vect{s}}_{(x,y)}$ are real unit-norm vectors. By virtue of Cauchy-Schwarz's inequality (i.e., $-\left|{\vect{u}}\right|\left|{\vect{v}}\right| \leq \vect{u}\cdot\vect{v} \leq \left|{\vect{u}}\right| \left|{\vect{v}}\right|$ for any $\vect{u}$ and $\vect{v}$), we have $-1\leq\tilde{\vect{p}}_{\textup{rec}}\cdot\tilde{\vect{p}}_{\textup{ref}} \leq 1$, $-1\leq\hat{\vect{s}}_{(x,y)}\cdot\tilde{\vect{p}}_{\textup{rec}} \leq 1$, and $-1\leq\hat{\vect{s}}_{(x,y)}\cdot\tilde{\vect{p}}_{\textup{ref}} \leq 1$. Hence, we obtain $-2\frac{k^2p_{\textup{dm}}\mathcal{E}\left(\hat{\vect{p}}_{\textup{inc}},\hat{\vect{p}}_{\textup{tran}}\right)}{\epsilon_0}\leq \Omega_{\textup{ref}}(x,y;\hat{\vect{p}}_{\textup{ref}},\hat{\vect{p}}_{\textup{rec}}) \leq 2\frac{k^2p_{\textup{dm}}\mathcal{E}\left(\hat{\vect{p}}_{\textup{inc}},\hat{\vect{p}}_{\textup{tran}}\right)}{\epsilon_0}$ for $(x,y) \in \SS$. Thus: \vspace{-0.15cm}	
		\begin{equation}
		\left|F_R(\vect{r}_{\textup{Rx}})\right|
		\leq
		\frac{2k^3p_{\textup{dm}}\Gamma_{\textup{ref}}\mathcal{E}\left(\hat{\vect{p}}_{\textup{inc}},\hat{\vect{p}}_{\textup{ref}}\right)}{16\pi^2\epsilon_0} 
		\int_{-L_y}^{L_y}\int_{-L_x}^{L_x}  \frac{\left(\cos\theta_\textup{inc}(x,y) + \cos\theta_\textup{rec}(x,y)\right)}{d_\textup{Tx}(x,y)d_{\textup{Rx}}(x,y)}dxdy \vspace{-0.15cm}	
		\end{equation}
		Since $d_{\textup{P}1}(x,y) \leq d_{\textup{P}2}(x,y)$, we have $\frac{\left(\cos\theta_\textup{inc}(x,y) + \cos\theta_\textup{rec}(x,y)\right)}{d_{\textup{P}1}(x,y)d_{\textup{P}2}(x,y)} \leq \frac{z_{\textup{Tx}}+z_{\textup{Rx}}}{(d_{\textup{P}1}(x,y))^{3}}$. Using the notable integral $\int_{-L_y}^{L_y}\int_{-L_x}^{L_x}		\frac{1}{(d_{\textup{P}1}(x,y))^{3}}dxdy 		=		z_{\textup{P}1}^{-1} {\tan^{-1}\left[\frac{(x_{\textup{P}1}-x)(y_{\textup{P}1}-y)}{z_{\textup{P}1}\sqrt{(x_{\textup{P}1}-x)^2+(y_{\textup{P}1}-y)^2+z_{\textup{P}1}^2}}\right]\Bigr|^{x=L_x}_{x=-L_x}\Bigr|^{y=L_y}_{y=-L_y}}$, the proof follows.
\end{proof}
		\vspace{-0.6cm}	
	\begin{remark}
	From Corollary \ref{corollary:bound-Lx-Ly}, we observe that $\left|F_R(\vect{r}_{\textup{Rx}})\right| \lessapprox		\left(1+\frac{z_{\textup{P}2}}{z_{\textup{P}1}}\right)\frac{2k^3p_{\textup{dm}}\Gamma_{\textup{ref}}\mathcal{E}\left(\hat{\vect{p}}_{\textup{inc}},\hat{\vect{p}}_{\textup{ref}}\right)}{8\pi\epsilon_0}$ for $L_x,L_y \to \infty$. This implies that $\left|F_R(\vect{r}_{\textup{Rx}})\right|$ is upper-bounded if the size of $\SS$ increases without bound. Thus, the received power is bounded even for an infinitely large RIS. The scaling law as a function of the transmission distances is, in general, different from the weighted-sum distance obtained in \eqref{eq:short-approx-3D-anomalous-general-case}. This is because of the different optimization of $\angle \Gamma_{\textup{ref}}(x,y)$. In the electrically-large regime, an anomalous reflecting RIS and a focusing RIS behave, in general, differently.
	\end{remark}
\vspace{-0.6cm}	
	\begin{corollary}\label{corollary:electrically-small-lens-reflection}
		In the electrically-small regime, $F_R(\vect{r}_{\textup{Rx}})$ in \eqref{eq:Ex-focusing-lens} can be approximated as follows: \vspace{-0.15cm}	
		\begin{equation}\label{eq:long-approx-3D-beamforming}
		F_R(\vect{r}_{\textup{Rx}})
		\approx  \frac{jk e^{j(\phi_0+\phi_\textup{rec}+\phi_\textup{ref})}}{16\pi^2 d_{\textup{Tx}0}d_{\textup{Rx}0}}
		\Omega_{\textup{ref}}(0,0;\hat{\vect{p}}_{\textup{ref}},\hat{\vect{p}}_{\textup{rec}})  \left(\cos\theta_{\textup{inc}0}+\cos\theta_{\textup{rec}0}\right)\int_{-L_y}^{L_y}\int_{-L_x}^{L_x} |\Gamma_{\textup{ref}}(x,y)|dxdy \vspace{-0.25cm}	
		\end{equation}
		If $|\Gamma_{\textup{ref}}(x,y)| = \Gamma_{\textup{ref}} >0$ for $(x,y) \in \SS$, then $F_R(\vect{r}_{\textup{Rx}})$ can be further approximated as follows: \vspace{-0.15cm}	
		\begin{equation}\label{eq:long-approx-3D-beamforming-constant-reflection}
		F_R(\vect{r}_{\textup{Rx}}) 
		\approx \frac{jk\Gamma_{\textup{ref}}\Omega_{\textup{ref}}(0,0;\hat{\vect{p}}_{\textup{ref}},\hat{\vect{p}}_{\textup{rec}})L_xL_y\left(\cos\theta_{\textup{inc}0}+\cos\theta_{\textup{rec}0}\right)}{4\pi^2d_{\textup{Tx}0}d_{\textup{Rx}0}}
		e^{j(\phi_0+\phi_\textup{rec}+\phi_\textup{ref})} \vspace{-0.25cm}	
		\end{equation}
	\end{corollary}
	\vspace{-0.25cm}	
	\begin{proof}
		It follows directly from \eqref{eq:long-distance-approximation-3D-type-2}.
	\end{proof}
\vspace{-0.5cm}	
	\begin{remark}
The scaling laws of $\left|F_R(\vect{r}_{\textup{Rx}})\right|$ in \eqref{eq:long-approx-3D-beamforming-constant-reflection} as a function of the distances and the size of $\SS$ are the same as in \eqref{eq:electrically-small-anomalous-general-reflection-constant-reflection} for anomalous reflection. This can be justified by analyzing $\angle \Gamma_{\textup{ref}}(x,y)$ for focusing and anomalous reflection. As for focusing, $\angle \Gamma_{\textup{ref}}(x,y) = k \left(d_{\textup{Tx}}(x,y) + d_{\textup{Rx}}(x,y)\right) + \phi_0$. In the electrically-small regime, $d_{\textup{Tx}}(x,y)$ and $d_{\textup{Rx}}(x,y)$ can be approximated by using \eqref{eq:Taylor-Tx} and ignoring $R_2(x,y)$, which yields $\angle \Gamma_{\textup{ref}}(x,y) = k(\alpha_R + \beta_R) + ({d_{{\rm{Tx0}}}} + {d_{{\rm{Rx0}}}} + \phi_0)$ with $\alpha_R$ and $\beta_R$ as given in Remark 9. The obtained $\angle \Gamma_{\textup{ref}}(x,y)$ coincides with that of a surface that operates as an anomalous reflector towards the same direction as the focusing spot of a surface that operates as a focusing lens. In the electrically-small regime, hence, anomalous reflectors and focusing lenses are almost equivalent. This does not apply in the electrically-large regime.
\end{remark}

		\vspace{-0.75cm}
	\section{Electric Field In the Presence of a Transmitting Surface}\label{section:transmission} 	\vspace{-0.25cm}
In this section, we analyze $\vect{E}(\vect{r}_{\textup{Rx}})$ under the assumption that $\SS$ is a transmitting surface according to the definitions and assumptions given in Section \ref{section:system-model} (see Fig. \ref{fig:systemmodeltransmission} and Fig. \ref{fig:kirchhofftransmission}). Some analytical steps are similar to the setup of reflecting surfaces. Thus, only the final results and the most important steps of the analysis are reported. The same applies to the performance trends.
	\vspace{-0.25cm}
	
	\begin{proposition}\label{proposition:transmitted-field-general}		
	Let $\hat{\vect{s}}_{(x,y)} = \sin\theta_{\textup{inc}}(x,y)\cos\varphi_{\textup{inc}}(x,y)\hat{\vect{x}} + \sin\theta_{\textup{inc}}(x,y)\sin\varphi_{\textup{inc}}(x,y)\hat{\vect{y}} + \cos\theta_{\textup{inc}}(x,y)\hat{\vect{z}}$, be the  unit-norm propagation vector from $\vect{r}_{\textup{Tx}}$ to $\vect{s} = x \hat{\vect{x}} + y \hat{\vect{y}} \in \SS$. 
	Define $\Omega_{\textup{inc}}(x,y;\hat{\vect{p}}_{\textup{inc}},\hat{\vect{p}}_{\textup{rec}}) = (k^2/\epsilon_0) p_{\textup{dm}} \left(\tilde{\vect{p}}_{\textup{rec}}\cdot\tilde{\vect{p}}_{\textup{inc}} - \left(\hat{\vect{s}}_{(x,y)}\cdot\tilde{\vect{p}}_{\textup{rec}}\right)\left(\hat{\vect{s}}_{(x,y)}\cdot\tilde{\vect{p}}_{\textup{inc}}\right)\right) $ and $\Omega_{\textup{tran}}(x,y;\hat{\vect{p}}_{\textup{tran}},\hat{\vect{p}}_{\textup{rec}}) 
	=
	(k^2/\epsilon_0) p_{\textup{dm}} (\tilde{\vect{p}}_{\textup{rec}}\cdot\tilde{\vect{p}}_{\textup{tran}} - \left(\hat{\vect{s}}_{(x,y)}\cdot\tilde{\vect{p}}_{\textup{rec}}\right)\left(\hat{\vect{s}}_{(x,y)}\cdot\tilde{\vect{p}}_{\textup{tran}}\right) )\mathcal{E}\left(\hat{\vect{p}}_{\textup{inc}},\hat{\vect{p}}_{\textup{tran}}\right)$.
	Under the assumptions stated in Lemma \ref{lemma:incident-E-dipole}, the electric field $\vect{E}(\vect{r}_{\textup{Rx}})$ projected onto $\hat{\vect{p}}_{\textup{rec}}$ can be formulated as follows: 	\vspace{-0.25cm}
		\begin{align}\label{eq:transmitted-field-general}
		\vect{E}(\vect{r}_{\textup{Rx}}) \cdot \hat{\vect{p}}_{\textup{rec}}
		&\approx
		\hat{\vect{p}}_{\textup{rec}} \cdot \vect{E}_{0,\textup{inc}}\left(\vect{r}_{\textup{Rx}};\hat{\vect{p}}_{\textup{inc}}\right) G(\vect{r}_{\textup{Rx}},\vect{r}_{\textup{Tx}})  
		+ jk \int_{\SS}  \left[\Gamma_{\textup{tran}}(x,y)\Omega_{\textup{tran}}(x,y;\hat{\vect{p}}_{\textup{tran}},\hat{\vect{p}}_{\textup{rec}}) e^{j(\phi_{\textup{tran}}+\phi_{\textup{rec}})} \right. 
		\nonumber\\
		&\hspace{-0.5cm} \left.
		- \Omega_{\textup{inc}}(x,y;\hat{\vect{p}}_{\textup{inc}},\hat{\vect{p}}_{\textup{rec}})e^{j(\phi_{\textup{inc}}+\phi_{\textup{rec}})}\right] G\left(\vect{s},\vect{r}_{\textup{Tx}}\right)G(\vect{r}_{\textup{Rx}},\vect{s}) \left[\frac{z_{\textup{Rx}}}{|\vect{s} - \vect{r}_{\textup{Rx}}|}  - \frac{z_{\textup{Tx}}}{|\vect{s} - \vect{r}_{\textup{Tx}}|} \right] d\vect{s} %\nonumber
		\end{align} %\vspace{-0.25cm}
	\begin{equation}\label{eq:transmitted-field-general-A} 
	\begin{split}
		&\hspace{1.5cm} = \hat{\vect{p}}_{\textup{rec}} \cdot \vect{E}_{0,\textup{inc}}\left(\vect{r}_{\textup{Rx}};\hat{\vect{p}}_{\textup{inc}}\right)G(\vect{r}_{\textup{Rx}},\vect{r}_{\textup{Tx}}) \\ &\hspace{1.5cm} \quad 
		- \mathcal{I}_0 \int_{-L_y}^{L_y}\int_{-L_x}^{L_x} \mathcal{I}_D(x,y)e^{-jk\mathcal{P}_D(x,y)}dxdy
		+ \mathcal{I}_0 \int_{-L_y}^{L_y}\int_{-L_x}^{L_x} \mathcal{I}_T(x,y)e^{-jk\mathcal{P}_T(x,y)}dxdy \nonumber
		\end{split} \vspace{-0.15cm}
		\end{equation}
		where $\mathcal{I}_0 = jk/(16\pi^2)$, $\mathcal{P}_D(x,y)
		= 
		d_\textup{Tx}(x,y) + d_\textup{Rx}(x,y) -  (\angle \phi_{\textup{inc}} + \angle \phi_{\textup{rec}}) /k$, $\mathcal{P}_T(x,y)
		= 
		d_\textup{Tx}(x,y) + d_\textup{Rx}(x,y) -  (\angle\Gamma_{\textup{tran}}(x,y) + \angle \phi_{\textup{tran}} + \angle \phi_{\textup{rec}})/k$, and the following shorthands are used: \vspace{-0.25cm}
		\begin{align}
		\mathcal{I}_D(x,y) &= \frac{\Omega_{\textup{inc}}(x,y;\hat{\vect{p}}_{\textup{inc}},\hat{\vect{p}}_{\textup{rec}})\left(\cos\theta_{\textup{inc}}(x,y) + \cos\theta_{\textup{rec}}(x,y)\right) }{d_\textup{Tx}(x,y)d_\textup{Rx}(x,y)} 
		\\
		\mathcal{I}_T(x,y) 
		&= \frac{\left|\Gamma_{\textup{tran}}(x,y)\right|\Omega_{\textup{inc}}(x,y;\hat{\vect{p}}_{\textup{tran}},\hat{\vect{p}}_{\textup{rec}})\left(\cos\theta_{\textup{inc}}(x,y) + \cos\theta_{\textup{rec}}(x,y)\right)}{d_\textup{Tx}(x,y)d_\textup{Rx}(x,y)}
		\end{align}
	\end{proposition}
	\vspace{-0.60cm}
	\begin{proof}
		See Appendix G. %\ref{appendix:proof-of-transmitted-field-general}
	\end{proof}
\vspace{-0.5cm}
	\begin{remark}
		Consider $\hat{\vect{p}}_{\textup{tran}} = \hat{\vect{p}}_{\textup{inc}}$, $\angle \Gamma_{\textup{tran}}(x,y) = 0$, and $|\Gamma_{\textup{tran}}(x,y)| = 1$. By definition, we have $\mathcal{E}(\hat{\vect{p}}_{\textup{inc}},\hat{\vect{p}}_{\textup{inc}}) = 1$. Then, the integral terms in Proposition \ref{proposition:transmitted-field-general} coincide and their difference vanishes. This result is consistent with the fact that, under the considered special setup, the surface transmits the impinging wave without any modifications. Therefore, we retrieve the setup in the absence of $\SS$, and the received field coincides with the incident field in the absence of $\SS$.
	\end{remark}
\vspace{-0.25cm}

In \eqref{eq:transmitted-field-general}, the only term that depends on the design and properties of $\SS$ is the last one, which we denote by $F_T(\vect{r}_{\textup{Rx}}) = \mathcal{I}_0 \int_{-L_y}^{L_y}\int_{-L_x}^{L_x} \mathcal{I}_T(x,y)e^{-jk\mathcal{P}_T(x,y)}dxdy$. In the next sub-sections, therefore, we focus our attention only on the analysis of $F_T(\vect{r}_{\textup{Rx}})$. Similar to Section \ref{section:reflection}, $F_T(\vect{r}_{\textup{Rx}})$ is analyzed as a function of important design parameters and configurations for $\SS$, e.g., $\angle\Gamma_{\textup{tran}}(x,y)$. As illustrative examples, similar to reflecting surfaces, we consider phase gradient metasurfaces \cite{Marco-JSAC}.

\vspace{-0.5cm}
	\subsection{$\SS$ is Configured for Specular Transmission} \vspace{-0.25cm}
	This setup is obtained if $\angle \Gamma_{\textup{tran}}(x,y) = \phi_0$ for $(x,y) \in \SS$, where $\phi_0 \in [0,2\pi)$ is a fixed phase. \vspace{-1.00cm}
	\begin{corollary}\label{corollary:electrically-large-uniform-transmission}
		Let $(x_s,y_s) \in \SS$ be the solution of the following system of equations: \vspace{-0.15cm}
		\begin{equation}
		\frac{(x_s-x_\textup{Tx})}{d_\textup{Tx}(x_s,y_s)} + \frac{(x_s-x_\textup{Rx})}{d_\textup{Rx}(x_s,y_s)} = 0, \quad\quad
		\frac{(y_s-x_\textup{Tx})}{d_\textup{Tx}(x_s,y_s)} + \frac{(y_s-x_\textup{Rx})}{d_\textup{Rx}(x_s,y_s)} = 0 \vspace{-0.15cm} \label{eq:condition-uniform-transmission}
		\end{equation}
		In the electrically-large regime, $F_T(\vect{r}_{\textup{Rx}})$ can be approximated as follows: \vspace{-0.15cm}
		\begin{equation}\label{eq:electrically-large-anomalous-general-transmission}
		F_T(\vect{r}_{\textup{Rx}})
		\approx
		\frac{\left|\Gamma_{\textup{tran}}(x_s,y_s)\right|\Omega_{\textup{tran}}(x_s,y_s;\hat{\vect{p}}_{\textup{tran}},\hat{\vect{p}}_{\textup{rec}})}{4\pi(d_\textup{Tx}(x_s,y_s)+d_\textup{Rx}(x_s,y_s))}  e^{-jk\left(d_\textup{Tx}(x_s,y_s)+d_\textup{Rx}(x_s,y_s) - (\phi_0+\phi_{\textup{tran}}+\phi_{\textup{rec}})/k\right)} \vspace{-0.15cm}
		\end{equation}
	\end{corollary}
	\vspace{-0.2cm}
	\begin{proof}
		It is similar to the proof of Corollary \ref{corollary:electrically-large-uniform-reflection}. 
	\end{proof}

		\vspace{-0.5cm}
	\begin{corollary}\label{corollary:electrically-small-uniform-transmission}
		In the electrically-small regime, $F_T(\vect{r}_{\textup{Rx}})$ can be approximated as follows: \vspace{-0.15cm}
		\begin{equation}
		F_T(\vect{r}_{\textup{Rx}})
		\approx
		\frac{jk\Omega_{\textup{tran}}(0,0;\hat{\vect{p}}_{\textup{tran}},\hat{\vect{p}}_{\textup{rec}})\left(\cos\theta_{{\textup{inc}}0}+\cos\theta_{{\textup{rec}}0}\right)}{16\pi^2d_{\textup{Tx}0}d_{\textup{Rx}0}}e^{-jk\left(d_{\textup{Tx}0}+d_{\textup{Rx}0}-(\phi_0+\phi_{\textup{tran}}+\phi_{\textup{rec}})/k\right)}   \nonumber
		\end{equation}
		\begin{equation}
		\hspace{-3.0cm} \int_{-L_y}^{L_y}\int_{-L_x}^{L_x} |\Gamma_{\textup{tran}}(x,y)|e^{jk(\mathcal{D}_xx+\mathcal{D}_yy)}dxdy
		\end{equation}
where definitions and notation similar to Corollary \ref{corollary:electrically-small-uniform-reflection} are employed. If $|\Gamma_{\textup{tran}}(x,y)| = \Gamma_{\textup{tran}} > 0$ for $(x,y) \in \SS$, then $F_T(\vect{r}_{\textup{Rx}})$ can be simplified as follows: \vspace{-0.15cm}
		\begin{align}\label{eq:electrically-small-uniform-transmission-constant-transmission}
		F_R(\vect{r}_{\textup{Rx}})
		&\approx
		\frac{jk\Gamma_{\textup{tran}}\Omega_{\textup{tran}}(0,0;\hat{\vect{p}}_{\textup{tran}},\hat{\vect{p}}_{\textup{rec}})L_xL_y\left(\cos\theta_{{\textup{inc}}0}+\cos\theta_{{\textup{rec}}0}\right)}{4\pi^2d_{\textup{Tx}0}d_{\textup{Rx}0}} 
		\sinc\left(kL_x\mathcal{D}_x\right)\sinc\left(kL_y\mathcal{D}_y\right) 
		\nonumber\\
		& \quad
		e^{-jk\left(d_{\textup{Tx}0}+d_{\textup{Rx}0}-(\phi_0+\phi_{\textup{tran}}+\phi_{\textup{rec}})/k\right)}
		\end{align}
	\end{corollary}
	\vspace{-0.65cm}
	\begin{proof}
		It follows by direct application of \eqref{eq:long-distance-approximation-3D}. 
	\end{proof}

\vspace{-0.20cm}
The intensity of $F_T(\vect{r}_{\textup{Rx}})$ in Corollaries \ref{corollary:electrically-large-uniform-transmission} and \ref{corollary:electrically-small-uniform-transmission} is similar to that in Corollaries \ref{corollary:electrically-large-uniform-reflection} and \ref{corollary:electrically-small-uniform-reflection}, respectively. Therefore, similar scaling laws and performance trends are obtained. In the electrically-large regime, in particular, the law of transmission, i.e., $\varphi_{{\textup{inc}}}(x_s,y_s) = (\varphi_{{\textup{rec}}}(x_s,y_s) + \pi) \mod 2\pi$ and $\theta_{\textup{inc}}(x_s,y_s) = \theta_{\textup{rec}}(x_s,s_y)$ can be retrieved by direct inspection of \eqref{eq:condition-uniform-transmission}.

	\vspace{-0.4cm}
	\subsection{$\SS$ is Configured for Anomalous Transmission} \vspace{-0.25cm}
	This setup is obtained by setting $\angle \Gamma_{\textup{tran}}(x,y) = k(\alpha_T x + \beta_T y) + \phi_0$ for $(x,y) \in \SS$, where $\alpha_T \in \mathbb{R}$ and $\beta_T \in \mathbb{R}$ are design parameters, and $\phi_0 \in [0,2\pi)$ is a fixed phase. Similar to reflecting surfaces, the direction of anomalous transmission is determined by the setup of $\alpha_T$ and $\beta_T$. \vspace{-0.35cm}

	\begin{corollary}\label{corollary:electrically-large-anomalous-general-transmission}
		Let $(x_s,y_s)\in \SS$ be the solution of the following system of equations: \vspace{-0.15cm}
		\begin{equation}
		\frac{(x_s-x_\textup{Tx})}{d_\textup{Tx}(x_s,y_s)} + \frac{(x_s-x_\textup{Rx})}{d_\textup{Rx}(x_s,y_s)} = \alpha_T, \quad\quad
		\frac{(y_s-x_\textup{Tx})}{d_\textup{Tx}(x_s,y_s)} + \frac{(y_s-x_\textup{Rx})}{d_\textup{Rx}(x_s,y_s)} = \beta_T \label{eq:condition-anomalous-general-transmission} \vspace{-0.15cm}
		\end{equation}
Assume the same notation and definitions as in Corollary \ref{corollary:electrically-large-anomalous-general-reflection} and Corollary \ref{corollary:electrically-large-anomalous-general-reflection-scaling}. In the electrically-large regime, $F_T(\vect{r}_{\textup{Rx}})$ can be approximated as follows: \vspace{-0.15cm}
		\begin{align}\label{electrically-small-anomalous-general-transmission}
		&F_T(\vect{r}_{\textup{Rx}})
		\approx
		\frac{\left|\Gamma_{\textup{tran}}(x_s,y_s)\right|\Omega_{\textup{tran}}(x_s,y_s;\hat{\vect{p}}_{\textup{tran}},\hat{\vect{p}}_{\textup{rec}})e^{-jk(d_\textup{Tx}(x_s,y_s)+d_\textup{Rx}(x_s,y_s) - (\alpha_T x_s + \beta_T y_s) -(\phi_0+\phi_{\textup{tran}}+\phi_{\textup{rec}})/k)}}{8\pi{\sqrt{\mathcal{R}_{1} (d_{\textup{Tx}}(x_s,y_s))^2+\mathcal{R}_{2} (d_{\textup{Rx}}(x_s,y_s))^2 + \mathcal{R}_{3} d_{\textup{Tx}}(x_s,y_s)d_{\textup{Rx}}(x_s,y_s)}}} \nonumber \\ 
		&\approx
		\frac{\left|\Gamma_{\textup{tran}}(x_s,y_s)\right|\Omega_{\textup{tran}}(x_s,y_s;\hat{\vect{p}}_{\textup{tran}},\hat{\vect{p}}_{\textup{rec}})}{8\pi(K_1{d_{\textup{Tx}}(x_s,y_s)} + K_2{d_{\textup{Rx}}(x_s,y_s))}}
		e^{-jk(d_\textup{Tx}(x_s,y_s)+d_\textup{Rx}(x_s,y_s) - (\alpha_T x_s + \beta_T y_s) -(\phi_0+\phi_{\textup{tran}}+\phi_{\textup{rec}})/k)} \vspace{-0.15cm}
		\end{align} 
	\end{corollary}
\vspace{-0.25cm}
	\begin{proof}
		It is similar to the proofs of Corollary \ref{corollary:electrically-large-anomalous-general-reflection} and Corollary \ref{corollary:electrically-large-anomalous-general-reflection-scaling}.
	\end{proof}
	
\vspace{-0.5cm}	
	\begin{corollary}\label{corollary:electrically-small-anomalous-general-transmission}
		In the electrically-small regime, $F_T(\vect{r}_{\textup{Rx}})$ can be approximated as follows: \vspace{-0.25cm}	
		\begin{align}
		F_T(\vect{r}_{\textup{Rx}})
		&\approx
		\frac{jk\Omega_{\textup{tran}}(0,0;\hat{\vect{p}}_{\textup{tran}},\hat{\vect{p}}_{\textup{rec}})\left(\cos\theta_{{\textup{inc}}0}+\cos\theta_{{\textup{rec}}0}\right)}{16\pi^2d_{\textup{Tx}0}d_{\textup{Rx}0}}e^{-jk(d_{\textup{Tx}0}+d_{\textup{Rx}0}-(\phi_0+\phi_{\textup{tran}})/k)}  \\
		& \quad
		\int_{-L_y}^{L_y}\int_{-L_x}^{L_x} |\Gamma_{\textup{tran}}(x,y)|e^{jk(\mathcal{D}_{\alpha_T}x + \mathcal{D}_{\beta_T}y)}dxdy \nonumber
		\end{align}
		where the shorthand notation $\mathcal{D}_{\alpha_T} = \alpha_T + \mathcal{D}_x$ and $\mathcal{D}_{\beta_T} = \beta_T + \mathcal{D}_y$ is used. If $\left| \Gamma_{\textup{tran}}(x,y) \right| = \Gamma_{\textup{tran}}$ for $(x,y) \in \SS$, then $F_T(\vect{r}_{\textup{Rx}})$  can further be approximated as follows: \vspace{-0.15cm}	
		\begin{align}
		F_T(\vect{r}_{\textup{Rx}})
		&\approx 
		\frac{jk\Gamma_{\textup{tran}}\Omega_{\textup{tran}}(0,0;\hat{\vect{p}}_{\textup{tran}},\hat{\vect{p}}_{\textup{rec}})L_xL_y\left(\cos\theta_{{\textup{inc}}0}+\cos\theta_{{\textup{rec}}0}\right)}{4\pi^2d_{\textup{Tx}0}d_{\textup{Rx}0}}  
		\sinc\left(kL_x\mathcal{D}_{\alpha_T}\right) \sinc\left(kL_y\mathcal{D}_{\beta_T}\right) 
		\nonumber\\
		&\quad 
		e^{-jk(d_{\textup{Tx}0}+d_{\textup{Rx}0}-(\phi_0+\phi_{\textup{tran}})/k)}
		\end{align} 
	\end{corollary}	
\vspace{-0.5cm}	
	\begin{proof}
		It is similar to the proof of Corollary \ref{corollary:electrically-small-anomalous-general-reflection}.
	\end{proof}

\vspace{-0.20cm}
Once again, we observe that the intensity of $F_T(\vect{r}_{\textup{Rx}})$ in Corollaries \ref{corollary:electrically-large-anomalous-general-transmission} and \ref{corollary:electrically-small-anomalous-general-transmission} is similar to that in Corollaries \ref{corollary:electrically-large-anomalous-general-reflection} and \ref{corollary:electrically-small-anomalous-general-reflection}, respectively. In particular, the angles of transmission can be optimized through the setup of $\alpha_T$ and $\beta_T$, similar to the optimization of $\alpha_R$ and $\beta_R$ for reflecting surfaces.

\vspace{-0.5cm}
	\subsection{$\SS$ is Configured for Focusing} \label{sec:focusing-lens-transmission} \vspace{-0.25cm} 
	This setup is obtained by setting $\angle \Gamma_{\textup{tran}}(x,y) = k \left(d_{\textup{Tx}}(x,y) + d_{\textup{Rx}}(x,y)\right) + \phi_0$	for $(x,y) \in \SS$, where $\phi_0 \in [0,2\pi)$ is a fixed phase. With this setup, $F_T(\vect{r}_{\textup{Rx}})$ in \eqref{eq:transmitted-field-general} simplifies as follows: \vspace{-0.25cm}	
	\begin{align}\label{eq:Ex-focusing-lens-transmission}
	F_T(\vect{r}_{\textup{Rx}})
	&\approx \frac{jk   e^{j(\phi_0+\phi_\textup{rec}+\phi_\textup{tran})}}{16\pi^2}
	\int_{-L_y}^{L_y}\int_{-L_x}^{L_x} \Omega_{\textup{tran}}(x,y;\hat{\vect{p}}_{\textup{tran}},\hat{\vect{p}}_{\textup{rec}}) \nonumber\\
	&\quad   \frac{|\Gamma_{\textup{tran}}(x,y)|\left(\cos\theta_\textup{inc}(x,y) + \cos\theta_\textup{rec}(x,y)\right)}{d_\textup{Tx}(x,y)d_{\textup{Rx}}(x,y)}dxdy
	\end{align}
Similar to reflecting surfaces, the following corollaries provide an upper-bound and an asymptotic approximation for \eqref{eq:Ex-focusing-lens-transmission} in the electrically-large and electrically-small regimes, respectively.

		\vspace{-0.25cm}
	\begin{corollary}\label{corollary:bound-Lx-Ly-transmission}
	Assume $d_{\textup{P}1}(x,y) \leq d_{\textup{P}2}(x,y)$, where $(\textup{P}1,\textup{P}2) = (\textup{Tx},\textup{Rx})$ or $(\textup{P}1,\textup{P}2) = (\textup{Rx},\textup{Tx})$, $|\Gamma_{\textup{tran}}(x,y)| = \Gamma_{\textup{tran}} > 0$ for $(x,y) \in \SS$, and $z_{\textup{P}1}\neq0$. Define $C_{\textup{tran}} = \frac{2k^3p_{\textup{dm}}\Gamma_{\textup{tran}}\mathcal{E}\left(\hat{\vect{p}}_{\textup{inc}},\hat{\vect{p}}_{\textup{tran}}\right)}{16\pi^2\epsilon_0}$. Then: 	\vspace{-0.15cm}	
		\begin{equation}\label{eq:upperbound-dTx-transmission}
		\left|F_T(\vect{r}_{\textup{Rx}})\right|
		\leq
		C_{\textup{tran}} \left(1-\frac{z_{\textup{P}2}}{z_{\textup{P}1}}\right)  \tan^{-1}\left[\frac{(x_{\textup{P}1}-x)(y_{\textup{P}1}-y)}{|z_{\textup{P}1}|\sqrt{(x_{\textup{P}1}-x)^2+(y_{\textup{P}1}-y)^2+z_{\textup{P}1}^2}}\right]\Bigr|^{x=L_x}_{x=-L_x}\Bigr|^{y=L_y}_{y=-L_y}
		\end{equation}
	\end{corollary}
\vspace{-0.35cm}
	\begin{proof}
		It is the same as for Corollary \ref{corollary:bound-Lx-Ly}.
	\end{proof}
	\vspace{-0.5cm}
	\begin{corollary}\label{corollary:electrically-small-lens-transmission}
		In the electrically-small regime, $F_T(\vect{r}_{\textup{Rx}})$ can be approximated as follows: \vspace{-0.15cm}
		\begin{equation}\label{eq:long-approx-3D-beamforming-transmission}
		F_T(\vect{r}_{\textup{Rx}})
		\approx  \frac{jk e^{j(\phi_0+\phi_\textup{rec}+\phi_\textup{tran})}}{16\pi^2}
		\Omega_{\textup{tran}}(0,0;\hat{\vect{p}}_{\textup{tran}},\hat{\vect{p}}_{\textup{rec}})  \left(\cos\theta_{\textup{inc}0}+\cos\theta_{\textup{rec}0}\right)\int_{-L_y}^{L_y}\int_{-L_x}^{L_x} |\Gamma_{\textup{tran}}(x,y)|dxdy
		\end{equation}
		If $|\Gamma_{\textup{tran}}(x,y)| = \Gamma_{\textup{tran}} >0$ for $(x,y) \in \SS$, then $F_T(\vect{r}_{\textup{Rx}})$ can be further approximated as follows: \vspace{-0.15cm}
		\begin{equation}\label{eq:long-approx-3D-beamforming-constant-transmission}
		F_T(\vect{r}_{\textup{Rx}}) 
		\approx \frac{jk\Gamma_{\textup{tran}}|\Omega_{\textup{tran}}(0,0;\hat{\vect{p}}_{\textup{tran}},\hat{\vect{p}}_{\textup{rec}})|L_xL_y\left(\cos\theta_{\textup{inc}0}+\cos\theta_{\textup{rec}0}\right)}{4\pi^2d_{\textup{Tx}0}d_{\textup{Rx}0}}
		e^{j(\phi_0+\phi_\textup{rec}+\phi_\textup{tran})}
		\end{equation}
	\end{corollary}
	\vspace{-0.4cm}
	\begin{proof}
		It follows by direct application of \eqref{eq:long-distance-approximation-3D-type-2}.
	\end{proof}
	
	In conclusion, we evince that $\left| F_T(\vect{r}_{\textup{Rx}}) \right|$ in Corollaries \ref{corollary:bound-Lx-Ly-transmission} and \ref{corollary:electrically-small-lens-transmission} is similar to $\left| F_R(\vect{r}_{\textup{Rx}}) \right|$ in Corollaries \ref{corollary:bound-Lx-Ly} and \ref{corollary:electrically-small-lens-reflection}, respectively. As for the performance trends as a function of the size of $\SS$ and the transmission distances, reflecting and transmitting surfaces have a similar behavior.

	\begin{figure}[!t]
		\centering
		\begin{subfigure}[!t]{0.48\columnwidth}
		\footnotesize
		\centering
		%Table 3(a): Simulation setup 
		\caption{Simulation setup} \vspace{-0.15cm}
		\label{table:parameters}
		\newcommand{\tabincell}[2]{\begin{tabular}{@{}#1@{}}#2\end{tabular}}
			\begin{tabular}{l} \hline
			\hspace{2.75cm} Settings \\ 
			\hline
			$f = 28$ GHz, $\lambda = 10.71$ mm\\
			$\epsilon_0 = 8.85 \cdot 10^{-12}$ Farad/meter\\
			$p_{\textup{dm}} = (k^2/\epsilon_0)^{-1}$ \\
			$\tilde{\vect{p}}_{\textup{inc}}=
			\tilde{\vect{p}}_{\textup{ref}}=
			\tilde{\vect{p}}_{\textup{tran}}=
			\tilde{\vect{p}}_{\textup{rec}} = \hat{\vect{y}}$ (transverse electric) \\
			$\phi_0 = \phi_\textup{inc} = \phi_\textup{ref} = \phi_\textup{tran} = \phi_\textup{rec} = 0$ \\
			$\mathcal{E}\left(\hat{\vect{p}}_{\textup{inc}},\hat{\vect{p}}_{\textup{ref}}\right) = \mathcal{E}\left(\hat{\vect{p}}_{\textup{inc}},\hat{\vect{p}}_{\textup{tran}}\right) = 1$ \\
			$|\Gamma_{\textup{ref}}(x,y)| = |\Gamma_{\textup{tran}}(x,y)| = 1~\forall(x,y)\in \SS$ \\
			$\theta_{\textup{inc}0} = \pi/4$, $\varphi_{\textup{inc}0} = \pi/3$ \\
			$\theta_{\textup{rec}0} = \pi/6$, $\varphi_{\textup{rec}0} = \pi$ (reflecting $\SS$) \\
			$\theta_{\textup{rec}0} = \pi/3$, $\varphi_{\textup{rec}0} = 5\pi/4$ (transmitting $\SS$) \\
			$\alpha_R = \alpha_T = -\sin\theta_{\textup{inc}0}\cos\varphi_{\textup{inc}0} - \sin\theta_{\textup{rec}0}\cos\varphi_{\textup{rec}0}$ \\
			$\beta_R = \beta_T = -\sin\theta_{\textup{inc}0}\sin\varphi_{\textup{inc}0} - \sin\theta_{\textup{rec}0}\sin\varphi_{\textup{rec}0}$ \\
			\hline
		\end{tabular}
		\end{subfigure}
		\begin{subfigure}[!t]{0.48\columnwidth}
			\includegraphics[width=0.98\linewidth]{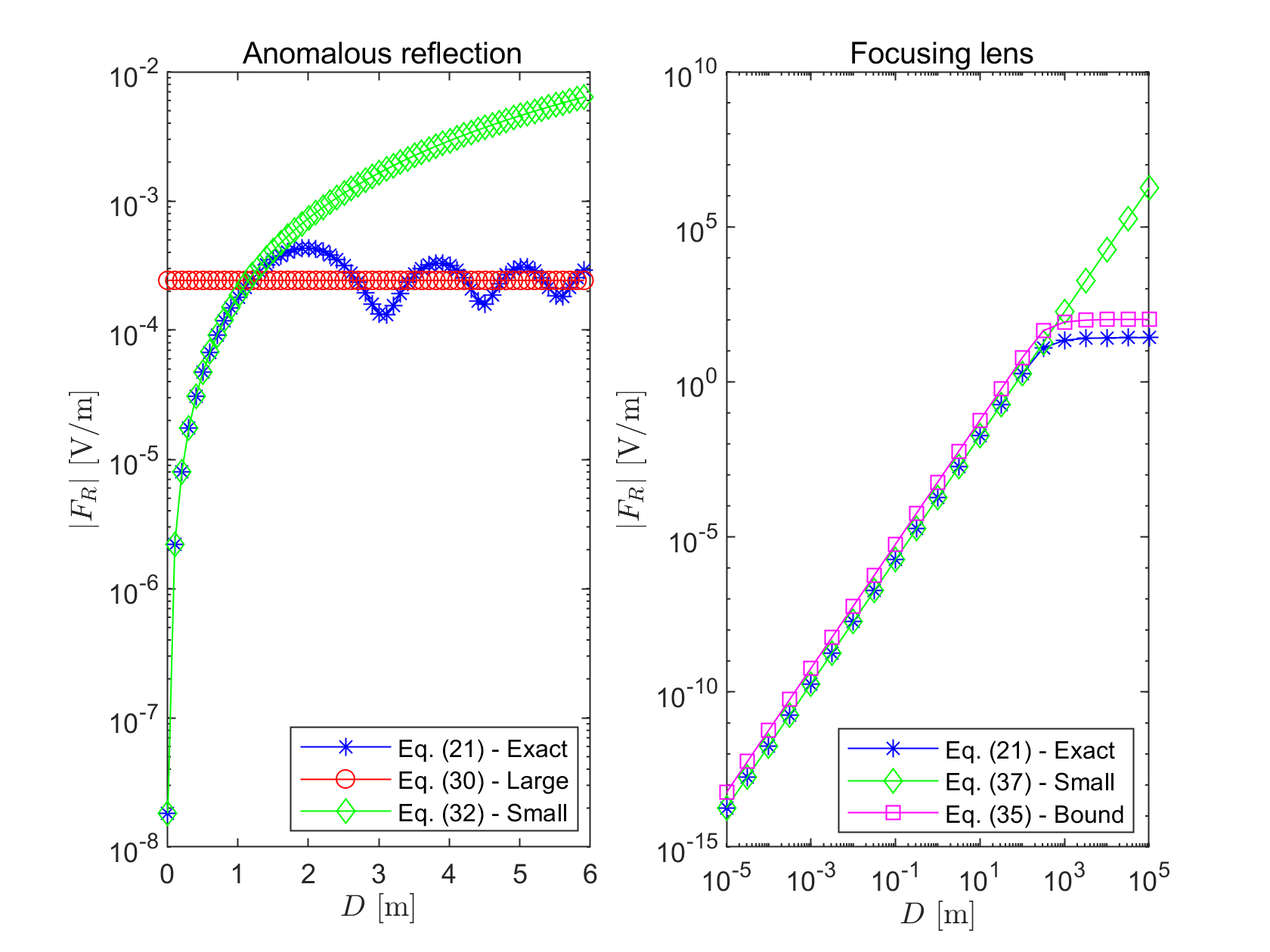}
			\vspace{-0.25cm} \caption{Reflecting surface. Setup: $d_{\textup{Tx}0} = d_{\textup{Rx}0} = 100$ m}
			\label{fig:SurfaceSize}
			\vspace{0.15cm}
		\end{subfigure}
		%\caption{Simulation setup (a) and impact of surface size (b).}
		\footnotesize{Table 2(a): Simulation setup and Fig. 3(b) impact of surface size.}
		\label{fig:Setup_SurfaceSize} \vspace{-0.5cm}
	\end{figure}
	
	\vspace{-0.5cm}
	\section{Numerical Results} \vspace{-0.25cm}
In this section, we	illustrate some numerical examples in order to shed light on the behavior of the path-loss in the presence of RISs. In addition, we aim to analyze the conditions under which the considered asymptotic regimes hold true, and whether the considered phase gradient metasurfaces allow us to realize anomalous reflection/transmission and focusing as elaborated in Sections \ref{section:reflection} and \ref{section:transmission}. Unless otherwise stated, we use the simulation setup in Table \ref{table:parameters}. The simulation results illustrate $F_R(\vect{r}_{\textup{Rx}})$ and $F_T(\vect{r}_{\textup{Rx}})$ obtained in Sections \ref{section:reflection} and \ref{section:transmission}, respectively.

	\begin{figure}[!t]
		\centering
		\begin{subfigure}[!t]{0.48\columnwidth}
		\centering
			\includegraphics[width=0.7\columnwidth]{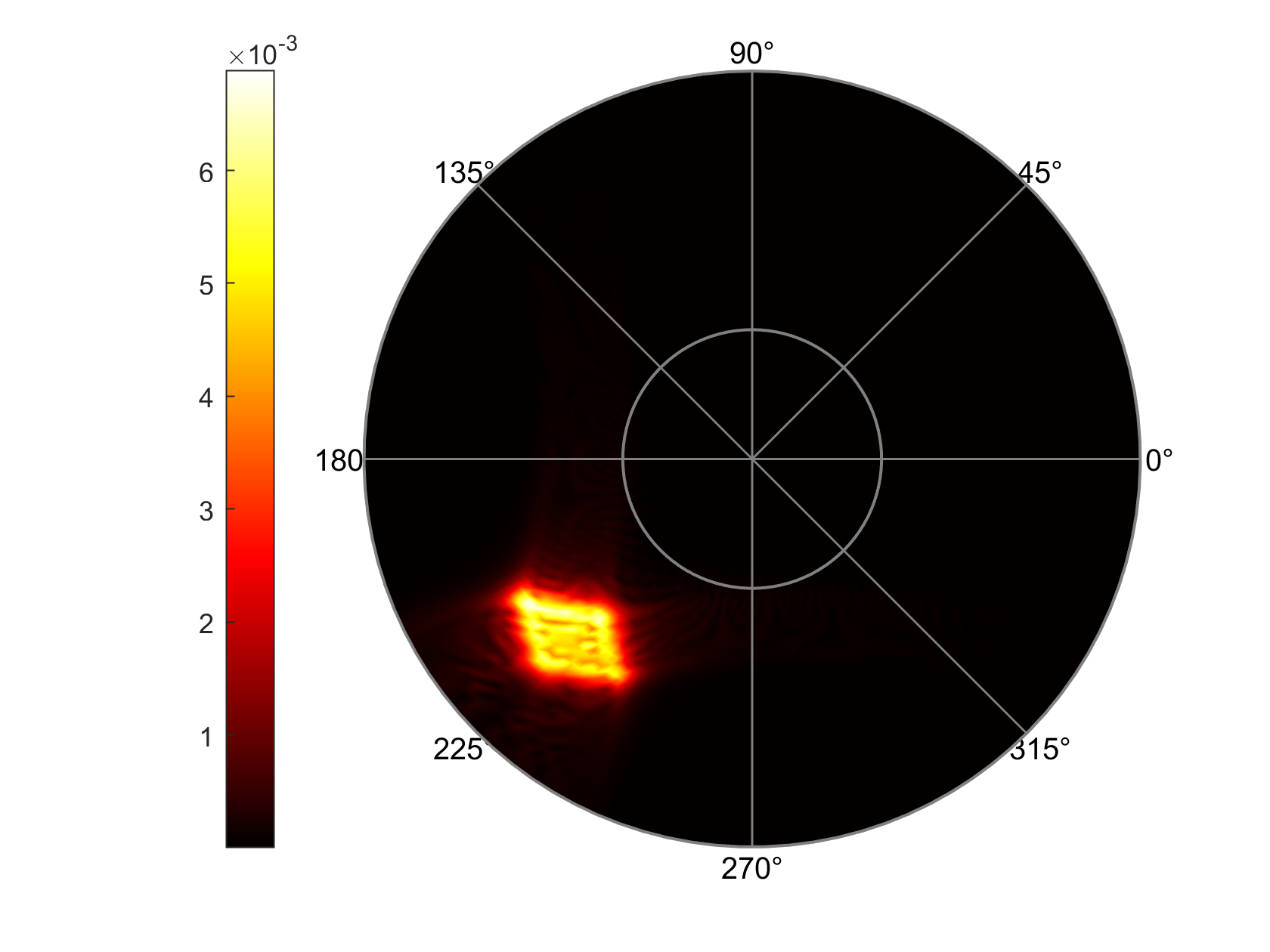}
			\vspace{-0.3cm} \caption{Anomalous transmission, $d_{\textup{Tx}0} = d_{\textup{Rx}0} = 5$ m}
		\end{subfigure}
		\begin{subfigure}[!t]{0.48\columnwidth}
		\centering
			\includegraphics[width=0.7\columnwidth]{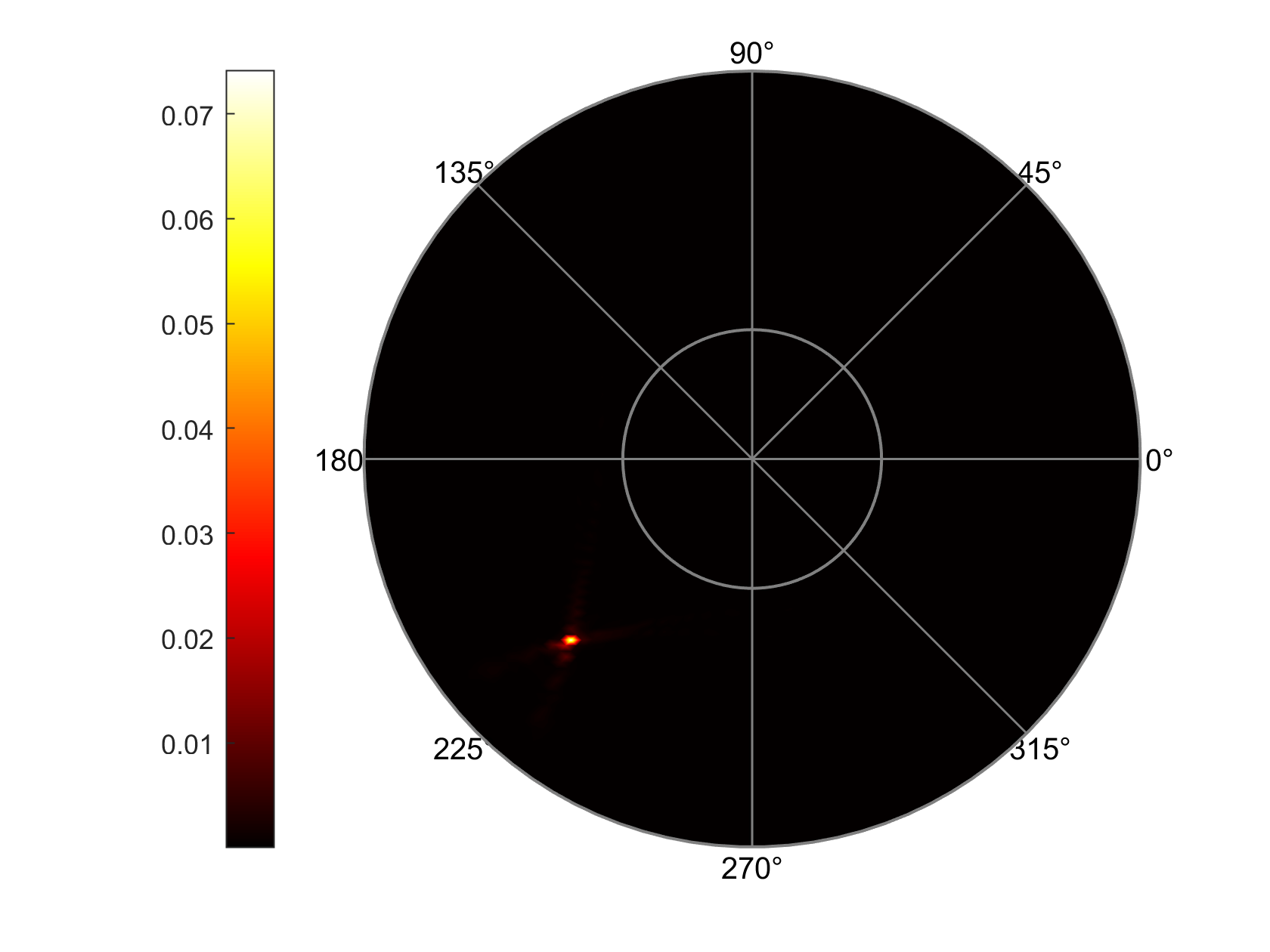}
			\vspace{-0.3cm} \caption{Focusing lens, $d_{\textup{Tx}0} = d_{\textup{Rx}0} = 5$ m}
		\end{subfigure}
		\begin{subfigure}[!t]{0.48\columnwidth}
		\centering
			\includegraphics[width=0.7\columnwidth]{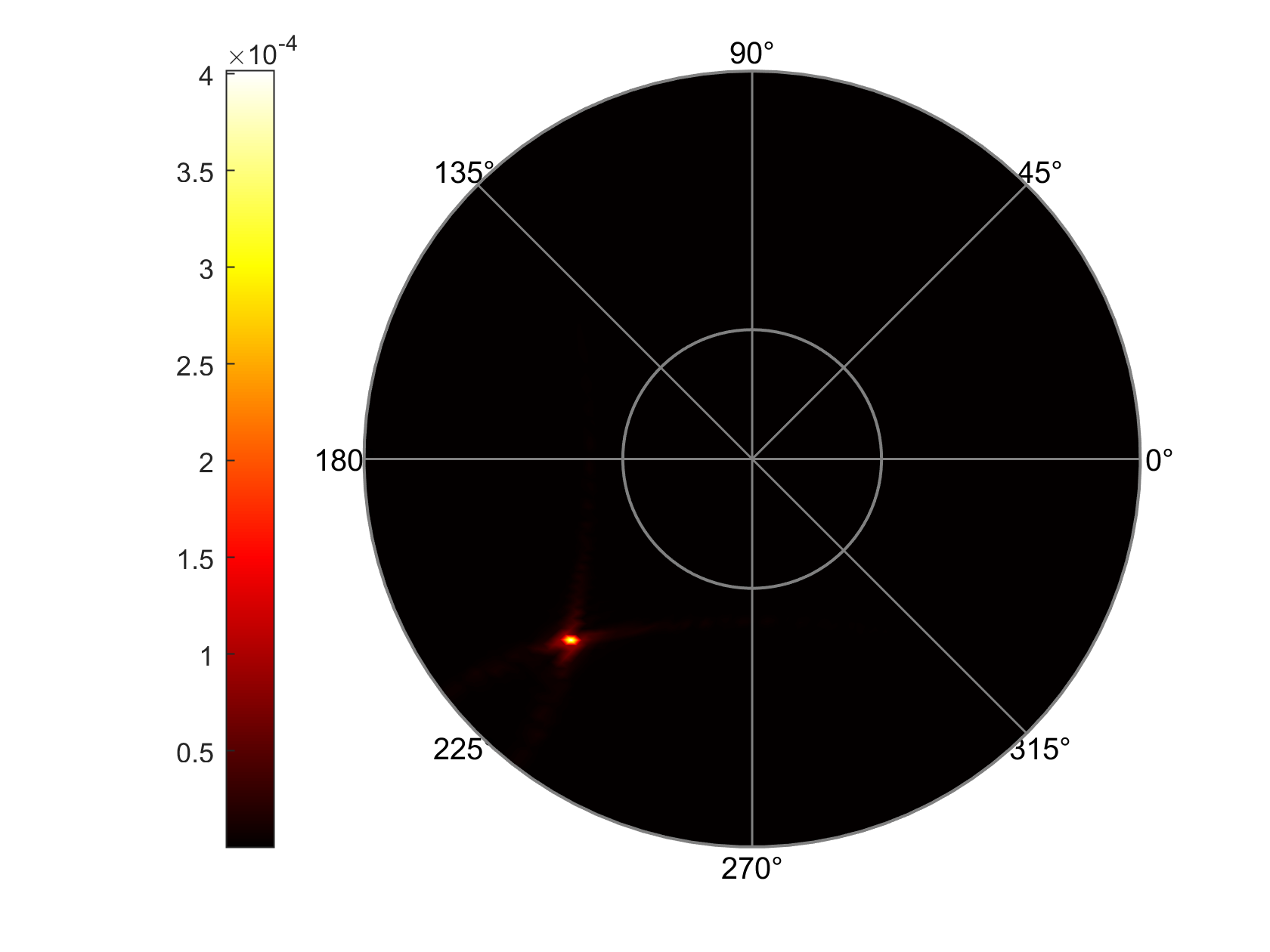}
			\vspace{-0.3cm} \caption{Anomalous transmission, $d_{\textup{Tx}0} = d_{\textup{Rx}0} = 50$ m}
		\end{subfigure}
		\begin{subfigure}[!t]{0.48\columnwidth}
		\centering
			\includegraphics[width=0.7\columnwidth]{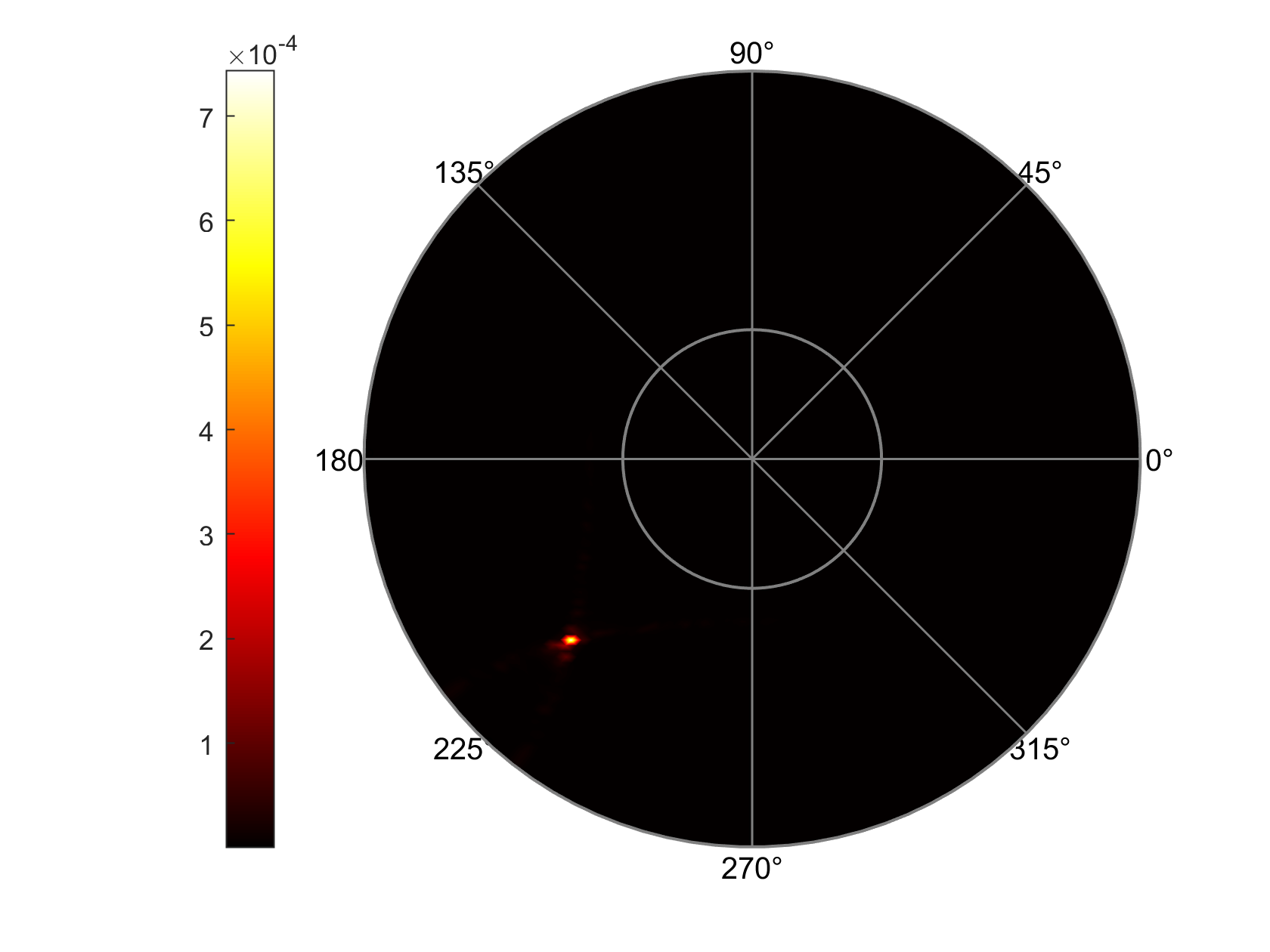}
			\vspace{-0.3cm} \caption{Focusing lens, $d_{\textup{Tx}0} = d_{\textup{Rx}0} = 50$ m}
		\end{subfigure}
		\caption{Anomalous transmission vs. focusing lens (transmitting surface). Setup: $2L_x=2L_y=1$ m.}
		\label{fig:Colormap-Transmission} \vspace{-0.75cm}
	\end{figure}

		\begin{figure}[!t]
		\centering
		\begin{subfigure}[!t]{0.48\columnwidth}
		\centering
			\includegraphics[width=0.7\columnwidth]{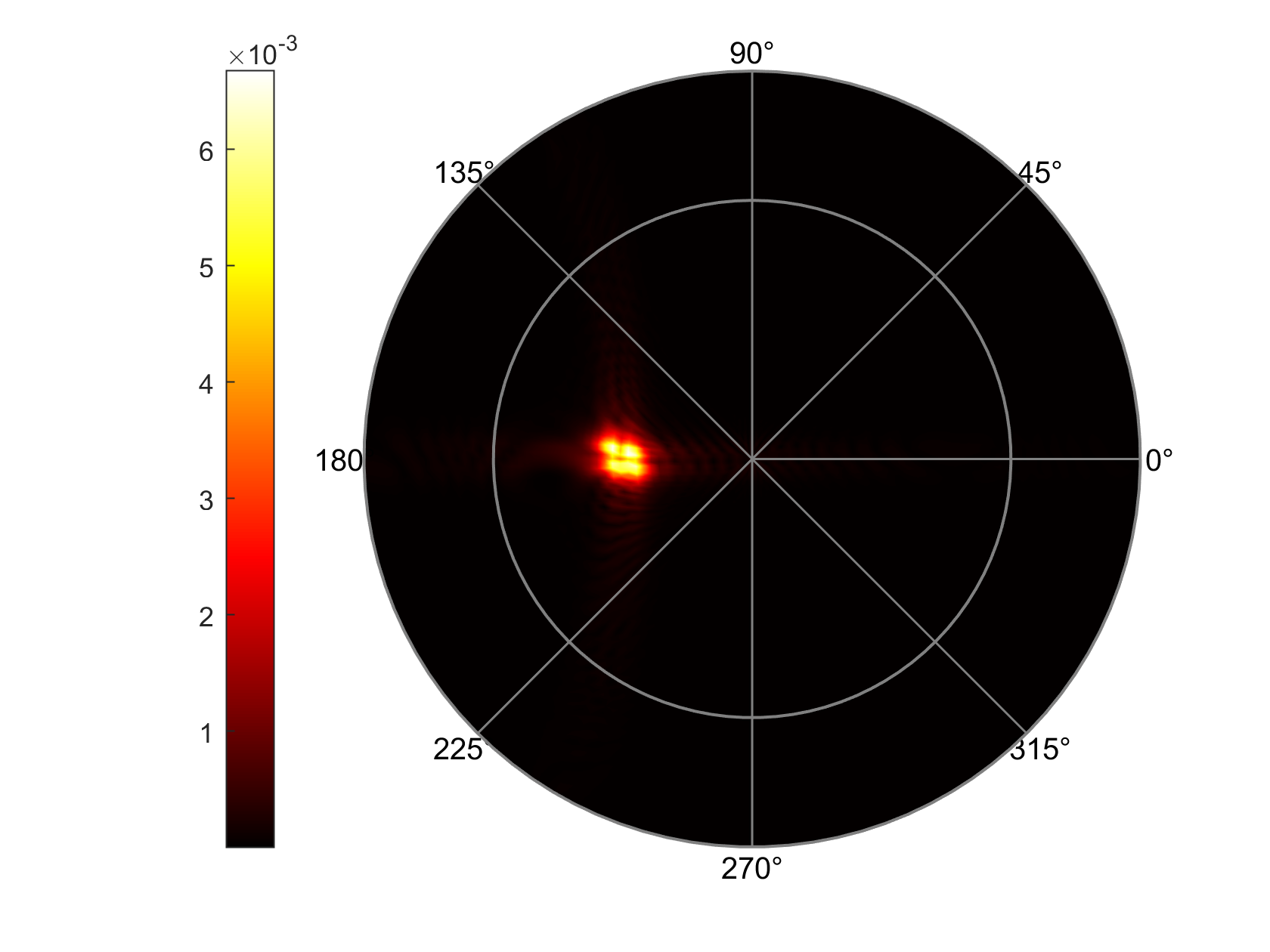}
			\vspace{-0.3cm} \caption{Anomalous reflection, no discretization}
		\end{subfigure}
		\begin{subfigure}[!t]{0.48\columnwidth}
		\centering
			\includegraphics[width=0.7\columnwidth]{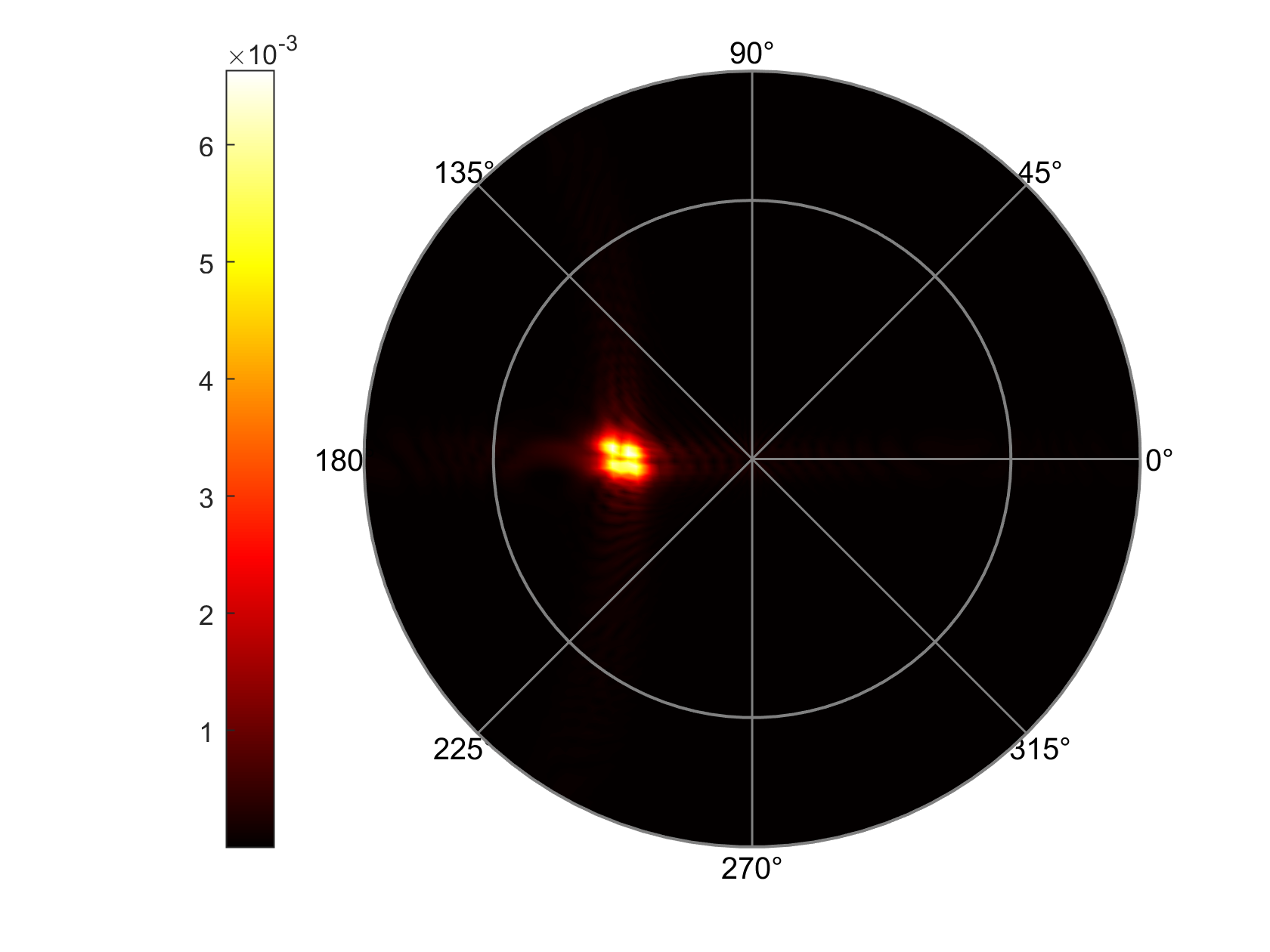}
			\vspace{-0.3cm} \caption{Anomalous reflection, discretization step = $0.25 \lambda$}
		\end{subfigure}
		\begin{subfigure}[!t]{0.48\columnwidth}
		\centering
			\includegraphics[width=0.7\columnwidth]{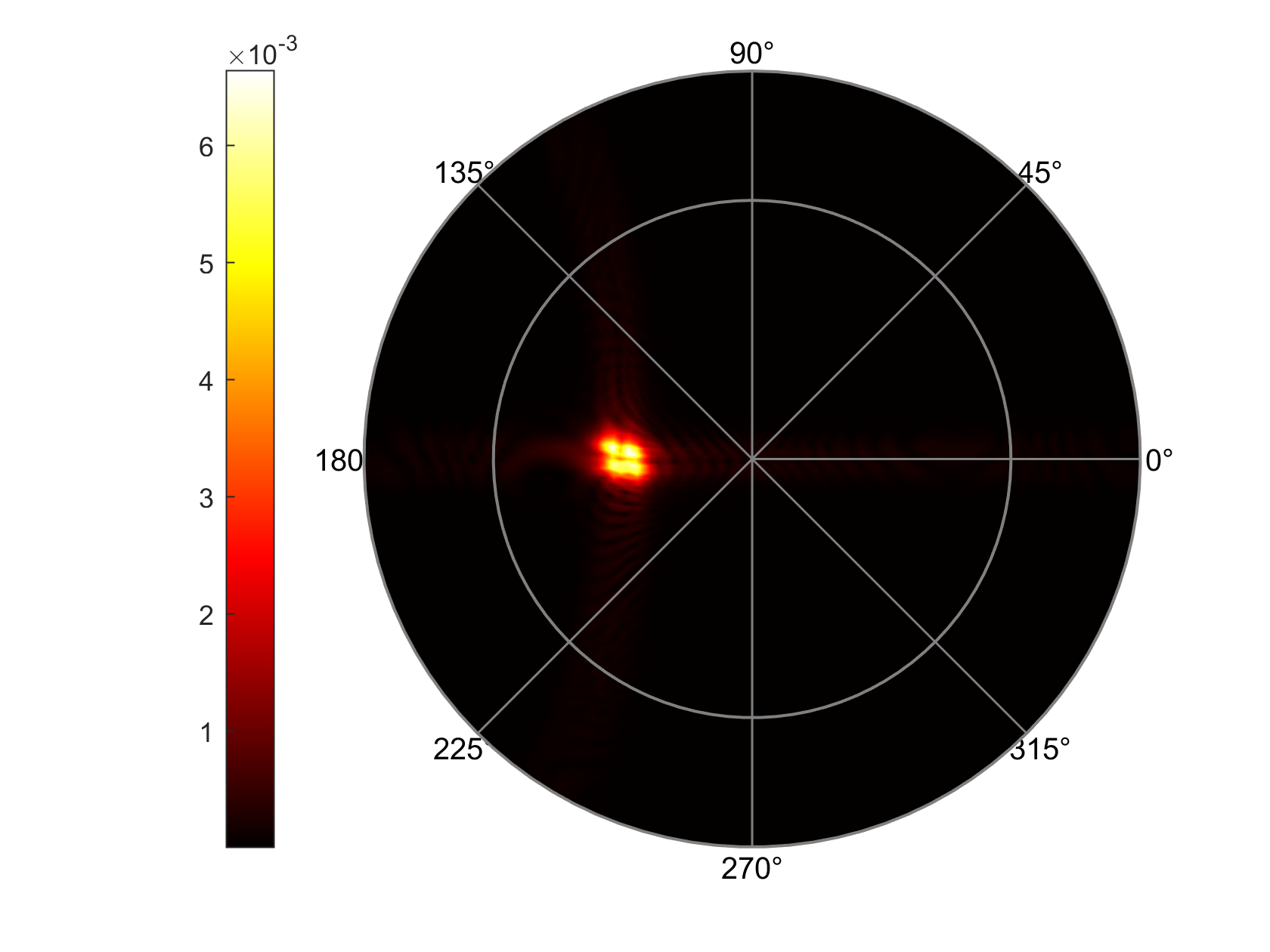}
			\vspace{-0.3cm} \caption{Anomalous reflection, discretization step = $0.5 \lambda$}
		\end{subfigure}
		\begin{subfigure}[!t]{0.48\columnwidth}
		\centering
			\includegraphics[width=0.7\columnwidth]{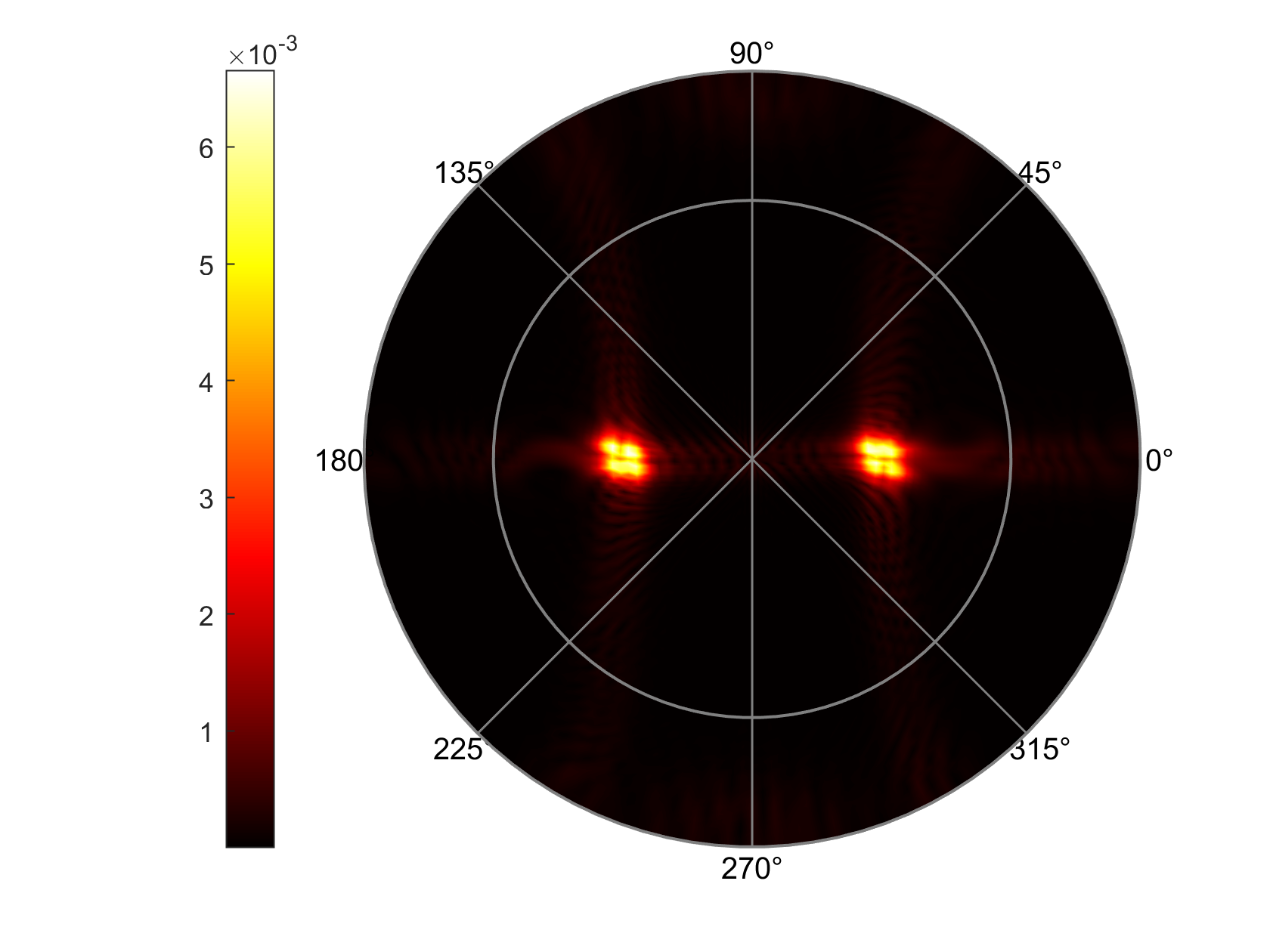}
			\vspace{-0.3cm} \caption{Anomalous reflection, discretization step = $\lambda$}
		\end{subfigure}
		\vspace{-0.3cm} \caption{Anomalous reflection: Impact of discretization. Setup: $2L_x=2L_y=0.5$ m; $d_{\textup{Tx}0} = d_{\textup{Rx}0} = 5$ m.}
		\label{fig:Colormap-Reflection} \vspace{-0.5cm}
	\end{figure}

		\begin{figure}[!t]
		\centering
		\begin{subfigure}[!t]{0.31\columnwidth}
		\centering
			\includegraphics[width=1.10\columnwidth]{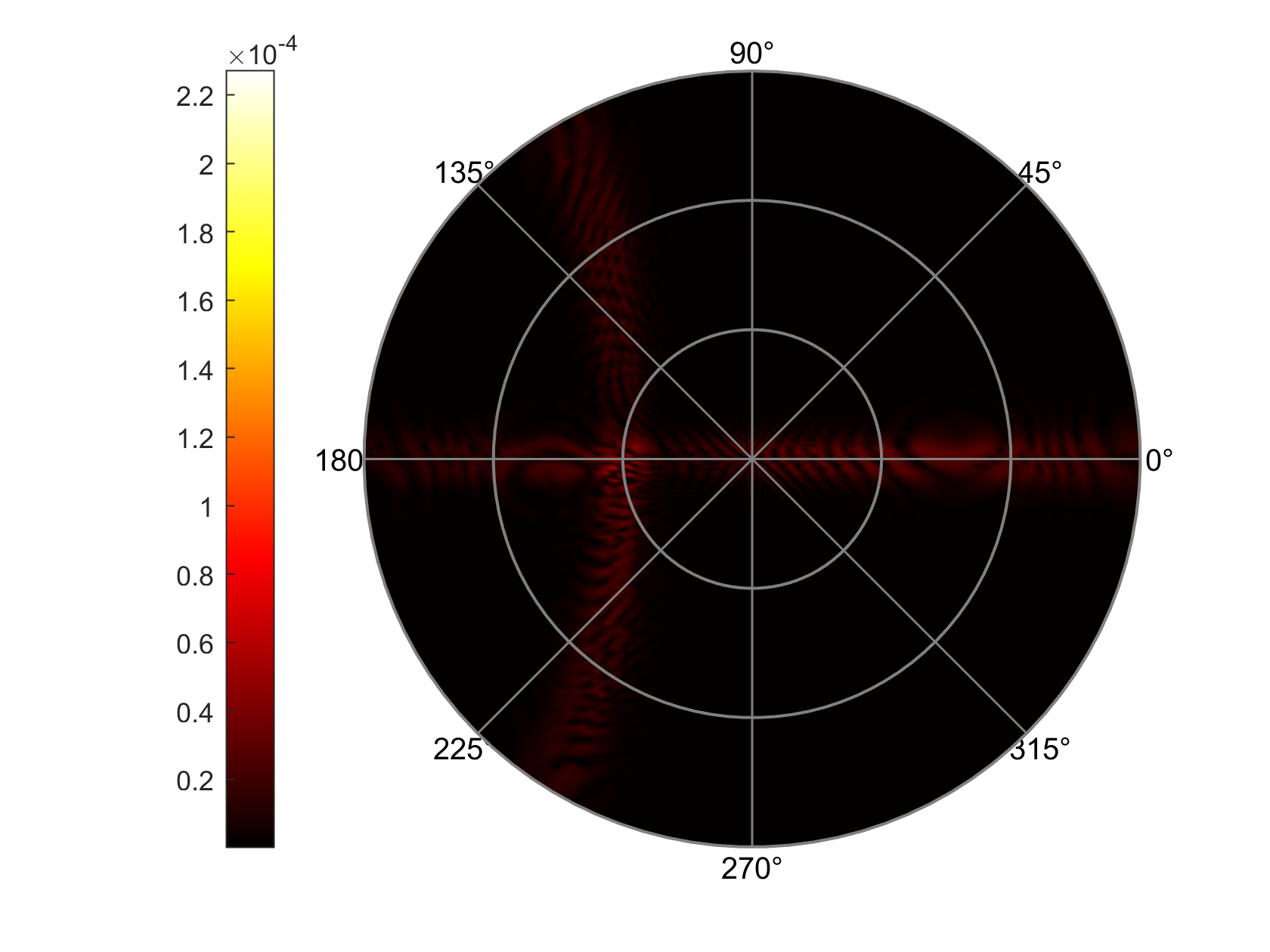}
			\vspace{-0.6cm} \caption{Absolute error of Figs. \ref{fig:Colormap-Reflection}(b), \ref{fig:Colormap-Reflection}(a)}
		\end{subfigure}
		\begin{subfigure}[!t]{0.31\columnwidth}
		\centering
			\includegraphics[width=1.10\columnwidth]{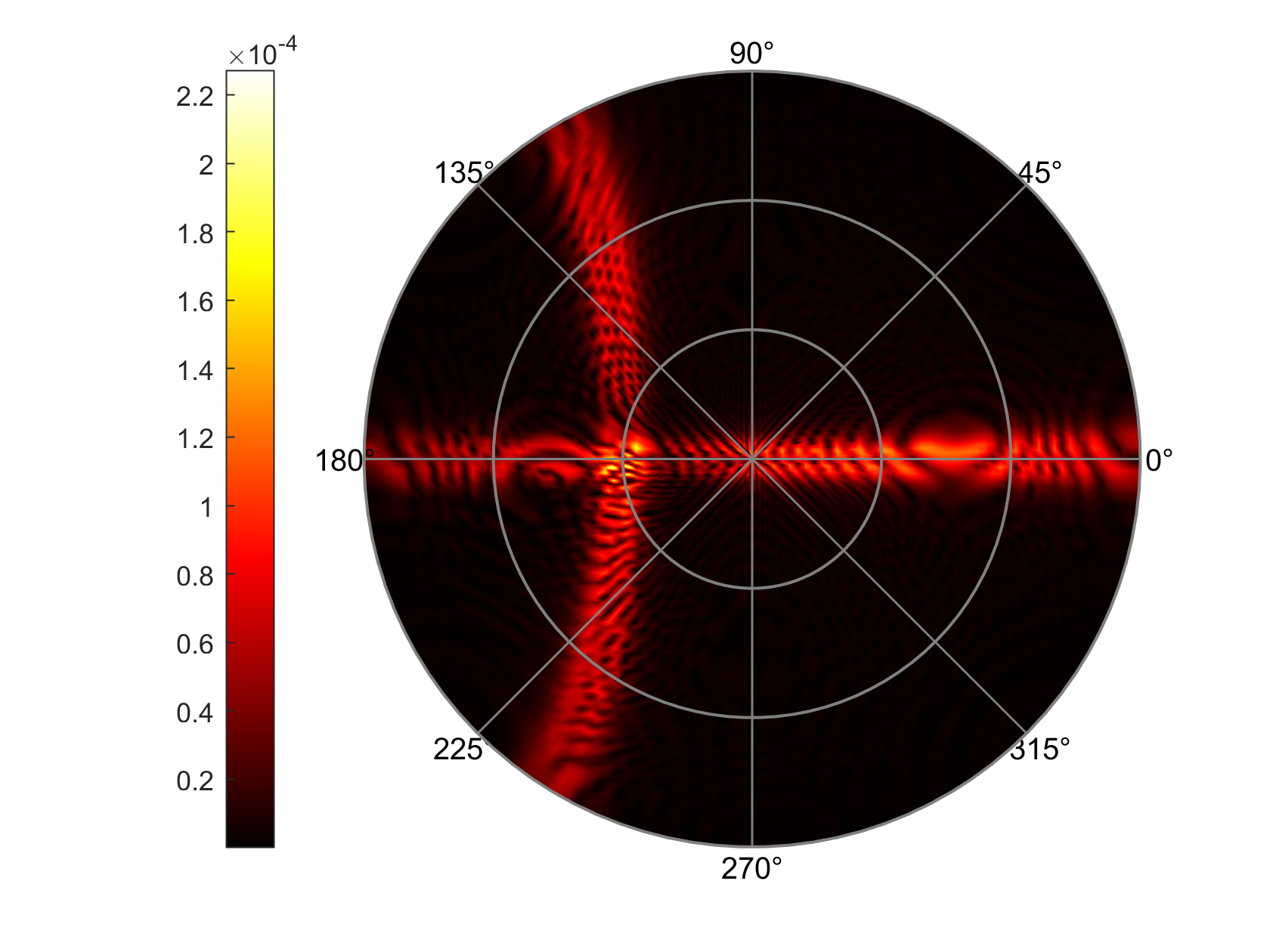}
			\vspace{-0.6cm} \caption{Absolute error of Figs. \ref{fig:Colormap-Reflection}(c), \ref{fig:Colormap-Reflection}(a)}
		\end{subfigure}
		\begin{subfigure}[!t]{0.31\columnwidth}
		\centering
			\includegraphics[width=1.10\columnwidth]{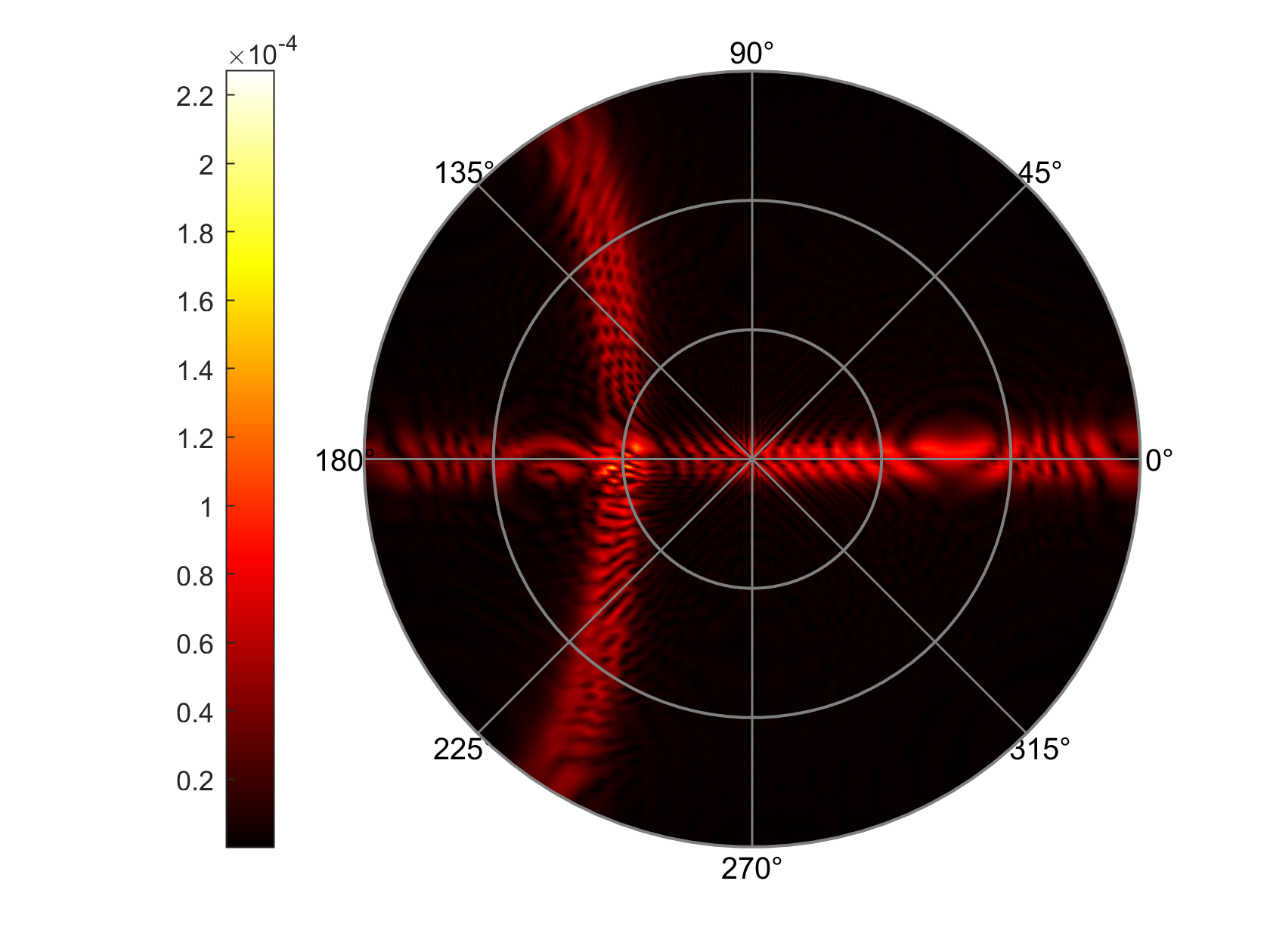}
			\vspace{-0.6cm} \caption{Absolute error of Figs. \ref{fig:Colormap-Reflection}(c), \ref{fig:Colormap-Reflection}(b)}
		\end{subfigure}
		\vspace{-0.2cm} \caption{Absolute error difference corresponding to Fig. \ref{fig:Colormap-Reflection}.}
		\label{fig:Colormap-ReflectionError} \vspace{-0.5cm}
	\end{figure}

	\vspace{-0.5cm}
	\subsection{Anomalous Reflection and Focusing} \vspace{-0.25cm}
In Fig. \ref{fig:Colormap-Transmission}, we analyze anomalous transmission and focusing (transmitting surface) by using Proposition \ref{proposition:transmitted-field-general}. The radial lines spaced by $45$ degrees denote the angle $\varphi$ and the three inner circles spaced by $30$ degrees denote the angle $\theta$ (some lines and circles are removed for clarity). We observe that the correct angles of transmission are obtained. We note that anomalous transmitting surfaces yield a larger coverage area  than focusing lenses. This is obtained, however, only for short transmission distances (near-field of the RIS). On the other hand, the larger coverage area is not apparent for long transmission distances (far-field of the RIS). This confirms Remark 12.

In Fig. \ref{fig:Colormap-Reflection}, we analyze anomalous reflection and focusing (reflecting surface) by using Proposition \ref{proposition:reflected-field-general}. In particular, we consider a discretized version of the integral in \eqref{eq:Ex-surface-contribution-reflection}, which corresponds to a practical implementation of the RIS based on (discrete) scattering elements. Provided that the scattering elements are spaced less than half of the wavelength apart (i.e., the discretization step is $\lambda/2$), we evince that no significant differences can be observed at the naked eye. If the discretization step is greater than $\lambda/2$, e.g., it is $\lambda$, we observe the presence of grating lobes (spurious reflections) in unwanted directions. To better appreciate the impact of discretization, Fig. \ref{fig:Colormap-ReflectionError} reports the absolute error difference that corresponds to the setups in Fig. \ref{fig:Colormap-Reflection}. We observe that some differences are indeed apparent and that more closely spaced scattering elements yield more accurate estimates of the electric field (especially in the considered near-field regime).

\begin{figure}[!t]
		\centering
\begin{subfigure}[!t]{0.48\columnwidth}
			\includegraphics[width=0.98\linewidth]{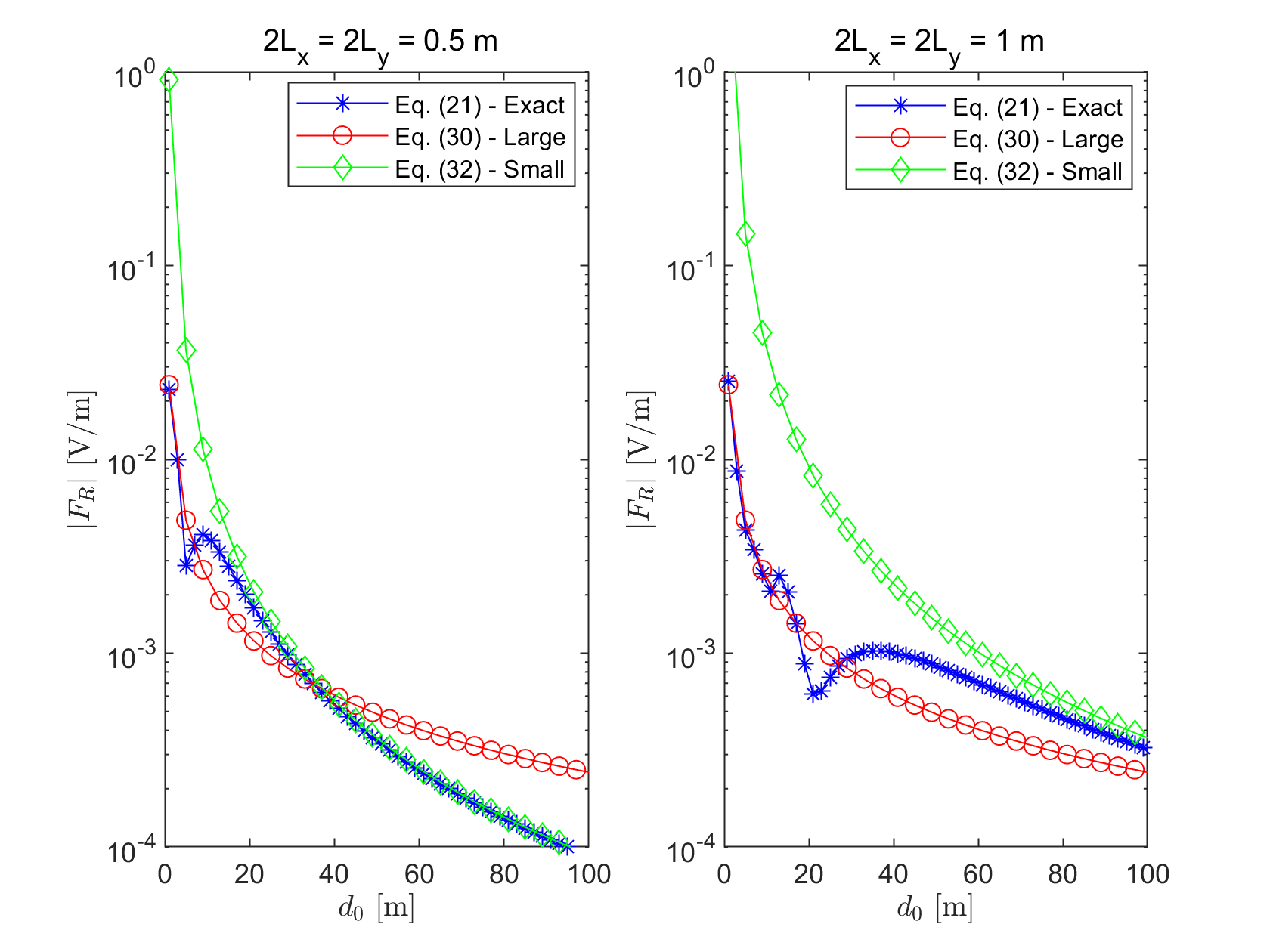}
			\vspace{-0.25cm} \caption{Anomalous reflection}
			\label{fig:TransmissionDistanceAnomalous}
		\end{subfigure}
		\begin{subfigure}[!t]{0.48\columnwidth}
			\includegraphics[width=0.98\linewidth]{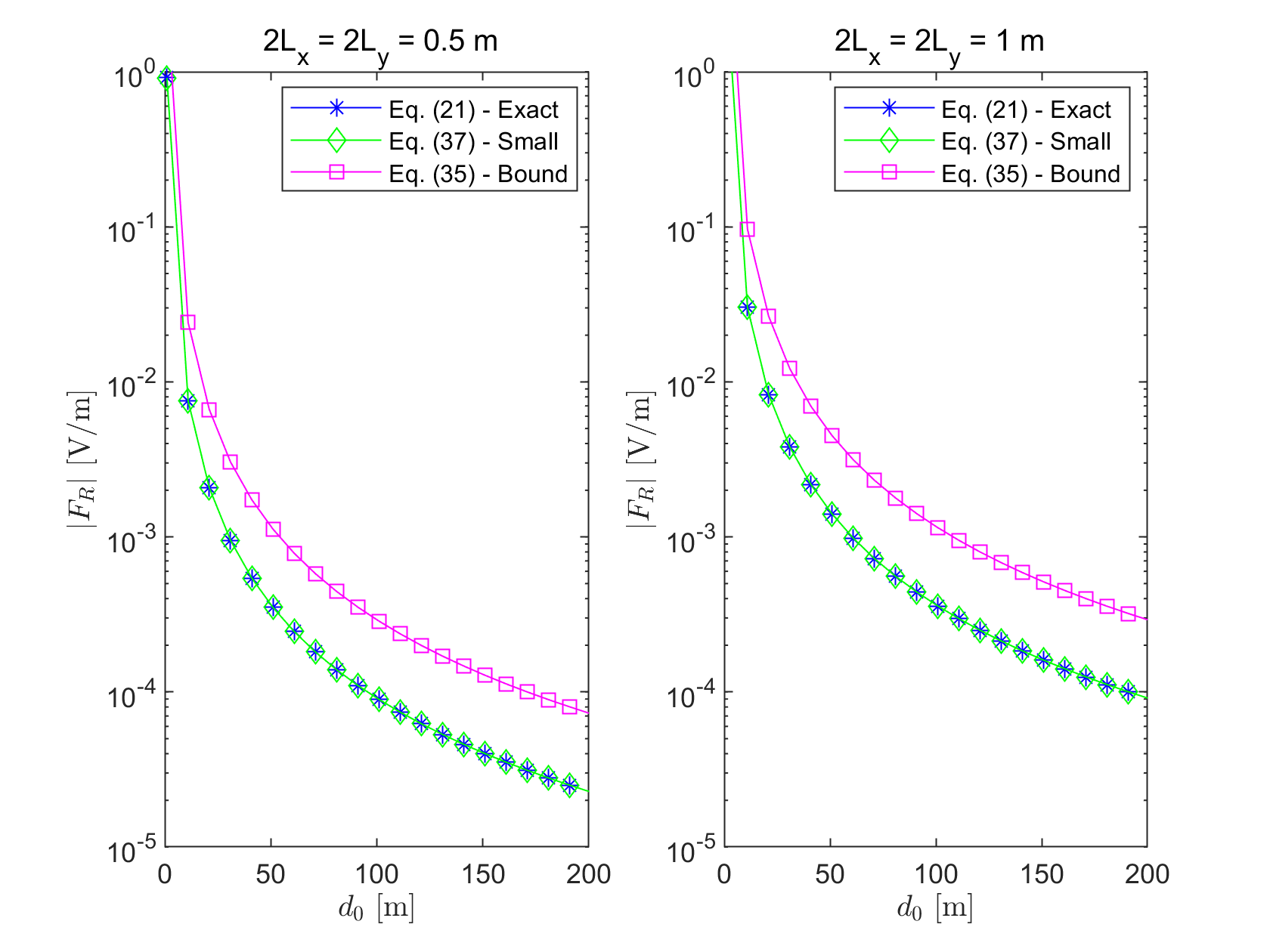}
			\vspace{-0.25cm} \caption{Focusing lens}
			\label{fig:TransmissionDistanceLens}
		\end{subfigure}
		\vspace{-0.25cm} \caption{Anomalous reflection vs. focusing lens: Impact of transmission distance.}
		\label{fig:TransmissionDistance} \vspace{-0.75cm}
	\end{figure}
	
	\vspace{-0.5cm}
	\subsection{Transmission Distance} \vspace{-0.25cm}
In Fig. \ref{fig:TransmissionDistance}, we analyze the impact of the transmission distance in the context of anomalous reflection and focusing. In particular, the distances from the transmitter to the center of the RIS, and from the center of the RIS to the receiver are denoted by $d_{\textup{Tx}0} = d_{\textup{Rx}0} = d_0$. The angles of observation computed with respect to the center of the RIS are kept fixed as $d_0$ increases or decreases. We observe that the analytical frameworks obtained in the electrically-large and electrically-small asymptotic regimes well overlap, in the regions of interest, with the integral representation of the electric field. In particular, we note a major difference between anomalous reflectors and focusing lenses. As for anomalous reflectors, we observe two scaling laws as a function of the distance: (i) the weighted-sum path-loss model for short distances and (ii) the product path-loss model for long distances. As for focusing lenses, on the other hand, we observe a single scaling law: the product path-loss model that is sufficiently accurate for short and long distances. This highlights that the impact of the distance depends on the setup of the RIS. A focusing lens co-phases all the contributions scattered from the RIS and this yields a different scaling law as compared with an anomalous reflector. As for anomalous reflectors, it is worth noting that the weighted-sum path-loss model may be accurate up to a few tens of meters, which may be important in indoor scenarios and for local coverage enhancement in outdoor scenarios.

	\vspace{-0.5cm}
		\subsection{Surface Size} \vspace{-0.25cm}
In Fig. \ref{fig:SurfaceSize}, we analyze the impact of the size of the RIS for anomalous reflection and focusing (reflecting surface). In particular, the figure reports the intensity of the electric field as a function of the diagonal, $D$, of the RIS. We observe that the intensity of the electric field is bounded even if the size of the surface increases without bound. The closed-form analytical frameworks and the bounds obtained in the electrically-large and electrically-small regimes well predict the scaling law. It is worth noting that the electrically-small approximation may significantly overestimate the intensity of the electric field even for relatively small surfaces and for long transmission distances (100 meters in the figure). These results confirm that the proposed path-loss model is compliant with the power conservation law and that the received power is always bounded, regardless of the size of the surface, in the far-field of the RIS microstructure.

\vspace{-0.5cm}	
\section{Conclusion} \vspace{-0.25cm}
We have introduced a physics-compliant path-loss model for RIS-aided wireless transmission. The proposed path-loss model is general enough for application to various operating regimes, which include near-field and far-field asymptotic regimes. The impact of several design parameters has been analyzed. In particular, we have proved that the scaling laws of the received power as a function of the transmission distance and the size of the RIS are different in the near-field and far-field regimes, and they depend on the wave transformations applied by the RIS. Notably, the received power scattered by an RIS is bounded as its size increases without bound.

In the context of wireless communications, in general terms, one should always use the integral representation of the path-loss in order to make sure that the received power is physically meaningful as a function of every design parameter, e.g., the surface size and the transmission distance. The simple analytical expressions obtained in the near-field and far-field asymptotic regimes can be employed provided that the considered system setup is compliant with their regime of validity. For application to the performance evaluation and optimization of wireless networks, one may consider the use of a two-law path-loss model (in analogy with two-slope path-loss models), which combines together the closed-form analytical expressions obtained in the near-field and far-field regimes. This approach may avoid the analytical intractability of using two-fold integrals while ensuring compliance with physics-based constraints.

	%\begin{appendices}
\vspace{-0.35cm}		
	\section*{Appendix A -- Proof of Theorem \ref{theorem:Stratton-Chu-equivalence}}\label{appendix:proof-of-Stratton-Chu-equivalence} \vspace{-0.25cm}
	
By inserting $\vect{H}_{\partial V}(\vect{r'}) = -\nabla_{\vect{r'}} \times \vect{E}_{\partial V}(\vect{r'})/(j\omega\mu_0)$ in \eqref{eq:Stratton-Chu-electric}, we obtain the following: \vspace{-0.25cm}
	\begin{align}\label{eq:Stratton-Chu-electric-alternative-1}
	&\vect{E}_{\partial V}(\vect{r}_{\textup{Rx}})
	= 
	\mathbbm{1}_{(\vect{r}_{\textup{Tx}} \in V)}\vect{E}_{\textup{inc}}(\vect{r}_{\textup{Rx}};\hat{\vect{p}}_{\textup{inc}})
	- \int_{\partial V} \left[ \left(\hat{\vect{n}}_{\textup{out}}\times \nabla_{\vect{r'}} \times \vect{E}_{\partial V}(\vect{r'})\right)G(\vect{r}_{\textup{Rx}},\vect{r'}) \right.\\
	&\left.
	+ \left(\hat{\vect{n}}_{\textup{out}}\cdot\vect{E}_{\partial V}(\vect{r'})\right)\nabla_{\vect{r'}} G(\vect{r}_{\textup{Rx}},\vect{r'})
	+ (\hat{\vect{n}}_{\textup{out}}\times \vect{E}_{\partial V}(\vect{r'}))\times \nabla_{\vect{r'}} G(\vect{r}_{\textup{Rx}},\vect{r'})
	\right]d\vect{r'} \nonumber\\
	&\mystepA
	\mathbbm{1}_{(\vect{r}_{\textup{Tx}} \in V)}\vect{E}_{\textup{inc}}(\vect{r}_{\textup{Rx}};\hat{\vect{p}}_{\textup{inc}})
	- \int_{\partial V} \left[\left(\hat{\vect{n}}_{\textup{out}}\times \nabla_{\vect{r'}} \times \vect{E}_{\partial V}(\vect{r'})\right) G(\vect{r}_{\textup{Rx}},\vect{r'})
	+ 
	\left(\hat{\vect{n}}_{\textup{out}} \cdot \nabla_{\vect{r'}} G(\vect{r}_{\textup{Rx}},\vect{r'})\right) \vect{E}_{\partial V}(\vect{r'}) 
	\right]d\vect{r'} \nonumber\\
	&\mystepB 
	\mathbbm{1}_{(\vect{r}_{\textup{Tx}} \in V)}\vect{E}_{\textup{inc}}(\vect{r}_{\textup{Rx}};\hat{\vect{p}}_{\textup{inc}})
	- \int_{\partial V} \left[\left(\hat{\vect{n}}_{\textup{out}} \cdot \vect{E}_{\partial V}(\vect{r'})\right) \nabla_{\vect{r'}}G(\vect{r}_{\textup{Rx}},\vect{r'})
	-
	G(\vect{r}_{\textup{Rx}},\vect{r'})\left(\hat{\vect{n}}_{\textup{out}} \cdot \nabla_{\vect{r'}}\right)\vect{E}_{\partial V}(\vect{r'})
	\right]d\vect{r'} \nonumber
	\end{align}
	where $(a)$ and $(b)$ follow by applying the identity $\vect{a} \times \vect{b} \times \vect{c} = (\vect{a}\cdot\vect{c})\vect{b} - (\vect{a}\cdot\vect{b})\vect{c}$, for any vectors $\vect{a}$, $\vect{b}$, $\vect{c}$ \cite[Eq. (VII-37)]{balanis2016antenna}, to the fourth and second addends, respectively, i.e.: \vspace{-0.25cm}
	\begin{equation}\label{eq:triple-roduct-identity-1}
		\hat{\vect{n}}_{\textup{out}}\times \vect{E}(\vect{r'})\times \nabla_{\vect{r'}} G(\vect{r}_{\textup{Rx}},\vect{r'})
		=
		\left(\hat{\vect{n}}_{\textup{out}} \cdot \nabla_{\vect{r'}} G(\vect{r}_{\textup{Rx}},\vect{r'})\right) \vect{E}(\vect{r'}) 
		-
		\left(\hat{\vect{n}}_{\textup{out}} \cdot \vect{E}(\vect{r'})\right) \nabla_{\vect{r'}} G(\vect{r}_{\textup{Rx}},\vect{r'}) \vspace{-0.55cm}
\end{equation}
\begin{align}\label{eq:triple-roduct-identity-2} \vspace{-0.25cm}
		& \left(\hat{\vect{n}}_{\textup{out}}\times \nabla_{\vect{r'}} \times \vect{E}(\vect{r'})\right) G(\vect{r}_{\textup{Rx}},\vect{r'})
		= \left(\hat{\vect{n}}_{\textup{out}} \cdot \vect{E}(\vect{r'})\right) \nabla_{\vect{r'}}G(\vect{r}_{\textup{Rx}},\vect{r'})
		-
		\left(\hat{\vect{n}}_{\textup{out}} \cdot \nabla_{\vect{r'}}\right)\vect{E}(\vect{r'})G(\vect{r}_{\textup{Rx}},\vect{r'}) \nonumber\\
		&\mystepC
		\left(\hat{\vect{n}}_{\textup{out}} \cdot \vect{E}(\vect{r'})\right) \nabla_{\vect{r'}}G(\vect{r}_{\textup{Rx}},\vect{r'})
		-
		G(\vect{r}_{\textup{Rx}},\vect{r'})\left(\hat{\vect{n}}_{\textup{out}} \cdot \nabla_{\vect{r'}}\right)\vect{E}(\vect{r'})
		-
		\left(\hat{\vect{n}}_{\textup{out}} \cdot \nabla_{\vect{r'}} G(\vect{r}_{\textup{Rx}},\vect{r'})\right)\vect{E}(\vect{r'})
	\end{align}
where $(c)$ comes from the product rule of derivatives. The proof follows by scalar-multiplying both sides of \eqref{eq:Stratton-Chu-electric-alternative-1} with $\hat{\vect{p}}_{\textup{rec}}$ and from the identity $\left(\hat{\vect{n}}_{\textup{out}} \cdot \nabla_{\vect{r'}}\right)\vect{E}_{\partial V}(\vect{r'}) \cdot \hat{\vect{p}}_{\textup{rec}} = \hat{\vect{n}}_{\textup{out}} \cdot \left( \nabla_{\vect{r'}} (\vect{E}_{\partial V}(\vect{r'}) \cdot \hat{\vect{p}}_{\textup{rec}}) \right)$.

		\vspace{-0.35cm}
	\section*{Appendix B -- Proof of Theorem \ref{theorem:Stratton-Chu-equivalence-reduced}}\label{appendix:proof-of-Stratton-Chu-equivalence-reduced} \vspace{-0.25cm}
	The total electric field at any point $\vect{r'} \in \partial V \setminus \SS$ 
	(i.e., not including $\SS$) 
	is equal to the incident field, i.e., $\vect{E}_{\partial V}(\vect{r'}) = \vect{E}_{\textup{inc}}(\vect{r'},\hat{\vect{p}}_{\textup{inc}})$. On the other hand, the electric field at any point $\vect{s} \in \SS$ is equal to $\vect{E}_{\partial V}(\vect{r'})= \vect{E}_{\SS}(\vect{s})$, where $\vect{E}_{\SS}(\vect{s})$ is given in \eqref{eq:E-surface-reflection} or \eqref{eq:E-surface-transmission}. By denoting $M(\vect{r}') = \vect{E}_{\textup{inc}}(\vect{r'},\hat{\vect{p}}_{\textup{inc}}) \cdot \hat{\vect{p}}_{\text{\textup{rec}}}$ and $N(\vect{s}) = \vect{E}_{\SS}(\vect{s}) \cdot \hat{\vect{p}}_{\text{\textup{rec}}}$, \eqref{eq:Stratton-Chu-equivalence} can be written, with the aid of some algebra, as follows: \vspace{-0.25cm}
	\begin{align}\label{eq:Stratton-Chu-equivalence-unified-2}
	\vect{E}(\vect{r}_{\textup{Rx}}) \cdot \hat{\vect{p}}_{\textup{rec}}
	&\mystepA   \mathbbm{1}_{(\vect{r}_{\textup{Tx}} \in V)} M(\vect{r}_{\textup{Rx}})   
	-  \int_{\partial V} \left[M(\vect{r}') \nabla_{\vect{r'}} G(\vect{r}_{\textup{Rx}},\vect{r'})
	- G(\vect{r}_{\textup{Rx}},\vect{r'}) \nabla_{\vect{r'}} M(\vect{r}')\right]\cdot \hat{\vect{n}}_{\textup{out}} d\vect{r'} \nonumber\\
	&
	-  \int_{\SS} \left[(N(\vect{s}) - M(\vect{s})) \nabla_{\vect{s}} G(\vect{r}_{\textup{Rx}},\vect{s}) - G(\vect{r}_{\textup{Rx}},\vect{s}) \nabla_{\vect{s}} (N(\vect{s}) - M(\vect{s})) \right]\cdot \hat{\vect{n}}_{\textup{out}} d\vect{s} \vspace{-0.25cm}
	\end{align}	 
where $(a)$ is obtained by taking into account that: (i) 	$\int_{\partial V} = \int_{ \partial V \setminus \SS} +   \int_{\SS}$, and (ii) $N(\vect{s}) - M(\vect{s})$ is the difference between the total electric field $\vect{E}_{\SS}(\vect{s})$ and the incident field $\vect{E}_{\textup{inc}}(\vect{s},\hat{\vect{p}}_{\textup{inc}})$ on $\SS$.

Let us consider $I_{\partial V} = -  \int_{\partial V} \left[M(\vect{r}') \nabla_{\vect{r'}} G(\vect{r}_{\textup{Rx}},\vect{r'}) - G(\vect{r}_{\textup{Rx}},\vect{r'}) \nabla_{\vect{r'}} M(\vect{r}')\right]\cdot \hat{\vect{n}}_{\textup{out}} d\vect{r'}$. By applying the divergence theorem \cite[Eq. (C.25)]{orfanidis2016electromagnetic} and the identity $\nabla_{\vect{r}} \cdot \left(f\nabla_{\vect{r}} g\right) = f\nabla_{\vect{r}}^2 g + \nabla_{\vect{r}}f\cdot\nabla_{\vect{r}}g$ \cite[Eq. (VII-46)]{balanis2016antenna} to generic scalar functions $f$ and $g$, $I_{\partial V}$ can be simplified as follows: \vspace{-0.25cm}
	\begin{align}\label{eq:aux-divergence-3}
	I_{\partial V} 
	&= - \int_{V}  \left[ M(\vect{r}) (\nabla^2_{\vect{r}} + k^2) G(\vect{r}_{\textup{Rx}},\vect{r}) - G(\vect{r}_{\textup{Rx}},\vect{r}) (\nabla^2_{\vect{r}} + k^2) M(\vect{r})\right] d\vect{r}
	\\
	&\mystepA 
	M(\vect{r}_{\textup{Rx}}) 
	+ \int_V G(\vect{r}_{\textup{Rx}},\vect{r}) (\nabla^2_{\vect{r}} + k^2) M(\vect{r}) d\vect{r} \nonumber \\
	&\mystepB
	M(\vect{r}_{\textup{Rx}}) 
	+ \int_V G(\vect{r}_{\textup{Rx}},\vect{r}) \left((\vec{\nabla}_{\vect{r}}^2 + k^2) \vect{E}_{\textup{inc}}(\vect{r};\hat{\vect{p}}_{\textup{inc}})\right)\cdot\hat{\vect{p}}_{\textup{rec}} d\vect{r} \nonumber
	\end{align} 
	where $(a)$ follows from \eqref{eq:Greens-function-Helmholtz} and from the fact that $\vect{r}_{\textup{Rx}}$ is always contained in $V$, and (b) follows from the identity $(\nabla_{\vect{r}}^2 + k^2)M(\vect{r}) = \left((\vec{\nabla}_{\vect{r}}^2 + k^2) \vect{E}_{\textup{inc}}(\vect{r};\hat{\vect{p}}_{\textup{inc}})\right)\cdot\hat{\vect{p}}_{\textup{rec}}$. 
	
	Consider the integral $I_V = \int_V G(\vect{r}_{\textup{Rx}},\vect{r}) \left((\vec{\nabla}_{\vect{r}}^2 + k^2) \vect{E}_{\textup{inc}}(\vect{r};\hat{\vect{p}}_{\textup{inc}})\right)\cdot\hat{\vect{p}}_{\textup{rec}} d\vect{r}$. From \eqref{eq:Helmholtz-equation-E} and by virtue of the identities $\nabla_{\vect{r}} \times \nabla_{\vect{r}} \times \vect{E}_{\textup{inc}}(\vect{r};\hat{\vect{p}}_{\textup{inc}}) = \nabla_{\vect{r}}(\nabla_{\vect{r}}\cdot \vect{E}_{\textup{inc}}(\vect{r};\hat{\vect{p}}_{\textup{inc}})) - \vec{\nabla}_{\vect{r}}^2\vect{E}_{\textup{inc}}(\vect{r};\hat{\vect{p}}_{\textup{inc}})$ \cite[Eq. (VII-51)]{balanis2016antenna} and $\nabla \cdot \vect{E}_{\textup{inc}}(\vect{r};\hat{\vect{p}}_{\textup{inc}}) = \rho(\vect{r},\vect{r}_{\textup{Tx}})/\epsilon_0$ \cite[Sec. 1.1]{orfanidis2016electromagnetic}, the integral $I_V$ simplifies to $I_V = \int_V G(\vect{r}_{\textup{Rx}},\vect{r}) (\nabla_{\vect{r}}\rho(\vect{r},\vect{r}_{\textup{Tx}})/\epsilon_0 + j\omega\mu_0 \vect{J}(\vect{r},\vect{r}_{\textup{Tx}}))\cdot\hat{\vect{p}}_{\textup{rec}} d\vect{r}$. By definition: (i) $I_V=0$ if Tx is not contained in $V$, and (ii) $I_V= -\vect{E}_{\textup{inc}}(\vect{r}_{\textup{Rx}};\hat{\vect{p}}_{\textup{inc}})\cdot\hat{\vect{p}}_{\textup{rec}}$ if Tx is contained in $V$ \cite[Eqs. (15.3.3), (15.3.6)]{orfanidis2016electromagnetic}. Thus, we have $I_V = -\mathbbm{1}_{(\vect{r}_{\textup{Tx}} \in V)} M(\vect{r}_{\textup{Rx}})$, and, from \eqref{eq:aux-divergence-3}, $I_{\partial V} = M(\vect{r}_{\textup{Rx}}) -\mathbbm{1}_{(\vect{r}_{\textup{Tx}} \in V)} M(\vect{r}_{\textup{Rx}})$. With the aid of some simplifications, the proof follows.

	\vspace{-0.5cm}
	\section*{Appendix C -- Proof of Lemma \ref{lemma:short-approximation-SPM}}\label{appendix:proof-of-SPM} \vspace{-0.25cm}
The proof is based on the application of the stationary phase method \cite[Appendix VIII]{balanis2016antenna}, \cite{Borovikov} to	\eqref{eq:integral-type-1} under the assumption of operating in the electrically-large regime, as stated in Definition \ref{definition:short-distance-3D}. Let $(x_s,y_s) \in \Psi$ be the stationary points of $\mathcal{P}(x,y) =d_{\textup{Tx}}(x,y)+d_{\textup{Rx}}(x,y) - \mathcal{C}(x,y)$ for $(x,y) \in \SS$. By invoking the stationary phase method, the integral $I_1$ in \eqref{eq:integral-type-1} oscillates very quickly outside a small region centered at $(x_s,y_s) \in \Psi$, and, thus, the contributions outside the small region around the stationary points cancel out when computing the integral \cite[pg. 923]{balanis2016antenna}. Under these conditions, $I_1$ can be well approximated by (i) replacing $\mathcal{P}(x,y)$ with its Taylor approximation evaluated at $(x_s,y_s) \in \Psi$, i.e., $\mathcal{P}(x,y) \approx \mathcal{P}(x_s,y_s) + A(x-x_s)^2  + B(y-y_s)^2 + C(x-x_s)(y-y_s)$ where $A = \frac{\partial^2}{\partial x^2}\mathcal{P}(x,y)|_{(x,y) = (x_s,y_s)}$, $B = \frac{\partial^2}{\partial y^2}\mathcal{P}(x,y)|_{(x,y) = (x_s,y_s)}$, and $C = \frac{\partial^2}{\partial x \partial y}\mathcal{P}(x,y)|_{(x,y) = (x_s,y_s)}$ and (ii) by letting the extremes of integration go to infinity, since $I_1$ is dominated by a small region around the stationary points and the contributions to the integral outside that small region cancel out. For simplicity, let us assume that a single stationary point exists. The case study with multiple stationary points is obtained by summing up the contributions from all the stationary points \cite[Sec. 1.3, pg. 15]{Borovikov}. Accordingly $I_1$, can be approximated as follows: \vspace{-0.25cm}
	\begin{align}\label{eq:integral-extended-2}
	I_1 
	&\approx \mathcal{A}_1(d_{\textup{Tx}}(x_s,y_s),d_{\textup{Rx}}(x_s,y_s))\mathcal{B}_1(x_s,y_s)e^{-jk\mathcal{P}(x_s,y_s)}\nonumber \\ & \quad \int_{-\infty}^{\infty} \int_{-\infty}^{\infty}  e^{-jk\left(A(x-x_s)^2 + C(x-x_s)(y-y_s) + B(x-x_s)^2\right)}dxdy
	\end{align} 

From \eqref{eq:integral-extended-2}, the proof	 follows from \cite[Eqs. (VIII-10)-(VIII-22)]{balanis2016antenna}.

	\vspace{-0.5cm}
	\section*{Appendix D -- Proof of Lemma \ref{lemma:short-approximation-nonSPM}}\label{appendix:proof-of-short-distance-approximation-nonSPM} \vspace{-0.25cm}
Define $\mathcal{Q}(x,y) = \mathcal{A}_1(d_{\textup{Tx}}(x,y),d_{\textup{Rx}}(x,y))\mathcal{B}_1(x,y)$. Since no stationary points lie in $\SS$, we can divide and multiply the integrand of \eqref{eq:integral-type-1} by ${\partial\mathcal{P}(x,y)}/{\partial x} \ne 0$. Thus,\eqref{eq:integral-type-1} can be written as: \vspace{-0.25cm}
	\begin{align}
	I_1
	&=  \int_{-L_y}^{L_y}\int_{-L_x}^{L_x} \frac{\mathcal{Q}(x,y)e^{-jk\mathcal{P}(x,y)}}{\partial\mathcal{P}(x,y)/\partial x}\frac{\partial\mathcal{P}(x,y)}{\partial x} dxdy \nonumber\\
	&\mystepA \frac{1}{(-jk)} \int_{-L_y}^{L_y} \left( \frac{\mathcal{Q}(x,y)e^{-jk\mathcal{P}(x,y)}}{\partial\mathcal{P}(x,y)/\partial x}\Big|^{x=L_x}_{x=-L_x} - \int_{-L_x}^{L_x} \frac{\partial}{\partial x}\left(\frac{\mathcal{Q}(x,y)}{\partial\mathcal{P}(x,y)/\partial x}\right)e^{-jk\mathcal{P}(x,y)}dx\right)dy \nonumber\\
	&\myapproxB \frac{1}{(-jk)}  \int_{-L_y}^{L_y} \left(\frac{\mathcal{Q}(L_x,y)e^{-jk\mathcal{P}(L_x,y)}}{\mathcal{P}_x(L_x,y)} \right)dy - \frac{1}{(-jk)}  \int_{-L_y}^{L_y} \left(\frac{\mathcal{Q}(-L_x,y)e^{-jk\mathcal{P}(-L_x,y)}}{\mathcal{P}_x(-L_x,y)} \right)dy \nonumber\\
	&\myapproxC \frac{1}{(-jk)^2}  \left[\frac{\mathcal{Q}(L_x,y)e^{-jk\mathcal{P}(L_x,y)}}{\mathcal{P}_x(L_x,y)\mathcal{P}_y(L_x,y)} - \frac{\mathcal{Q}(-L_x,y)e^{-jk\mathcal{P}(-L_x,y)}}{\mathcal{P}_x(-L_x,y)\mathcal{P}_y(-L_x,y)}\right] \Big|^{y=L_y}_{y=-L_y}
	\end{align}
where	$(a)$ is obtained by using integration by parts, $(b)$ follows by virtue of Riemann-Lebesgue's lemma, which states that the integral over $x$ decays with $1/k^2$, and, therefore, it can be ignored as compared with the first term \cite[Eqs. (3.21), (3.22)]{Wong}, \cite[Eq. (4.2)]{Lopez}, and $(c)$ follows by applying again the same procedure but by multiplying and dividing the two integrands in $(b)$ by ${\partial\mathcal{P}(\pm L_x,y)}/{\partial y} \ne 0$. The proof follows by iterating the same procedure once more.

	\vspace{-0.5cm}
	\section*{Appendix E -- Proof of Proposition \ref{proposition:reflected-field-general}} \label{appendix:proof-of-reflected-field-general} \vspace{-0.25cm}
	Consider \eqref{eq:Stratton-Chu-equivalence-reduced}. From Lemma \ref{lemma:incident-E-dipole}, $\vect{E}_{\textup{inc}}(\vect{r}_{\textup{Rx}};\hat{\vect{p}}_{\textup{inc}})\approx\vect{E}_{0,\textup{inc}}\left(\vect{r}_{\textup{Rx}};\hat{\vect{p}}_{\textup{inc}}\right) G\left(\vect{r}_{\textup{Rx}},\vect{r}_{\textup{Tx}}\right)$. From \eqref{eq:E-surface-reflection} and Lemma \ref{lemma:incident-E-dipole}, $\vect{E}_{\SS}(\vect{s}) =	\vect{E}_{\textup{inc}}(\vect{s};\hat{\vect{p}}_{\textup{inc}}) + \Gamma_{\textup{ref}}(\vect{s}) \mathcal{E}_{\textup{ref}}(\hat{\vect{p}}_{\textup{inc}}, \hat{\vect{p}}_{\textup{ref}}) \vect{E}_{\textup{inc}}(\vect{s};\hat{\vect{p}}_{\textup{ref}})$, with $\vect{E}_{\textup{inc}}(\vect{s};\hat{\vect{p}}_{\textup{inc}}) \approx \vect{E}_{0,\textup{inc}}\left(\vect{s};\hat{\vect{p}}_{\textup{inc}}\right)$ $G\left(\vect{s},\vect{r}_{\textup{Tx}}\right)$ and $ \vect{E}_{\textup{inc}}(\vect{s};\hat{\vect{p}}_{\textup{ref}}) \approx \vect{E}_{0,\textup{inc}}\left(\vect{s};\hat{\vect{p}}_{\textup{ref}}\right) G\left(\vect{s},\vect{r}_{\textup{Tx}}\right)$. By inserting them in \eqref{eq:Stratton-Chu-equivalence-reduced}, we obtain: \vspace{-0.25cm}
		\begin{align}\label{eq:Stratton-Chu-equivalence-reflection}
	\vect{E}(\vect{r}_{\textup{Rx}}) \cdot \hat{\vect{p}}_{\textup{rec}}
	&\approx
	\hat{\vect{p}}_{\textup{rec}} \cdot \vect{E}_{0,\textup{inc}}\left(\vect{r}_{\textup{Rx}};\hat{\vect{p}}_{\textup{inc}}\right) G(\vect{r}_{\textup{Rx}},\vect{r}_{\textup{Tx}}) \\
	&\quad  -  \int_{\SS} \left[ \mathcal{F}_{\textup{ref}}(\vect{s}) G\left(\vect{s},\vect{r}_{\textup{Tx}}\right) \nabla_{\vect{s}} G(\vect{r}_{\textup{Rx}},\vect{s})  - G(\vect{r}_{\textup{Rx}},\vect{s}) \nabla_{\vect{s}} \left(\mathcal{F}_{\textup{ref}}(\vect{s}) G\left(\vect{s},\vect{r}_{\textup{Tx}}\right)\right)\right] \cdot \hat{\vect{n}}_{\textup{out}} d\vect{s} \nonumber
	\end{align}
	where $\mathcal{F}_{\textup{ref}}(\vect{s}) =  \Gamma_{\textup{ref}}(\vect{s}) \mathcal{E}_{\textup{ref}}(\hat{\vect{p}}_{\textup{inc}}, \hat{\vect{p}}_{\textup{ref}}) \vect{E}_{0,\textup{inc}}\left(\vect{s};\hat{\vect{p}}_{\textup{ref}}\right) \cdot \hat{\vect{p}}_{\textup{rec}} = \Gamma_{\textup{ref}}(\vect{s})\Omega_{\textup{ref}}(x,y;\hat{\vect{p}}_{\textup{ref}},\hat{\vect{p}}_{\textup{rec}})e^{j(\phi_{\textup{ref}}+\phi_{\textup{rec}})}$ for $(x,y) \in \SS$, where $\Omega_{\textup{ref}}(\cdot)$ is defined in the statement of Proposition \ref{proposition:reflected-field-general}.
	
In the reflection case, $\hat{\vect{n}}_{\textup{out}} = - \hat{\vect{z}}$. Thus, by using the product rule of derivatives, we have 
$\hat{\vect{n}}_{\textup{out}} \cdot \nabla_{\vect{s}} \left(\mathcal{F}_{\textup{ref}}(\vect{s}) G\left(\vect{s},\vect{r}_{\textup{Tx}}\right)\right) = \mathcal{Z}_1 + \mathcal{Z}_2$, where $\mathcal{Z}_1 = - \mathcal{F}_{\textup{ref}}(\vect{s}) \frac{\partial}{\partial z} G\left(\vect{s},\vect{r}_{\textup{Tx}}\right)$ and $\mathcal{Z}_2 = - G\left(\vect{s},\vect{r}_{\textup{Tx}}\right) \frac{\partial}{\partial z} \mathcal{F}_{\textup{ref}}(\vect{s})$. By computing the derivatives, it can be shown that $\mathcal{Z}_1 \propto \frac{k^3}{|\vect{s}-\vect{r}_{\textup{Tx}}|}G\left(\vect{s},\vect{r}_{\textup{Tx}}\right)$ and $\mathcal{Z}_2 \propto \frac{k^2}{|\vect{s}-\vect{r}_{\textup{Tx}}|}G\left(\vect{s},\vect{r}_{\textup{Tx}}\right)$. Under the assumption $k \gg {1}/{\left|\vect{s} - \vect{r}_{\textup{Tx}}\right|}$, $\mathcal{Z}_1$ dominates $\mathcal{Z}_2$ and hence $\hat{\vect{n}}_{\textup{out}} \cdot \nabla \left(\mathcal{F}_{\textup{ref}}(\vect{s}) G\left(\vect{s},\vect{r}_{\textup{Tx}}\right)\right) \approx \mathcal{Z}_1 = - \mathcal{F}_{\textup{ref}}(\vect{s}) \frac{\partial}{\partial z} G\left(\vect{s},\vect{r}_{\textup{Tx}}\right)$. Therefore, \eqref{eq:Stratton-Chu-equivalence-reflection} can be simplified as follows: \vspace{-0.25cm}
	\begin{align}\label{eq:prefinal-reflection-1}
	&\vect{E}(\vect{r}_{\textup{Rx}})\cdot \hat{\vect{p}}_{\textup{rec}} 
	\approx
	\hat{\vect{p}}_{\textup{rec}} \cdot \vect{E}_{0,\textup{inc}}\left(\vect{r}_{\textup{Rx}};\hat{\vect{p}}_{\textup{inc}}\right) G(\vect{r}_{\textup{Rx}},\vect{r}_{\textup{Tx}}) 
	\\
	&- \int_{\SS} \mathcal{F}_{\textup{ref}}(\vect{s}) \left[  G\left(\vect{s},\vect{r}_{\textup{Tx}}\right) \nabla_{\vect{s}} G(\vect{r}_{\textup{Rx}},\vect{s}) - G(\vect{r}_{\textup{Rx}}, \vect{s}) \nabla_{\vect{s}} G\left(\vect{s},\vect{r}_{\textup{Tx}}\right)\right] \cdot \hat{\vect{n}}_{\textup{out}} d\vect{s} \nonumber \\
	&\myapproxA
	\hat{\vect{p}}_{\textup{rec}} \cdot \vect{E}_{0,\textup{inc}}\left(\vect{r}_{\textup{Rx}};\hat{\vect{p}}_{\textup{inc}}\right)G(\vect{r}_{\textup{Rx}},\vect{r}_{\textup{Tx}}) + jk \int_{\SS}  \mathcal{F}_{\textup{ref}}(\vect{s})G\left(\vect{s},\vect{r}_{\textup{Tx}}\right)G(\vect{r}_{\textup{Rx}},\vect{s}) \left[\frac{z_{\textup{Rx}}}{|\vect{s} - \vect{r}_{\textup{Rx}}|}  + \frac{z_{\textup{Tx}}}{|\vect{s} - \vect{r}_{\textup{Tx}}|} \right] d\vect{s} \nonumber
	\end{align}  
	where $(a)$ is obtained by taking into account that $\hat{\vect{n}}_{\textup{out}}= -\hat{\vect{z}}$, and, hence, by definition:\vspace{-0.35cm}
\begin{equation}\label{eq:dot-product-Green-1} 
	\nabla_{\vect{s}} G(\vect{s},\vect{r}_{\textup{Tx}}) \cdot \hat{\vect{n}}_{\textup{out}} \mystepB + \frac{\partial}{\partial z} G(\vect{s},\vect{r}_{\textup{Tx}})_{|z=0} \myapproxD jkG(\vect{s},\vect{r}_{\textup{Tx}})\frac{z_{\textup{Tx}}}{|\vect{s} - \vect{r}_{\textup{Tx}}|} \vspace{-0.35cm}
	\end{equation} \vspace{-0.35cm}
	\begin{equation}\label{eq:dot-product-Green-2}
	\nabla_{\vect{s}} G(\vect{r}_{\textup{Rx}},\vect{s}) \cdot \hat{\vect{n}}_{\textup{out}} \mystepC - \frac{\partial}{\partial z} G(\vect{r}_{\textup{Rx}},\vect{s})_{|z=0} \myapproxD  -jkG(\vect{r}_{\textup{Rx}},\vect{s}) \frac{z_{\textup{Rx}}}{|\vect{r}_{\textup{Rx}}- \vect{s}|} \vspace{-0.35cm}
		\end{equation}
where the ``+'' sign in $(b)$ and the ``-'' sign in $(c)$ take into account that the direction of propagation of the incident and reflected signals point towards the same and the opposite directions with respect to $\hat{\vect{n}}_{\textup{out}}$, respectively, and the approximations in $(d)$ take into account that ${1}/{|\vect{s} - \vect{r}_{\textup{Tx}}|} \ll k$ and ${1}/{|\vect{r}_{\textup{Rx}} - \vect{s}|} \ll k$. This completes the proof.

	\vspace{-0.5cm}
	\section*{Appendix F -- Proof of Corollary \ref{corollary:electrically-large-uniform-reflection}}
	\label{appendix:proof-of-electrically-large-uniform-reflection} \vspace{-0.25cm}
Consider \eqref{eq:Ex-surface-contribution-reflection}. The proof is based on the stationary phase method stated in Lemma \ref{lemma:short-approximation-SPM}. According to Definition \ref{definition:stationarypoints}, the stationary points of $\mathcal{P}(x,y)= \mathcal{P}_R(x,y)$ in \eqref{eq:Ex-surface-contribution-reflection-A} correspond to the solutions of \eqref{eq:condition-uniform-reflection}. Due to the monotonicity of \eqref{eq:condition-uniform-reflection} with respect to $x_s$ and $y_s$, either a single or no stationary point exists. More precisely, \eqref{eq:condition-uniform-reflection} can be equivalently re-written as follows:\vspace{-0.25cm}
\begin{align}
&\sin\theta_{\textup{inc}}(x_s,y_s)\cos\varphi_{\textup{inc}}(x_s,y_s) = -\sin\theta_{\textup{rec}}(x_s,y_s)\cos\varphi_{\textup{rec}}(x_s,y_s)  \\
& \sin\theta_{\textup{inc}}(x_s,y_s)\sin\varphi_{\textup{inc}}(x_s,y_s) = -\sin\theta_{\textup{rec}}(x_s,y_s)\sin\varphi_{\textup{rec}}(x_s,y_s) \nonumber \vspace{-0.35cm}
	\end{align}
 which, using some algebra, yields $\varphi_{{\textup{inc}}}(x_s,y_s) = (\varphi_{{\textup{rec}}}(x_s,y_s) + \pi) \mod 2\pi$ and $\theta_s = \theta_{\textup{inc}}(x_s,y_s) = \theta_{\textup{rec}}(x_s,y_s)$ (i.e., the law of reflection). Based on Lemma \ref{lemma:short-approximation-SPM} with $\mathcal{P}(x,y)= \mathcal{P}_R(x,y)$, the determinant of $\A(x_s,y_s)$ is $\det(\A(x_s,y_s)) = P_{xx}P_{yy} - (P_{xy})^2 = \cos^2\theta_s\left(1/d_{\textup{Tx}}(x_s,y_s) + 1/d_{\textup{Rx}}(x_s,y_s)\right)^2$, where the derivatives are $P_{xx} = \frac{\partial^2}{\partial x^2}\mathcal{P}_R(x,y)\Huge|_{(x,y)=(x_s,y_s)}$, $P_{yy} = \frac{\partial^2}{\partial y^2}\mathcal{P}_R(x,y)\Huge|_{(x,y)=(x_s,y_s)}$, and $P_{xy} = \frac{\partial^2}{\partial x \partial y}\mathcal{P}_R(x,y)\Huge|_{(x,y)=(x_s,y_s)}$. In addition, it can proved that $(P_{xy})^2 < P_{xx}P_{yy}$, which implies that the two eigenvalues of $\A(x_s,y_s)$ are positive and distinct. Therefore, we obtain $\sign(\A(x_s,y_s)) = 2$. The proof follows by inserting $\det(\A(x_s,y_s))$ and $\sign(\A(x_s,y_s))$ in \ref{lemma:short-approximation-SPM}.

%	\vspace{-0.5cm}
%	\section*{Appendix G -- Proof of Corollary \ref{corollary:electrically-large-anomalous-general-reflection-scaling}}	
%	\label{appendix:proof-of-electrically-large-anomalous-general-reflection-scaling} \vspace{-0.25cm}

	\vspace{-0.5cm}
	\section*{Appendix G -- Proof of Proposition \ref{proposition:transmitted-field-general}} \label{appendix:proof-of-transmitted-field-general} \vspace{-0.25cm}
	Consider \eqref{eq:Stratton-Chu-equivalence-reduced}. From Lemma \ref{lemma:incident-E-dipole}, $\vect{E}_{\textup{inc}}(\vect{r}_{\textup{Rx}};\hat{\vect{p}}_{\textup{inc}})\approx\vect{E}_{0,\textup{inc}}\left(\vect{r}_{\textup{Rx}};\hat{\vect{p}}_{\textup{inc}}\right) G\left(\vect{r}_{\textup{Rx}},\vect{r}_{\textup{Tx}}\right)$. From \eqref{eq:E-surface-transmission} and Lemma \ref{lemma:incident-E-dipole}, $\vect{E}_{\SS}(\vect{s}) = \Gamma_{\textup{tran}}(\vect{s}) \mathcal{E}_{\textup{tran}}(\hat{\vect{p}}_{\textup{inc}}, \hat{\vect{p}}_{\textup{tran}}) \vect{E}_{\textup{inc}}(\vect{s};\hat{\vect{p}}_{\textup{tran}})$, with $ \vect{E}_{\textup{inc}}(\vect{s};\hat{\vect{p}}_{\textup{tran}}) \approx \vect{E}_{0,\textup{inc}}\left(\vect{s};\hat{\vect{p}}_{\textup{tran}}\right) G\left(\vect{s},\vect{r}_{\textup{Tx}}\right)$. By inserting them in \eqref{eq:Stratton-Chu-equivalence-reduced}, we obtain: \vspace{-0.25cm}
	\begin{align}\label{eq:Stratton-Chu-equivalence-transmission}
	\vect{E}(\vect{r}_{\textup{Rx}}) \cdot \hat{\vect{p}}_{\textup{rec}}
	&\approx
	\hat{\vect{p}}_{\textup{rec}} \cdot \vect{E}_{0,\textup{inc}}\left(\vect{r}_{\textup{Rx}};\hat{\vect{p}}_{\textup{inc}}\right) G(\vect{r}_{\textup{Rx}},\vect{r}_{\textup{Tx}}) \\
	&\quad  -  \int_{\SS} \left[ \mathcal{F}_{\textup{tran}}(\vect{s}) G\left(\vect{s},\vect{r}_{\textup{Tx}}\right) \nabla_{\vect{s}} G(\vect{r}_{\textup{Rx}},\vect{s})  - G(\vect{r}_{\textup{Rx}},\vect{s}) \nabla_{\vect{s}} \left(\mathcal{F}_{\textup{tran}}(\vect{s}) G\left(\vect{s},\vect{r}_{\textup{Tx}}\right)\right)\right] \cdot \hat{\vect{n}}_{\textup{out}} d\vect{s} \nonumber
	\end{align}
	where $\mathcal{F}_{\textup{tran}}(\vect{s}) =  (\Gamma_{\textup{tran}}(\vect{s}) \mathcal{E}_{\textup{tran}}(\hat{\vect{p}}_{\textup{inc}}, \hat{\vect{p}}_{\textup{tran}}) \vect{E}_{0,\textup{inc}}\left(\vect{s};\hat{\vect{p}}_{\textup{tran}}\right) - \vect{E}_{0,\textup{inc}}\left(\vect{s};\hat{\vect{p}}_{\textup{inc}}\right) ) \cdot \hat{\vect{p}}_{\textup{rec}}$, which can be formulated in terms of $\Omega_{\textup{inc}}(\cdot)$ and $\Omega_{\textup{tran}}(\cdot)$ as defined in the statement of Proposition \ref{proposition:transmitted-field-general}.
	
The rest of the proof is similar to Appendix E. The difference is that $\hat{\vect{n}}_{\textup{out}}=\hat{\vect{z}}$, and, hence, the signs in $(b)$ and $(c)$ that correspond to \eqref{eq:dot-product-Green-1} and \eqref{eq:dot-product-Green-2} are both negative because the direction of propagation of the incident and transmitted signals is opposite to $\hat{\vect{n}}_{\textup{out}}$.

	%\end{appendices}


\vspace{-0.5cm}
	\begin{thebibliography}{99}\vspace{-0.35cm}
	
\bibitem{Marco-JSAC}
M. Di Renzo et al., ``Smart radio environments empowered by reconfigurable intelligent surfaces: How it works, state of research, and the road ahead'', \textit{IEEE J. Sel. Areas Commun.}, to appear. Available: ArXiv:2004.09352.

\bibitem{Docomo_Glass} NTT DOCOMO, ``DOCOMO conducts world's first successful trial of transparent dynamic metasurface'', Jan. 2020. [Online]. Available: https://www.nttdocomo.co.jp/english/info/media\_center/pr/2020/0117\_0 0.html.


\bibitem{Tang-RIS} 
W. Tang et al., ``Wireless communications with reconfigurable intelligent surface: Path loss modeling and experimental measurement'', \textit{IEEE Trans. Wireless Commun.}, submitted. Available: ArXiv:1911.05326.
	
\bibitem{Bucheli-RIS-bridging} 
J. C. Bucheli et al., ``Reconfigurable intelligent surfaces: Bridging the gap between scattering and reflection'', \textit{IEEE J. Sel. Areas Commun.}, to appear. Available: ArXiv:1912.05344.
	
\bibitem{Ellingson-Path-loss} 
S. W. Ellingson, ``Path loss in reconfigurable intelligent surface-enabled channels'', submitted. Available: ArXiv:1912.06759.
	
\bibitem{Khawaja-Coverage} 
W. Khawaja et al., ``Coverage enhancement for NLOS mmWave links using passive reflectors'', \textit{IEEE Open Access J. Commun. Society}, vol. 1, pp. 263-281, Jan. 2020.

\bibitem{Emil} 
O. Ozdogan et al., ``Intelligent reflecting surfaces: Physics, propagation, and pathloss modeling'', \textit{IEEE Wireless Commun. Lett.}, vol. 9, no. 5, pp. 581-585, May 2020.

\bibitem{Schober} 
M. Najafi et al., ``Physics-based modeling and scalable optimization of large intelligent reflecting surfaces'', submitted. Available: ArXiv:2004.12957.

\bibitem{Davide} 
D. Dardari, ``Communicating with large intelligent surfaces: Fundamental limits and models'', \textit{IEEE J. Sel. Areas Commun.}, to appear. Available: ArXiv:1912.01719.

\bibitem{Marco-SPAWC} 
M. Di Renzo et al., ``Analytical modeling of the path-loss for reconfigurable intelligent surfaces - Anomalous mirror or scatterer ?'', \textit{IEEE SPAWC 2020}. Available: ArXiv:2001.10862.

\bibitem{Green_1828} G. Green, ``An essay on the application of mathematical analysis to the theories of electricity and magnetism'', 1828.




\bibitem{stratton1939diffraction}
J. A. Stratton and L. J. Chu, ``Diffraction theory of electromagnetic waves'', \textit{Physical Review}, vol. 56, pp. 99-107, 1939.

\bibitem{orfanidis2016electromagnetic}
S. J. Orfanidis, \textit{Electromagnetic Waves and Antennas}, Rutgers University, 2016.

\bibitem{osipov2017modern}
A. V. Osipov and S. A. Tretyakov, \textit{Modern Electromagnetic Scattering Theory with Applications}, John Wiley, 2017.

\bibitem{tai1972kirchhoff}
C. T. Tai, ``Kirchhoff theory: Scalar, vector, or dyadic?'', \textit{IEEE Trans. Antennas Propag.}, vol. 20, pp. 114-115, Jan. 1972.

\bibitem{treuhaft2011formulating}
R. N. Treuhaft et al., ``Formulating a vector wave expression for polarimetric GNSS surface scattering'', \textit{PIER B}, 2011.
%\textit{Progress in Electromagnetics Research B}, vol. 33, pp. 257-276, 2011.

\bibitem{balanis2016antenna}
C. A. Balanis, \textit{Antenna Theory: Analysis and Design}, John Wiley, 2016.



\bibitem{Borovikov}
V. A. Borovikov, \textit{Uniform Stationary Phase Method}, Inst of Engineering \& Technology, 1994.

\bibitem{Wong}
R. Wong, \textit{Uniform Stationary Phase Method}, SIAM, 2001.

\bibitem{Lopez}
J. L. Lopez et al.,  ``Asymptotic approximations of integrals: An introduction, with recent developments and applications to orthogonal polynomials'', \textit{Electronic Trans. Numerical Analysis}, vol. 19, pp. 58-83, 2005.



\end{thebibliography}
\end{document}